\documentclass[sn-mathphys,Numbered]{sn-jnl}

\usepackage{graphicx}%
\usepackage{multirow}%
\usepackage{amsmath,amssymb,amsfonts}%
\usepackage{amsthm}%
\usepackage{mathrsfs}%
\usepackage[title]{appendix}%
\usepackage{xcolor}%
\usepackage{textcomp}%
\usepackage{manyfoot}%
\usepackage{booktabs}%
\usepackage{algorithm}%
\usepackage{algorithmicx}%
\usepackage{algpseudocode}%
\usepackage{listings}%
\RequirePackage{graphicx,subfigure,placeins,physics,bm,dcolumn}
\usepackage{amssymb}
\usepackage{amsbsy}
\usepackage{physics}
\usepackage{blindtext}
\usepackage{etoolbox}
\makeatletter
\def\@authorcontrib{}        
\def\@print@authorcontrib{}  
\let\@equalconttext\@empty   
\makeatother

\begin{document}

\title{\textbf{Cosmological Singularities in Brane Gravity}}

\author[1]{\fnm{R.} \sur{Jalalzadeh}}\email{r.jalalzadeh@uok.ac.ir}

\author[2,3]{\fnm{S.} \sur{Jalalzadeh}}\email{shahram.jalalzadeh@ufpe.br}

\author[4]{\fnm{Y.} \sur{Heydarzade}}\email{yheydarzade@bilkent.edu.tr}


\affil[1]{\orgdiv{Department of Physics}, \orgname{University of Kurdistan}, \orgaddress{\street{Pasdaran St.}, \city{Sanandaj},   \country{Iran}}}

\affil[2]{\orgdiv{Departamento de F\'{i}sica}, \orgname{Universidade Federal de Pernambuco}, \orgaddress{\city{Recife}, \postcode{50670-901}, \state{PE}, \country{Brazil}}}

\affil[3]{\orgdiv{Department of Physics and Technical Sciences}, \orgname{Western Caspian University}, \orgaddress{
\street{AZ 1001},
\city{Baku}, 
\country{Azerbaijan}}}

\affil[4]{\orgdiv{Department of Mathematics, Faculty of Sciences}, \orgname{Bilkent University}, \orgaddress{ \city{Ankara}, \postcode{06800}, \country{Turkey}}}


\abstract{We present a comprehensive study of cosmological singularities within the framework of Covariant Extrinsic Gravity (CEG), addressing both the initial Big Bang singularity and potential finite-time future singularities. Through detailed analysis of the emergent universe scenario, we systematically examine homogeneous and inhomogeneous perturbations (encompassing scalar, vector, and tensor modes) in a 4D FLRW brane geometry. Our work establishes rigorous existence criteria and stability conditions for a nonsingular Einstein static initial state, demonstrating that such a configuration remains stable for well-defined parameter ranges in CEG - thereby providing a compelling resolution to the long-standing initial singularity problem. Extending our analysis to late-time cosmology, we perform a complete classification of future singularity types following Barrow et al.'s formalism, deriving precise conditions that determine whether the universe in CEG evolves toward or avoids these singular states. 
}

\maketitle


\section{Introduction}

The unresolved matter of the singularity problem within general relativity (GR) is a subject of utmost significance. Following the Penrose--Hawking theorem \cite{Penrose:1964wq, Hawking:1967ju}, the existence of singularities in space-time is indicated by the incomplete nature of geodesics in GR manifolds. Singularities in GR materialize in various forms, encompassing the singularities at the beginning and end of a closed universe, and the singularity at the core of a black hole. These singularities pose substantial challenges to researchers due to the inability to anticipate the occurrences that transpire after the big crunch, or the existence of anything before the Big Bang in the first scenario. It has been postulated that there is a boundary to classical space-time beyond which standard GR cannot be employed, as evidenced by research \cite{Borde:1996pt, Borde:2001nh}. In this setting, singularities frequently recognized by divergences in scalar invariants of curvature or torsion tensors, or by the collapse of geodesics at a particular point--bear enormous significance. Examining the Singularity at the beginning can reveal information about whether or not our universe has a beginning.

In the framework of the standard model of cosmology, which incorporates the concept of inflation, the fundamental problem of the initial singularity remains unresolved. However, numerous pre-inflationary scenarios have been introduced to tackle this issue. These scenarios encompass emergent universes (EU) \cite{Ellis:2002we,Ellis:2003qz} and cyclic/ekpyrotic scenarios \cite{Khoury:2001wf,Steinhardt:2001st,Khoury:2003rt,Barrow:2004ad}, among others. Many of these scenarios are characterized by a non-singular or past eternal nature. The considered scenarios possess three notable attributes. Firstly, the absence of an initial singularity implies the absence of a temporal origin and an infinite past. Secondly, the horizon problem is eliminated. Lastly, the quantum gravity era is non-existent. An intriguing cosmological model was developed by Ellis et al. \cite{Ellis:2002we, Ellis:2003qz} to remove the initial singularity within the framework of GR. Their model suggests that the universe originated from an initially static state known as the Einstein Static Universe (ESU) in the eternal past. This model adheres to the standard energy conditions while evading the singularity of the Big Bang. According to this model, the closed static universe existed indefinitely but eventually underwent a phase of inflation \cite{Mukherjee:2006ds}. Two conditions must be met for a gravitational theory to align with the EU paradigm. Firstly, the presence and stability of the ESU are necessary. Secondly, an acceptable transition away from the ESU is required, allowing for a connection with the history of the standard model of cosmology via an elegant pathway.
The potential resolution of the instability of ESU, as remarked upon in \cite{Barrow:2004ad}, can conceivably be addressed through the utilization of modified gravity models. A variety of models, such as loop quantum gravity \cite{Mulryne:2005ef,Parisi:2007kv}, $f(R)$ \cite{Goheer:2008tn,Khodadi:2015fav}, $f(T)$ \cite{Wu:2011xa}, Einstein--Cartan theory \cite{Boehmer:2003iv}, Massive gravity \cite{Parisi:2012cg}, and braneworld models inspired by string theory \cite{Lidsey:2006md,Atazadeh:2014xsa,Heydarzade:2015tua}, have been thoroughly examined within the framework of ESU.

On the other hand, further investigation into the characteristics of dark energy has resulted in the discovery of distinct forms of singularities that go beyond the initial  Big Bang singularity. These future singularities are distinguished by violations of specific energy conditions. Notable among them are finite-time singularities that are anticipated to occur in the future, such as the Big Rip singularity \cite{Nojiri:2005sx}, Sudden singularity \cite{Lake:2004fu}, Big Freeze singularity \cite{Bouhmadi-Lopez:2006fwq}, and Big Brake singularity \cite{Chimento:2015gum}. There are also classifications for potential infinite-time singularities, see for instance \cite{Boko:2021opx}.

Our aim in this study is to investigate the predicament concerning cosmic singularities in both the past and future within the framework of Covariant Extrinsic Gravity CEG \cite{Maia:2004fq, Jalalzadeh:2013wza, Rostami:2015ixa, Jalalzadeh:2023upb, Jalalzadeh:2024mip}.
{Theoretical models with large extra dimensions have garnered considerable attention in recent years. These models suggest that the observable universe is situated within a brane in a higher-dimensional space-time bulk. The concept of extra dimensions originated from the influential works of Kaluza and Klein, published in 1921 and 1926, respectively \cite{Kaluza:1921tu, 1926ZPhy95K}. Their pioneering proposals implied the existence of compact spatial dimensions beyond the familiar three dimensions of space and one dimension of time, revolutionizing theoretical physics. Notably, the Space-Time-Matter theory introduced by Wesson \cite{wesson-1998, Doroud:2009zza, Jalalzadeh:2008xu} presents a model where the fifth dimension is non-compact, and the geometry of this extra dimension induces matter fields. Various classifications of brane universes have been put forward, including those by Arkani--Hamed, Dimopoulos, and Dvali, to address the hierarchy problem \cite{Arkani-Hamed:1998jmv, Arkani-Hamed:1998sfv}. Additionally, the Randall--Sundrum model developed a warped brane world to localize gravity \cite{Randall:1999vf}, while another braneworld model with an extra dimension of the infinite volume provides a solution to the recent accelerated expansion of the universe \cite{Dvali:2000hr}.} {Despite the numerous accomplishments of various brane models, they are also accompanied by significant limitations. For instance, these models rely on junction conditions to establish a connection between the extrinsic curvature and the matter fields on the brane. However, varying junction conditions can result in diverse physical consequences \cite{Battye:2001pb}. Furthermore, the compression of internal space can lead to issues with the masses of fermions in nonzero modes and pose challenges to vacuum stability \cite{Arkani-Hamed:1999ylh}. Additionally, concerns regarding the thickness of these models introduce further complexities \cite{Dzhunushaliev:2009va}.}

{Within the CEG braneworld model \cite{Maia:2004fq, Jalalzadeh:2013wza, Rostami:2015ixa, Jalalzadeh:2023upb, Jalalzadeh:2024mip}, our $4D$ universe is embedded in a bulk space that has an arbitrary number of non-compact extra dimensions. As such, there is no application of the junction condition. After consulting Ref. \cite{Jalalzadeh:2023upb}, the characteristics of this model can be summarized as follows:
\begin{enumerate}
    \item The Yang-Mills fields have a geometric origin, just as the original Kaluza-Klein theories. Like its forebears, the CEG investigates the complex interplay between geometry and physical phenomena, offering an extensive perspective on the fundamental processes at work in the universe.  
    \item The direct correlation between the brane thickness and the muon Compton wavelength highlights the intricate relationship between bulk dimensions and particle physics.
    \item An additional geometric term, $Q_{\mu\nu}$ in Eq. (\ref{eq17}), is present in the induced gravitational field equations and is posited to play a role in the current acceleration of the universe. This term is denoted as geometric dark energy (GDE). Moreover, it has been demonstrated in Refs. \cite{Jalalzadeh:2005ax, Jalalzadeh:2004uv} that this geometric term holds significance as it functions as an induced potential in the Klein--Gordon equation for particles confined to the brane.
    \item The energy density of GDE at the present epoch is determined by several factors, including the number of extra dimensions, $\mathfrak N$, $4D$ gravitational constant $G_N$, the fine structure constant $\alpha$, and the muon mass, $m_\mu$. This relationship is represented by the equation $\rho^\text{(GDE)}_0=gG_Nm_\mu^6/\alpha^2$. It not only helps to explain Zeldovich's \cite{Zeldovich:1967gd}  and Weinberg's \cite{Weinberg:1972kfs} well-known empirical formulas regarding vacuum energy and fundamental particle mass but also leads to a deeper comprehension of the cosmic landscape and its underlying principles. By intertwining geometry, field theory, and cosmology, these theories provide a unified framework that enriches our understanding of the universe's intricate evolution and the forces that govern its dynamics over cosmic scales.
    \item In general, the $4D$ gravitational `constant' is variable because it relies on the extrinsic curvature radii. Thus, it might vary with cosmic time within the isotropic and homogenous cosmological framework. 
    \item The CEG model predicts 22 extra dimensions, exactly as the original bosonic string theory did, even if these extra dimensions are non-compact. Important conclusions were also drawn from the model analysis, such as the prediction that the induced cosmological constant (CC) determines the graviton's mass. Remarkably, when the model is configured with an extra dimension count of $\mathfrak N=22$, the induced values for the present energy density of dark energy, the deceleration parameter at the current epoch, the redshift at the transition epoch, and the age of the Universe coincide with observable data. 
\end{enumerate}
}


In the present study, we shall examine the CEG theory by adopting $\mathfrak{N}=22$ non-compact extra dimensions, a choice motivated by the model's alignment with observational data as discussed in \cite{Jalalzadeh:2023upb}. However, this selection is not fixed---the GDE framework remains flexible, allowing for investigations of singularity problems with an arbitrary number of extra dimensions. Firstly, we investigate the ESU's stability versus homogeneous scalar perturbations, as well as inhomogeneous vector and tensor perturbations.
Next, we examine different types of finite-time future singularities in finite time, including the Big Rip, Sudden, Big Freeze, and Big Brake singularities, and discuss how CEG could potentially avoid these singularities. 
It is noteworthy to highlight that in order for the universe to transition from the initial static/quasi-static phase to progress into the radiation and subsequently to the matter-dominated phases, a superinflation phase must precede the onset of slow-roll inflation. This particular proposition is thoroughly explored in the research conducted by Labrana \cite{Labrana:2013oca} where they delve into a superinflation model situated within the emergent universe framework. The author explicitly emphasized that the presence of a superinflating phase occurring prior to the initiation of slow-roll inflation is a common occurrence in various emergent universe models. The rationale behind this concept is supported by the fact that the superinflationary period in the emergent universe scenario leads to a suppression of the Cosmic Microwave Background (CMB) anisotropies on a large scale, a phenomenon that could potentially explain the observed lack of power at large angular scales of the CMB. Furthermore, additional references such as \cite{Bastero-Gil:2003hfz, Biswas:2013dry, Huang:2022hye, Rios:2016trs} also contribute to the discourse surrounding this topic.

This paper is structured as follows: We will begin with a brief review of the CEG theory and its field equations in Section \ref{Geometry}. Next, in Section \ref{field}, we will derive the generalized Friedmann equations for an FLRW universe embedded isometrically in a bulk space with an arbitrary number of extra dimensions. In Section \ref{ESU}, we will investigate the initial singularity problem. In section \ref{Living}, we discuss how a simple inflation model by a scalar field provides the transition from the static state to the subsequent expanding era. In Section \ref{FS}, we shall study the theory versus the potential future singularities. Lastly, we will conclude the paper with a discussion summarized in Section \ref{Con}.

\section{Covariant extrinsic gravity Theory}\label{Geometry}

\begin{figure}[ht]
  \centering
   \includegraphics[width=8cm]{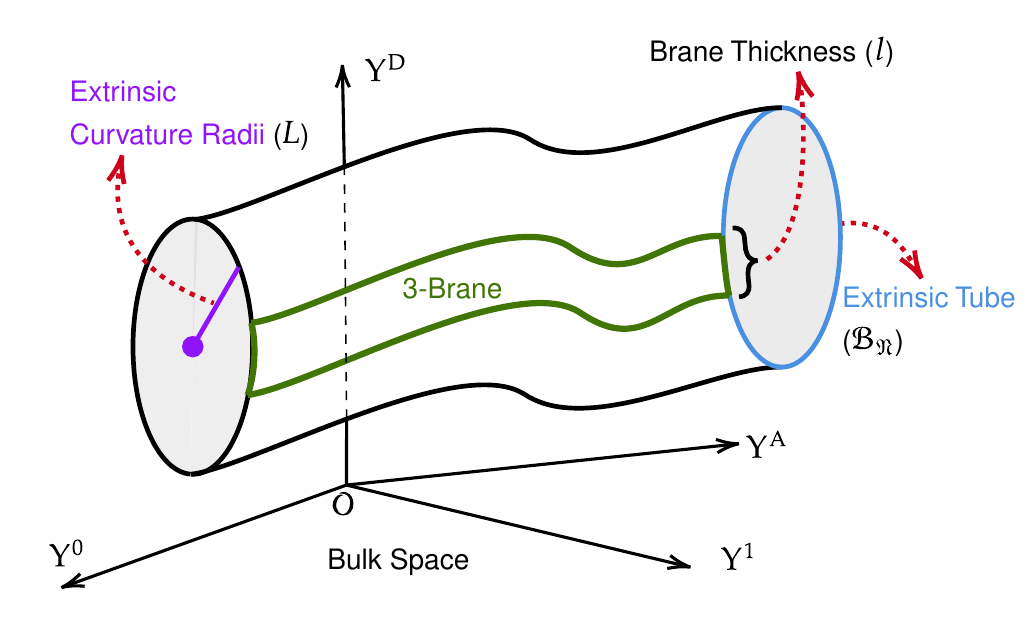}
  \caption{\small  Braneworld with thickness $l$ bounded by an extrinsic tube formed by the normal curvature radii $L$. Matter particles are confined to the line (representing the 3-brane) of length $l$ at each point.}\label{BraneR}
\end{figure}

Let us take a moment to briefly examine and analyze the fundamental concepts of CEG and the subsequent outcomes and implications that arise from it. The foundation of CEG, as outlined in Refs. \cite{Jalalzadeh:2013wza, Rostami:2015ixa, Jalalzadeh:2023upb}, begins with the assumption that the braneworld, characterized by its thickness denoted as $l$, is isometrically embedded in a flat bulk space of $D$ dimensions. Consequently, if we were to designate the local coordinates on the brane as $x^\mu$ (where $\mu$ ranges from 0 to 3) and the extra coordinates as $x^a$ (where $a$ spans from 4 to $D$), it becomes feasible to represent the metric of the bulk space, parameterized by Gaussian coordinates $\{x^\mu,x^a\}$, in a concise matrix format
\begin{equation}\label{sh7}
  \mathcal{G}_{AB}=
\begin{pmatrix}
g_{\mu\nu}+\lambda^2 A_{\mu c}A^{c}_\nu & \lambda A_{\mu j} \\
\lambda A_{\nu i} & \delta_{ij}
\end{pmatrix},
\end{equation}
where $g_{\mu\nu}$ is the metric of the brane with thickness $l$ \cite{Jalalzadeh:2013wza, Maia:1983zh} and $A_{\mu c}=x^bA_{\mu bc}$, and $A_{\mu bc}$ is the third fundamental form. In addition, the coupling ``constant'' $\lambda$ is given by 
\begin{equation}
    \label{Lambda}
    \lambda^2=\frac{12\pi^\frac{3}{2}\Gamma(\frac{{\mathfrak N}+1}{2})LL_\text{Pl}^2}{d\Gamma(\frac{{\mathfrak N}+2}{2})l^3},
\end{equation}
where $d = 2, 4, 10$ for the Maxwell, weak, and strong interactions, respectively, $L_\text{P}=\sqrt{G_N}$ is the Planck length, $L$ is the curvature radii in the bulk space, and ${\mathfrak N}=D-4$ denotes the number of non-compact extra dimensions. The confinement hypothesis postulates that the matter and gauge fields are localized on the brane with thickness denoted by the parameter $l$. In contrast, gravity is allowed to propagate in the bulk space, extending up to the curvature radii $L$. This implies that the gravitational interactions are not confined to the brane and can propagate to the extent of the bulk geometry, which can be visualized as a higher-dimensional hypersphere denoted by $\mathbb S^{\mathfrak N}$. For a visual representation refer to Fig. \ref{BraneR}.

The $4D$ gravitational constant, $G_N$, in terms of the fundamental scale of the bulk space, $M_D$, the number of extra dimensions, and the normal curvature radii is given by
\begin{eqnarray}\label{sh11}
\frac{1}{16\pi G_N}=\frac{\pi^{\frac{{\mathfrak N}}{2}-1}}{16\Gamma(\frac{{\mathfrak N}}{2}+1)}L^{\mathfrak N}M_{D}^{{\mathfrak N}+2}.
\end{eqnarray}
Also, the thickness is given by the number of extra dimensions and the Compton wavelength of the muon
\begin{equation}
    \label{6-2}
    l=\frac{\sqrt{{\mathfrak N}}\pi}{m_\mu},
\end{equation}
where $m_\mu$ is the muon mass.
Eq. (\ref{Lambda}) gives the relation between brane thickness, extra dimensions, normal curvature radii, and $4D$ Planck's length 
\begin{eqnarray}\label{New11}
L=\frac{d\Gamma(\frac{{\mathfrak N}+2}{2})}{3\sqrt{\pi}\Gamma(\frac{{\mathfrak N}+1}{2})}\frac{\lambda^2}{4\pi}\left(\frac{l}{L_\text{Pl}} \right)^3L_\text{Pl}.
\end{eqnarray}

At any given point along the brane, it is observed that the extrinsic curvature radius gives rise to a closed space known as the extrinsic tube denoted as ${\mathcal B}_{\mathfrak N}$. This intriguing phenomenon is depicted in Fig. \ref{BraneR}, where the accessible region of the bulk for gravitons is confined to this extrinsic tube. The extra dimensions are all space-like, thus allowing us to locally consider ${\mathcal B}_{n}$ as the $n$-sphere denoted as $S_{n}=SO(\mathfrak N)/SO(\mathfrak N-1)$. Furthermore, it is crucial to highlight that this $n$-sphere possesses a radius of $L$ which can be represented as $L:=\min\{L^a_{(\mu)}\}$ at every single point within the spacetime continuum. This observation offers valuable insights into the intricate nature of the extrinsic tube and its relationship with the underlying brane structure.

 As shown in detail in Ref. \cite{Jalalzadeh:2013wza,Jalalzadeh:2023upb}, the field equations are given by
 \begin{equation}
\label{eq17}
{G}_{\alpha\beta}= -Q_{\alpha\beta}+8\pi G_N\left(T_{\alpha\beta}+T_{\alpha\beta}^\text{(YM)}\right),
\end{equation}
\begin{equation}
\label{eq18}
\nabla^{(tot)}_\beta{K}_a-\nabla^{(tot)}_\alpha{K}_{\beta a}^{\alpha}=8\pi
G_NT_{a\beta},
\end{equation}
\begin{equation}
\label{eq19}
\frac{G_N}{g_i^2}\left(F^{\alpha\beta}_{\,\,\,\,\,\,\,am}F_{\alpha\beta
b}^{\,\,\,\,\,\,\,\,\,\,m}+\frac{1}{2}\eta_{ab}F_{\alpha\beta}^{\hspace{.3cm}lm}F^{\alpha\beta}_{\hspace{.3cm}lm}\right)-\frac{1}{2}\delta_{ab}\left({R}+{K}_{\mu\nu m}{K}_{\mu\nu}^{\hspace{.3cm}m}-{K}_a{K}^a\right)=8\pi
G_NT_{ab},
\end{equation}
where $g_i=\lambda$, ${G}_{\alpha\beta}$ is the $4D$ Einstein tensor and  $G_{N}$ is the $4D$ induced gravitational constant. Also, the energy-momentum tensor's components
$T_{\alpha\beta}$, $T_{a\beta}$, and $T_{ab}$ are defined in a way that makes sense in light of the confinement hypothesis, $T_{\alpha\beta}^\text{(YM)}$ represents the Yang--Mills--Maxwell energy-momentum tensor and $F_{ab\mu\nu}$ is the curvature associated to extrinsic twist vector field $A_{\mu ab}$, defined as  \cite{twist}
\begin{eqnarray}
\label{1-10}
F_{ab\mu\nu}=A_{\mu ab,\nu}-A_{\nu ab,\mu}-A_{\nu a}^{\,\,\,\,\,\,\,c}A_{\mu
cb}+A_{\mu a}^{\,\,\,\,\,\,\,c}A_{\nu cb},
\end{eqnarray}
where $A_{\mu ab}$ has the role of the gauge potential \cite{Yang}. The quantity $Q_{\alpha \beta}$ is a conserved geometric quantity stated in terms of extrinsic curvature and its  trace, $K_a$
\begin{equation}\label{eq20}
Q_{\alpha\beta}={K}_{\alpha}^{\,\,\,\,\eta a}{K}_{\beta\eta a}-{K}^a{K}_{\alpha\beta a}-\frac{1}{2}{g}_{\alpha\beta}({K}^{\mu\nu
a}{K}_{\mu\nu a}-{K}_a{K}^a),
\end{equation}
and
\begin{equation}\label{eq21}
\nabla_{\beta}Q^{\alpha \beta}=0.
\end{equation}
Due to these results, the product of the induced energy-momentum tensor for confined matter and Yang-Mills-Maxwell fields and the $4D$ gravitational ``constant'' are conserved
\begin{equation}\label{eq22}
\big(G_N T^{\alpha \beta} \big)_{;\beta}=\big(G_N T^{\alpha \beta}_{(YM)} \big)_{;\beta}=0.
\end{equation}

The field equations mentioned above are general and apply to any number of extra dimensions. It may be interesting for the reader to understand the connection between this formalism and $5D$ braneworld models. In the case of $D=5$, the extrinsic twist vector field $A_{\mu ab}=0$ disappears, which means that the Yang--Mills--Maxwell $F_{\mu\nu ab}$ described in Eq. (\ref{1-10}) does not have a geometric origin in this scenario. However, the value of the extrinsic curvature can still be determined by using Israel's junction condition. Utilizing Israel's junction condition, the extrinsic curvature may be replaced by the energy-momentum tensor of the matter fields, $T_{\mu\nu}$, confined to the brane. In addition, if we assume mirror symmetry, or $\mathbb Z_2$-symmetry, throughout the brane \cite{Shiromizu:1999wj}, then
\begin{eqnarray}\label{1b}
K_{\mu\nu}=-\frac{\kappa^2_{(5)}}{2}\left\{T_{\mu\nu}-\frac{1}{3}(T-\sigma)g_{\mu\nu}\right\},
\end{eqnarray}
where $\kappa^2_{(5)}$ is Einstein’s gravitational
constant of the bulk and $\sigma$ is the tension of the brane in the bulk space. Then, $Q_{\mu\nu}$ defined by Eq. (\ref{eq20}) reduces to
\begin{equation}
    Q_{\mu\nu}=\frac{\kappa^4_{(5)}}{12}\left\{-\sigma^2g_{\mu\nu}+TT_{\mu\nu}-3T_{\mu\alpha}T^\alpha_{\,\,\nu}+\frac{3}{2}T_{\alpha\beta}T^{\alpha\beta}g_{\mu\nu}-\frac{1}{2}T^2g_{\mu\nu}\right\}.
\end{equation}
Therefore, in this case, the field equations (\ref{eq17}) will reduce to the  Shiromizu-Maeda-Sasaki (SMS) braneworld model \cite{Shiromizu:1999wj}.

\section{FLRW cosmology according to CEG}\label{field}

The space-time metric of a homogeneous and isotropic universe, as described by the standard FLRW line element, can be expressed as
\begin{equation}\label{24}
    ds^2=-dt^2 + a(t)^2 \left(\frac{dr^2}{1-kr^2}+r^2d\Omega ^2 \right),
\end{equation}
where $a(t)$ is the cosmic scale factor. The closed, open, and flat universes are represented by $k = +1, -1, 0$, respectively. To ensure consistency with the symmetries of the FLRW universe, the energy-momentum tensor needs to be diagonal. Thus, in the simplest form, it is
\begin{equation}\label{25}
  \begin{split}
      T_{\mu \nu}&=(\rho +p)u_\mu u_\nu +pg_{\mu \nu},~~~~u_\mu=-\delta_\mu^0,\\
       T_{\mu a}&=0,\\ 
       T_{ab}&=p_{ext} \eta _{ab},
\end{split}      
  \end{equation}
where $p_{ext}$ represents the pressure of the cosmic fluid along the extra dimensions. Using Eq. (\ref{eq18}), we can find extrinsic curvature components
\begin{equation}\label{eq26}
   \begin{split}
     K_{\alpha\beta m}&=\frac{f_m(t)}{a^2(t)}g_{\alpha\beta}, ~~~~\alpha,\beta=1,2,3,  \\
     K_{00m}&=-\frac{1}{a(t)H}\frac{d}{dt}\left(\frac{f_m (t)}{a(t)}\right),
   \end{split}   
  \end{equation}
where $H$ represents the Hubble
parameter and $f_m(t)$ are $D-4$ functions of cosmic time. Due to the homogeneity and isotropy on the brane, we assume all these functions are equal, i.e., $f_m(t)=f(t)$. If we define $\phi(t):= \frac{a(t)^2}{f(t)}$ and $h:= \frac{\dot{\phi}(t)}{\phi(t)}$, the $Q_{\alpha\beta}$ defined in (\ref{eq20}) will include the following components 
\begin{equation}\label{eq27}
\begin{split}
 Q_{\alpha\beta}&=-\frac{3\mathfrak R}{\phi^2}\left(1-\frac{2h}{3H}\right)g_{\alpha\beta},~~~\alpha,\beta=1,2,3,\\
Q_{00}&=\frac{3\mathfrak R}{\phi^2},
\end{split}
\end{equation}
which $\mathfrak N:= D-4$. The induced gravitational constant is dependent on the local normal radii of the FLRW submanifold and is not a real constant \cite{Jalalzadeh:2013wza}
\begin{equation}\label{eq28}
G_N(t)=G_0\left(\frac{\phi(t)}{\phi_0} \right)^{-\mathfrak N},
\end{equation}
where $\phi_0$ represents the value at the current time of $\phi(t)$, and
$G_0$ is the $4D$ gravitational constant at the present epoch. Using Eq. (\ref{eq28}) we can write
\begin{equation}\label{eq29}
    \frac{\Dot{G}_N}{G_N}=-\mathfrak Nh.
\end{equation}
According to some findings \cite{Williams:2003wu,Merkowitz:2010kka,Gaztanaga:2001fh,1998ApJ871G,Damour:1988zz}, the relative change in $G_N$ relates to the Hubble parameter
\begin{equation}\label{eq30}
\frac{\dot G_N}{G_N} =\gamma H,~~~~|\gamma|<1,
\end{equation}
where $\gamma$ is a small dimensionless constant. Eq. (\ref{eq29}) and Eq. (\ref{eq30}) give
\begin{equation}\label{31}
L=\phi(t)=\phi_0\left(\frac{a(t)}{a_0}\right)^{-\frac{\gamma}{\mathfrak N}},~~~~~~
\frac{h}{H}=-\frac{\gamma}{ \mathfrak N}
\end{equation}
Like the $T_{\mu \nu}$ tensor, it is feasible to formulate the geometric energy-momentum tensor $Q_{\mu \nu}$ in the form presented below \cite{Jalalzadeh:2013wza}
\begin{equation}\label{eq32}
Q_{\mu\nu}=-8\pi G_N\Big\{\left(\rho^\text{(GDE)}+p^\text{(GDE)}\right)u_\mu u_\nu+p^\text{(GDE)}g_{\mu\nu}\Big\},
\end{equation}
where $\rho^\text{(GDE)}$ is the GDE density and $p^\text{(GDE)}$ is the GDE pressure (the suffix ``GDE'' stands for geometric dark energy which its origin is the extrinsic curvature). 
Comparing Eq. (\ref{eq27}) and Eq. (\ref{eq32}), we find out
\begin{equation}\label{eq33}
\begin{split}
\rho^\text{(GDE)}&=\frac{3{\mathfrak N}}{\phi^2 8\pi G_N}=\frac{3{\mathfrak N}}{8\pi G_0 \phi_0^2}\left(\frac{a}{a_0}\right)^{\left(\frac{2}{{\mathfrak N}}-1\right)\gamma},\\
p^\text{(GDE)}&=-\frac{{\mathfrak N}}{\phi^2 8\pi G_N}\left(3-2\frac{h}{H}\right)=-\frac{3{\mathfrak N}}{8\pi G_0 \phi_0^2}\left(1+\frac{2\delta}{3{\mathfrak N}}\right)\left(\frac{a}{a_0}\right)^{\left(\frac{2}{{\mathfrak N}}-1\right)\gamma}.
    \end{split}
\end{equation}
As one easily can see, in the above relations, the extrinsic part of the field equations, i.e., $\rho^\text{(GDE)}$ and $p^\text{(GDE)}$ are linearly proportional. This suggests we define the proportionality constant of them as an EoS parameter of GDE, given by
\begin{equation}\label{eq34}
\omega_\text{(GDE)}:=\frac{p^\text{(GDE)}}{\rho^\text{(GDE)}}=-\left(1+\frac{2\gamma}{3{\mathfrak N}}\right).
\end{equation}
This results in
\begin{equation}\label{35}
\frac{\gamma}{\mathfrak N}=-\frac{3}{2}\left(1+\omega_\text{GDE}\right).
\end{equation}
Therefore, equations (\ref{eq33}) can be rewritten as follows
\begin{equation}\label{eq36}
\begin{split}
\rho^\text{(GDE)}&=\frac{3{\mathfrak N}}{8\pi G_0 \phi_0^2}\left(\frac{a}{a_0}\right)^{-3\left(1+\omega_\text{GDE}\right)-\gamma},\\
p^\text{(GDE)}&=\frac{3{\mathfrak N}\omega_\text{GDE}}{8\pi G_0 \phi_0^2}\left(\frac{a}{a_0}\right)^{-3\left(1+\omega_\text{GDE}\right)-\gamma}.
    \end{split}
\end{equation}
Substituting Eq. (\ref{eq27}) and Eq. (\ref{eq28}) into Eq. (\ref{eq18}), we obtain the Friedmann equations
\begin{subequations}
    \label{eq37}
    \begin{align}
& H^2+\frac{k}{a^2}=\frac{8\pi G_N}{3}\left(\rho +\rho^\text{(GDE)} \right),\label{eq37a}\\
 &\frac{\ddot a}{a}=-\frac{4\pi G_N}{3}\left(\rho + \rho^\text{(GDE)} +3p+3p^\text{(GDE)}\right).\label{eq37b}
    \end{align}
\end{subequations}
Note that in this model the geometric part of these field equations has four free parameters: the value of the spatial curvature $k$, the extrinsic curvature $K_{\mu\nu m}$, EoS parameter $\omega_\text{(GDE)}$, and the number of extra dimensions $\mathfrak N$. While the spatial curvature's value can be determined through observational data, the same holds for the equation of state parameter of GDE, and $\mathfrak N$.

\section{Einstein static state  and its stability}\label{ESU}

{The aim of the present study is to analyze the introduced brane model, which is capable of explaining the late time accelerating expansion of the universe with a geometric origin \cite{Jalalzadeh:2013wza, Rostami:2015ixa, Jalalzadeh:2023upb}, versus both the past and future singularities. This study is motivated since after discovering the accelerating expansion of the universe, some investigations of the nature of the cosmic fluid responsible for this accelerating expansion showed a plethora of new types of singularities distinct from the initial Big Bang singularity. These new types of singularities are future singularities for a universe and can be classified by the violation of some of the energy conditions. This violation yields divergencies in some physical quantities, such as the scale factor, energy density, and pressure components. This study is important in the sense that any modified theory of gravity that can solve some issues in the standard GR may come with new types of issues, and this requires to be analyzed to check the viability of the theory. Here we show that this brane model, under some conditions on its parameters,
can be free of singularity issues in the standard model of cosmology based on GR.}

 The EU scenario within the context of GR is unstable and anticipated to decay rapidly in the early universe \cite{Barrow:2003ni}.
Because of that, this scenario was formulated on the modified gravity theories to ensure stability at high energy regimes. In the following, we will analyze whether a stable EU scenario can exist in CEG theory.

Based on ESU, where $a = a_0= a_{ES}=constant$, the GDE density and the GDE pressure are 
\begin{equation}\label{eq38}
    \begin{split}
{\rho_{ES}}^\text{(GDE)}={\rho}^\text{(GDE)}(a_{ES})&=\frac{3\mathfrak N}{8\pi G_0 \phi_0^2},\\
{p_{ES}}^\text{(GDE)}={p}^\text{(GDE)}(a_{ES})&=\frac{{3\mathfrak N\omega_\text{(GDE)}}}{8\pi G_0 \phi_0^2}.
    \end{split}
\end{equation}
Using Eq. (\ref{eq37a}), we can obtain the following equations for the conﬁned energy density and pressure
\begin{equation}\label{eq39}
\begin{split}
\rho=&\frac{3}{8\pi G_0}\left(\frac{a}{a_0}\right)^{-\gamma}\left(\frac{\dot a}{a}\right)^2+\frac{3k}{8\pi G_0 a^2}\left(\frac{a}{a_0}\right)^{-\gamma}- \frac{3\mathfrak N}{8\pi G_0 \phi_0^2}\left(\frac{a}{a_0}\right)^{-3(1+\omega_{\text{GDE}})-\gamma},\\
p=&-\frac{1}{4\pi G_0}\left(\frac{a}{a_0}\right)^{-\gamma}\frac{\ddot a}{a}-\frac{1}{8\pi G_0}\Bigg\{\left(\frac{\dot a}{a}\right)^2\left(\frac{a}{a_0}\right)^{-\gamma}+
\frac{k}{a^2}\left(\frac{a}{a_0}\right)^{-\gamma}+\frac{3\mathfrak N\omega_{\text{GDE}}}{\phi_0^2}\left(\frac{a}{a_0}\right)^{-3(1+\omega_{\text{GDE}})-\gamma}\Bigg\},
\end{split}    
\end{equation}
which leads to the following values for the ESU
\begin{equation}\label{eq40}
    \begin{split}
\rho_{ES}&=\frac{3k}{8\pi G_0a_{ES}^2}-\frac
{3\mathfrak N}{8\pi G_0\phi_0^2},\\
p_{ES}&=-\frac{k}{8\pi G_0 a_{ES}^2}-\frac{3\mathfrak N\omega_\text{(GDE)}}{8\pi G_0\phi_0^2}.
\end{split}
\end{equation}
These are known as the existence conditions for an ESU.
\subsection{Linear Homogeneous Scalar Perturbations}
\begin{figure*}[ht]
  \centering
  \subfigure[]{\includegraphics[width=0.24\textwidth]{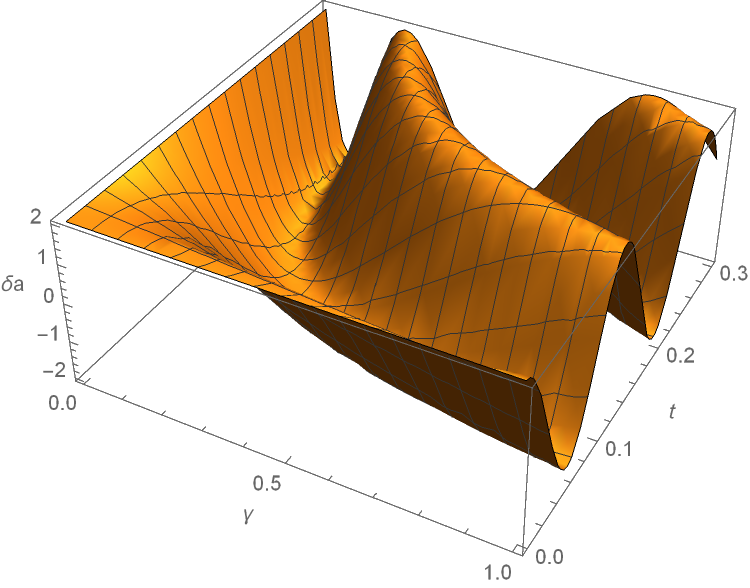}}
  \subfigure[]{\includegraphics[width=0.24\textwidth]{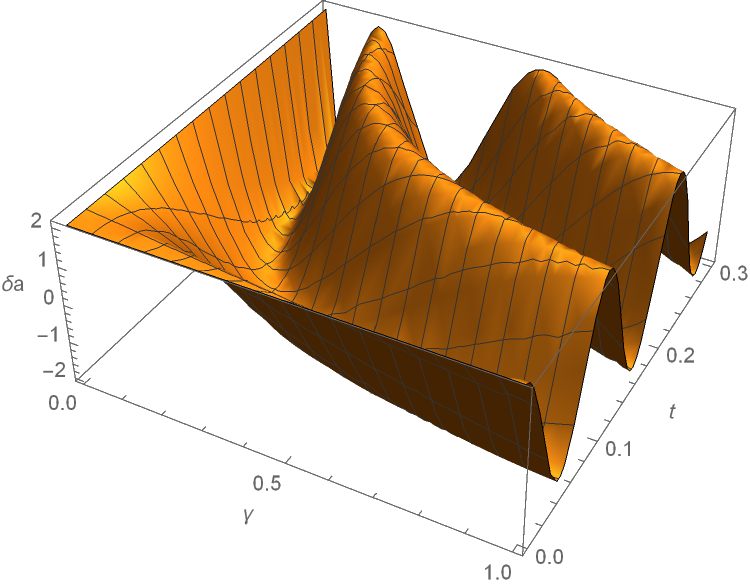}}
  \subfigure[]{\includegraphics[width=0.24\textwidth]{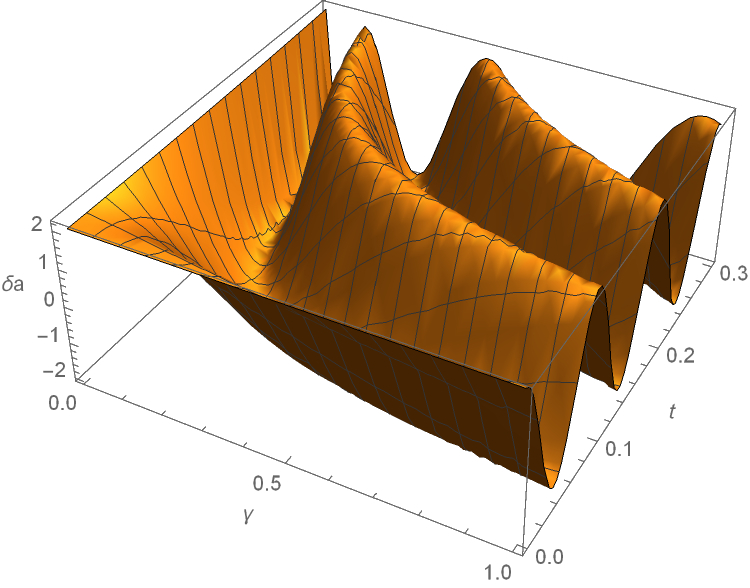}}
  \subfigure[]{\includegraphics[width=0.24\textwidth]{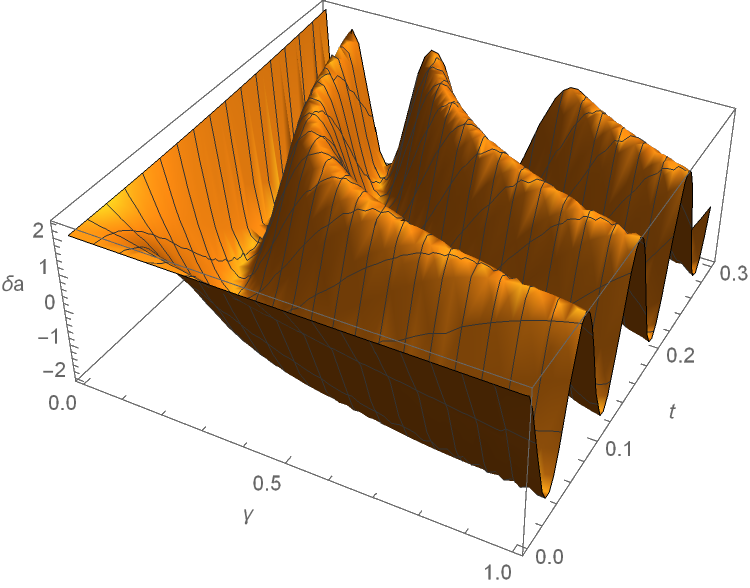}}
  \caption{\small  The oscillating behavior of the scale factor as a function of $t$ and $\gamma$, with different values of $\omega=-\frac{1}{3},~0,~\frac{1}{3},~1$, respectively and numerical values $k=1,~~\mathfrak N=22, ~~a_{ES}=0.1,~~ and~~~\phi _0=0.01$. These parameters satisfy the existence and stability conditions of ESU in TABLE \ref{table1}.}\label{fig1}
\end{figure*}

\begin{figure}[ht]
  \centering
  \includegraphics[width=6cm]{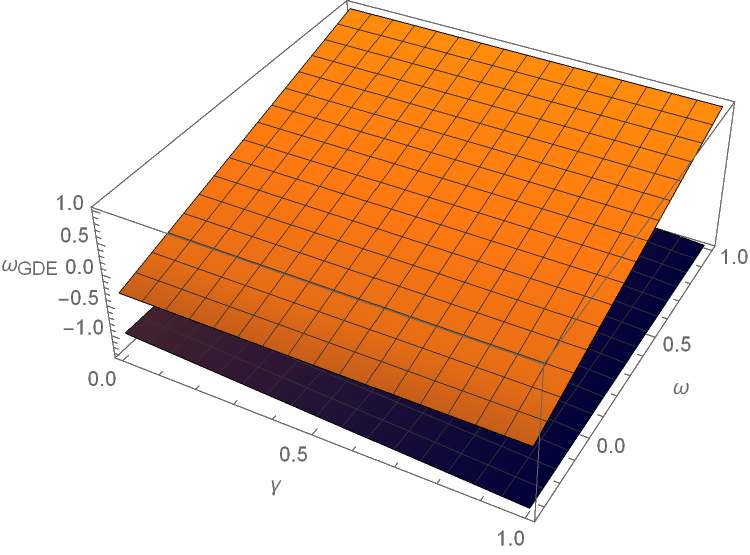}
  \caption{\small The upper plot is $\omega_\text{GDE}^{(2)}$ and the lower plot is $\omega_\text{GDE}^{(1)}$. The stable ESU region exists between $\omega_\text{GDE}^{(1)}$ and $ \omega_\text{GDE}^{(2)}$ plots.}\label{fig2}
\end{figure}

In this section, we consider linear homogeneous scalar perturbations around an ESU and examine its stability under such perturbations. Below are the perturbed forms of the scale factor and the matter-energy density used to derive equations governing homogeneous scalar perturbations
\begin{equation}\label{eq41}
\begin{split}
a(t)&=a_{ES}(1+\delta a(t)),\\
\rho (t)&=\rho _{ES} (1+\delta \rho (t)).
\end{split}
\end{equation}
One can derive the following equation by substituting Eq. (\ref{eq41}) into the confined energy density in Eq. (\ref{eq39}), followed by the deduction of Eq. (\ref{eq38}), and ultimately, the linearization of the resultant expression and assuming $a_0=a_{ES}$
\begin{equation}\label{eq42}
    \rho_{ES} \delta \rho (t)=\Biggl[\frac{-3k}{8\pi G_0a_{ES}^2}(\gamma+2)+\frac{3\mathfrak N}{8\pi G_0\phi_0^2}(3(1+\omega_\text{(GDE)})+\gamma)
    \Biggl]\delta a.
\end{equation}
By applying the above-mentioned method to Eq. (\ref{eq42}) and pressure in Eq. (\ref{eq39}), we obtain
\begin{equation}\label{eq43}
  \omega \rho_{ES} \delta \rho (t)= -\frac{1}{4\pi G_0}(1-\delta a (1+\gamma )) \delta \ddot a + \Biggl [\frac{k(2+\gamma)}{8\pi G_0 a_{ES} ^2} + \frac{3\mathfrak N\omega_\text{(GDE)}}{8\pi G_0\phi_0^2}(3(1+\omega_\text{(GDE)})+ \gamma)\Biggl]\delta a.
\end{equation}
Substituting Eq. (\ref{eq42}) into Eq. (\ref{eq43}) leads to the following evolutionary equation for the scale factor perturbation
\begin{multline}\label{eq44}
\delta \ddot a +\frac{1}{a_{ES} ^2}\Biggl[\frac{-k}{2}(\gamma+2)(1+3\omega)-\frac{9\mathfrak Na_{ES}^2}{2\phi_0^2}\Biggl(\omega_\text{(GDE)} ^2 - \omega_\text{(GDE)} (\omega - (1+\frac{\gamma}{3}))-\omega (1+\frac{\gamma}{3})\Biggl)\Biggl] \delta a =0,
\end{multline}
where we considered that the pressure $p$ and density $\rho$ satisfy the barotropic EoS
\begin{equation}\label{Baro}
    p=\omega\rho.
\end{equation}
The Eq. \eqref{eq44} has the solution
\begin{equation}\label{eq45}
    \delta a=\mathcal{C}_1e^{i\mathcal{M}t}+\mathcal{C}_2e^{-i\mathcal{M}t},
\end{equation}
where $\mathcal{C}_1$ and $\mathcal{C}_2$ are integration constants, and $\mathcal{M}$ is defined as
\begin{equation}\label{eq46}
    \mathcal{M}^2=\frac{-k}{2}(\gamma+2)(1+3\omega)- \frac{9\mathfrak Na_{ES}^2}{2\phi_0^2} \left(\omega_\text{(GDE)} ^2 -\omega_\text{(GDE)} (\omega - (1+\frac{\gamma}{3}))-\omega (1+\frac{\gamma}{3})\right).
\end{equation}
Consequently, the following condition must be satisfied to have oscillating perturbation modes, which implies the presence of a stable ESU
\begin{equation}\label{eq47}
   \omega_\text{(GDE)} ^2 -\omega_\text{(GDE)} (\omega - (1+\frac{\gamma}{3}))-\omega (1+\frac{\gamma}{3})+\frac{k\phi_0^2}{9\mathfrak Na_{ES} ^2}(\gamma+2)(1+3\omega)<0.
\end{equation}
 According to Eq. (\ref{eq47}), the acceptable range for $\omega_\text{(GDE)}$ is
\begin{equation}\label{eq48}
\omega_{\text{GDE}}^{(1)}<\omega_{\text{GDE}}<\omega_{\text{GDE}}^{(2)} ,
\end{equation}
where
\begin{equation}\label{eq49}
\omega_{\text{GDE}}^{(1)}=\frac{\omega}{2}-\frac{1}{2}-\frac{\gamma}{6}-\frac{1}{2}((\omega-(1+\frac{\gamma}{3}))^2 +4\omega (1+\frac{\gamma}{3})- \frac{4k\phi_0^2}{9\mathfrak Na_{ES}^2}(\gamma+2)(1+3\omega))^{\frac{1}{2}},
\end{equation}

\begin{equation}\label{eq50}
\omega_\text{GDE}^{(2)}=\frac{\omega}{2}-\frac{1}{2}-\frac{\gamma}{6}+\frac{1}{2}((\omega-(1+\frac{\gamma}{3}))^2 +4\omega (1+\frac{\gamma}{3})- \frac{4k\phi_0^2}{9\mathfrak Na_{ES}^2}(\gamma+2)(1+3\omega))^{\frac{1}{2}}.
\end{equation}
No stable ESU exists for $\omega_\text{GDE}$ values outside the above range. 
In Table \ref{table1}, the acceptable range of $\omega_\text{GDE}$ with respect to $\omega$ values, along with additional restrictive conditions, for having a stable ESU are given. We also note that according to Eq. (\ref{eq40}), the selection of $k$ is restricted to $k=1$. The energy density assumes a positive value only under these conditions, causing this limitation. In Fig. \ref{fig1}, we have illustrated the evolution of $\delta a$ with respect to cosmic time $t$ and  $\gamma$, using the typical parameter values satisfying the existence and stability conditions of ESU as listed in Table \ref{table1}. Fig. \ref{fig2} showcases the EoS of GDE as a function of $\gamma$ and $\omega$. The upper plot represents $\omega_\text{GDE}^{(2)}$, while the lower plot represents $\omega_\text{GDE}^{(1)}$. The region between the two graphs signifies the permissible range for the existence of an ESU. Hence, one can conclude that, under the discussed conditions, the CEG admits a stable ESU versus the homogeneous linear perturbations. 

\begin{center}
\begin{table*}[ht]
 \centering
\begin{tabular}{|c|c|c|c|}
 \hline
     $\omega$&$\omega_\text{GDE}^{(1)}$ &$\omega_\text{GDE}^{(2)}$&the restricting condition\\
      \hline
      $\omega=0$&$-\frac{1}{2}-\frac{\gamma}{6}-\frac{1}{6}\sqrt{f_1}$&$-\frac{1}{2}-\frac{\gamma}{6}+\frac{1}{6}\sqrt{f_1}$&$f_1>0$\\
      \hline
      $\omega=\frac{1}{3}$&$-\frac{1}{3}-\frac{\gamma}{6}-\frac{1}{6}\sqrt{f_2}$&$-\frac{1}{3}-\frac{\gamma}{6}+\frac{1}{6}\sqrt{f_2}$&$f_2>0$\\
      \hline
      $\omega=1$&$-\frac{\gamma}{6} -\frac{1}{6}\sqrt{f_3}$ & $-\frac{\gamma}{6} +\frac{1}{6}\sqrt{f_3}$&$f_3>0$\\
      \hline
    $\omega=-\frac{1}{3}$&$-\frac{2}{3}-\frac{\gamma}{6} -\frac{1}{6}\mid{\gamma+2}\mid$ & $-\frac{2}{3}-\frac{\gamma}{6} +\frac{1}{6}\mid{\gamma+2}\mid$& -\\ [1ex]
        \hline
          \end{tabular}
    \caption{The necessary conditions ensuring the existence of a stable ESU in CEG. In this table, we defined $f_1:=(\gamma+3)^2-\frac{4\phi_0^2}{\mathfrak N a_{ES}^2}(\gamma+2)$, $f_2:=(\gamma+4)^2-\frac{8\phi_0^2}{\mathfrak N a_{ES}^2}(\gamma+2)$, and $f_3:=(\gamma+6)^2-\frac{16\phi_0^2}{\mathfrak R a_{ES}^2}(\gamma+2)$. }
    \label{table1}
\end{table*}
\end{center}

\subsection{Inhomogeneous Perturbations}
We already show that defining the ordinary and geometric fluid sources as perfect fluids in \eqref{25} and \eqref{eq32} results in the induced field equations  \eqref{eq37}. Hence, one can treat the source of field equations \eqref{eq37} as an effective perfect fluid of the form $T^t_{\mu\nu}=(\rho_t + p_t)u_\mu u_\nu+p_t g_{\mu\nu}$ possessing the total energy density $\rho_t=\rho+\rho^\text{(GDE)}$, pressure $p_t=p+p^\text{(GDE)}$ and baraotropic equation of state $p_t=\omega_t \rho_t$. This formulation allows one to investigate the effect of inhomogeneous density perturbations on the simple one-component effective fluid models. Density perturbations of a general Friedmann universe on a $4D$-brane can be addressed by the $1+3$-covariant gauge-invariant approach \cite{Bruni:1992dg}. In this approach, letting $\Delta= a^2 D^2\rho_t/\rho_t$ where $D^2$ represents the covariant spatial Laplacian, the evolution of $\Delta$
follows 
\begin{eqnarray}\label{del}
\ddot \Delta &+&\left( 2-\omega_t +6c_s^2 \right)H\dot\Delta+\left[ 12(\omega_t-c_s^2)\frac{k}{a^2}+4\pi G(3\omega_t^2 +6c_s^2 -8\omega_t-1)\rho_t +(3c_s^2-5\omega_t)\Lambda\right]\Delta\nonumber\\
&-&c_s^2D^2\Delta-\omega_t(D^2+3\frac{k}{a^2})\mathcal{E}=0,
\end{eqnarray}
where  $c_s=\frac{dp_t}{d\rho_t}$  is the sound speed, $H=\frac{\dot a}{a}$ denotes the Hubble parameter, and $\mathcal{E}$ is the entropy perturbation of the effective source as 
\begin{equation}
 p_t  \mathcal{E}=a^2 D^2 p_t-\rho_t c_s^2 \Delta.  
\end{equation}
The ESU model with the effective perfect fluid has $\mathcal{E}=0$  \cite{Barrow:2003ni}. Hence, the equation \eqref{del} after decomposing into Fourier modes with the co-moving index $\kappa$ (such that $D^2 \to -\kappa^2/a^2_{ES}$) reduces to \cite{Barrow:2003ni}

\begin{equation}
 \ddot \Delta_\kappa -4\pi G_0(1+\omega_t)\left( 1+(3-\kappa^2)c_s^2 \right)\rho_{tES}\Delta_\kappa=0.
\end{equation}
Thus, the ESU remains stable against gravitational collapse if $\Delta_\kappa$ is oscillating. This is provided by
\begin{equation}
 (1+\omega_t)\left( 1+(3-\kappa^2)c_s^2 \right)<0.   
\end{equation}
If the total effective fluid violates the weak energy condition, i.e. $\omega_t<-1$, it is required $(\kappa^2-3)c_s^2<1$. This case can represent when the geometric fluid, as the dark energy profile, effectively dominates the Universe's ordinary matter content. If the total effective fluid satisfies the weak energy condition, i.e. $\omega_t>-1$, it is required $(\kappa^2-3)c_s^2>1$. This case is addressed in detail after Eq. (16) in \cite{Barrow:2003ni}. In particular, for physical modes with $n \geq 2$, the stability is assured for $c^2_s>1/5$. This condition was reported by Gibbons \cite{Gibbons:1987jt, Gibbons:1988bm} in the restricted case of conformal metric perturbations. In conclusion, an ESU with a fluid satisfying $c^2_s > 1/5$ is neutrally stable against adiabatic density perturbations of the fluid for all allowed inhomogeneous modes. 
\\

The vector perturbations of a perfect fluid are described by the comoving dimensionless vorticity, defined as $\bar\omega_a=a\bar\omega$,
 where its modes obey the propagation equation \cite{Barrow:2003ni}
\begin{equation}\label{vectorpert}
\dot{\bar\omega}_{\kappa}+ \left( 1-3c_s^2\right)H\bar\omega_\kappa=0.
\end{equation}
 This equation remains valid in our treatment of ESU in the present brane scenario with the effective perfect fluid source. Hence, for an ESU, where $H=0$, the above equation simplifies to
\begin{equation}
  \dot{\bar\omega}_{\kappa}=0.  
\end{equation}
This equation implies that the initial vector perturbations remain frozen for our brane model, and hence, there will be a neutral stability against these perturbations.

On the other hand, tensor perturbations, namely gravitational wave perturbations, of a perfect fluid with density $\rho_t$ and pressure $p_t = \omega_t \rho_t$ is given by the comoving dimensionless transverse-traceless shear $\Sigma_{ab} = a\sigma_{ab}$, whose modes in the presence of cosmological constant $\Lambda$ satisfy \cite{Dunsby:1997fyr, Challinor:1999xz, Maartens:2001zu}
\begin{equation}
\ddot\Sigma_\kappa +3H\dot\Sigma_\kappa +\left(  \frac{\kappa^2}{a^2} +\frac{2k}{a^2} -\frac{2\Lambda + (1+3\omega_t)\rho_t}{3}\right)\Sigma_\kappa=0.
\end{equation}    
 Using Eqs. \eqref{eq38}-\eqref{eq40} for the ESU, this equation reduces to
\begin{equation}
\ddot\Sigma_\kappa +\left(  \frac{\kappa^2}{2k}+1\right)\left( \rho_{ES} (1+\omega) +\rho_{ES}^\text{(GED)}(1+\omega_\text{GED})\right)\Sigma_\kappa=0.
\end{equation} 
Hence, one observes that neutral stability versus tensor perturbations, through the oscillatory modes, is also possible provided that values of parameters $k, \omega$ and $\omega_\text{GDE}$ keep the multiplication factor in front of $\Sigma_k$  positive. 

The findings above indicate that if the universe is initially close to an ESU,  it stays in that neighborhood. However, one notes that the ESU is not an attractor since its stability is neutral and characterized by undamped oscillations. A fall in matter pressure can initiate expansion away from the static state. Typically, this expansion will result in an inflation. It is also worth noting that nonlinear \cite{Langlois:2005ii, Unruh:2008zza, Mondal:2021pyz, Mondal:2022oyj} and non-perturbative effects \cite{Ijjas:2018cdm, Camara:2010zm, Shuhmaher:2005mf, McInnes:2004zm, Yoshida:2017swb}  are also anticipated to play a significant role in such cosmological models because of the initially envisaged infinite time scale. 
In addition, since the present model includes at least two matter sources (baryonic and geometric contributions), both adiabatic and isocurvature perturbations must be examined within this brane-world scenario. Such an analysis can be conducted by treating the total source as a system consisting of two scalar fields, as in the references \cite{Notari:2002yc, DeAngelis:2023fdu, Matsui:2018xwa, DiMarco:2002eb, Christopherson:2014eoa, Langlois:2008vk, Hwang:2001fb, Gordon:2000hv}.    
These critical aspects of the current braneworld scenario will be explored in our future research.

Finally, it is worth mentioning here that the stability of an ESU in other brane models like the Dvali--Gabadadze--Porrati  (DGP), Randall--Sundrum (RS), and Shtanov--Sahni (SS) braneworlds are also studied in \cite{Zhang:2016obw} and \cite{Zhang:2010qwa}, respectively. In the DGP model \cite{Zhang:2016obw}, two separate branches with $\epsilon=\pm 1$ and a perfect fluid of a constant equation of state parameter are considered. It is found that for the $\epsilon =1$ branch, there is an unstable Einstein static solution, while, for $\epsilon =-1$  a stable Einstein static universe exists. Hence, the DGP model is capable of solving the Big Bang singularity. In \cite{Zhang:2010qwa}, it is found that if the matter perfect fluid source on the brane has a phantom-like behavior and the projected ``Weyl fluid'' from the bulk Weyl tensor behaves like radiation with positive energy density, the ESU is stable in the SS braneworld, but unstable in the RS one. Moreover, it is shown that the ESU is also stable in bulk with a timelike extra dimension. Hence, in a brane model with a timelike extra dimension, the initial Big Bang singularity might be resolved.
\section{Leaving the static Einstein state}\label{Living}

In the previous section, we demonstrated the existence of a stable static Einstein state in our model, indicating that the universe could, in principle, persist indefinitely in this configuration. However, to account for the full evolutionary history of the universe, a transition from this steady state into an inflationary phase is necessary. In this section, we explore the mechanism by which this transition can occur. 
The stable ESU corresponds to a fixed point where the early universe could have undergone perpetual oscillations. Superinflationary effects \cite{Labrana:2013oca, Guendelman:2014bva, Huang:2020oqt, Huang:2022hye, Guendelman:2023jsk} are typically invoked to explain how the universe exited this state. Here, we demonstrate that in our framework, the escape from the ESU can instead be achieved using a conventional inflationary scalar field. To achieve this, we write the explicit form of the Friedmann equations (\ref{eq37}) for a closed universe filled with a perfect fluid with EoS $p=\omega \rho$ as
\begin{equation}\label{L1}
    \begin{split}
        H^2+\frac{1}{a_0^2x^2}=\frac{8\pi G_0\rho_0}{3}x^{-3(1+\omega)}+\frac{8\pi G_0\rho_0^\text{(GDE)}}{3}x^{-3(1+\omega_\text{GDE})},\\
        \frac{\ddot x}{x}=-\frac{4\pi G_0}{3}(1+3\omega)\rho_0x^{-3(1+\omega)}-\frac{4\pi G_0}{3}(1+3\omega_\text{GDE})\rho_0^\text{(GDE)}x^{-3(1+\omega_\text{GDE})},
    \end{split}
\end{equation}
where $x=a/a_0$, and all quantities with subscript zero show their values at the static state. The Einstein static state ($x=1$ or $a=a_0=a_{ES}$) is given by $\dot x=0=\ddot x$. First, we need to establish the conditions for the existence of this solution. Inserting these conditions into (\ref{L1}) gives
\begin{equation}
    \label{L2}
    \frac{1}{a^2_0}=\frac{8\pi G_0\rho_0^\text{(GDE)}}{(1+3\omega)}(\omega-\omega_\text{GDE}),
\end{equation}
and
\begin{equation}
    \label{L3}
    \rho_0=-\frac{1+3\omega_\text{GDE}}{1+3\omega}\rho_0^\text{(GDE)}.
\end{equation}
In the case of ordinary matter, which is our assumption, $\omega\geq0$. The definition of the EoS of GDE in (\ref{eq34}) implies $\omega_\text{GDE}<0$. As a result, the r.h.s. of (\ref{L2}) is positive. Also, $\rho_0^\text{(GDE)}$ defined by (\ref{eq36}) is positive, which implies either $\rho_0$ is negative in Eq. (\ref{L3}) or $\omega_\text{GDE}<-1/3$.

By defining the phase-space variable $y=\dot x$, it is useful to convert the second Friedmann equation in (\ref{L1}) as a 2-dimensional dynamical system, which helps explore the stability of this solution. Utilizing the condition (\ref{L3}) into the second Friedmann equation, one finds
\begin{equation}
    \label{L4}
    \begin{split}
    \dot x&=y,\\
    \dot y&=\frac{4\pi G_0}{3}(1+3\omega_\text{GDE})\rho_0^\text{(GDE)}\left(x^{-3(1+\omega)}- x^{-3(1+\omega_\text{GDE})}\right)x,
 \end{split}
\end{equation}
with a fixed point at $(x=1,y=0)$.
The stability of the equilibrium point can be determined by examining the eigenvalues, denoted as $\lambda$, of the Jacobian matrix, $J$, given by
\begin{equation}
    J=\begin{pmatrix}
        \frac{\partial\dot x}{\partial x}& \frac{\partial\dot x}{\partial y}\\
         \frac{\partial\dot y}{\partial x}& \frac{\partial\dot y}{\partial y}
    \end{pmatrix},
\end{equation}
 evaluated at this point, which are found to be
\begin{equation}
    \lambda_\pm=\pm\sqrt{4\pi G_0\rho_0^\text{(GDE)}(1+3\omega_\text{GDE})(\omega_\text{GDE}-\omega)}.
\end{equation}

\begin{figure}[ht]
  \centering
   \includegraphics[width=6cm]{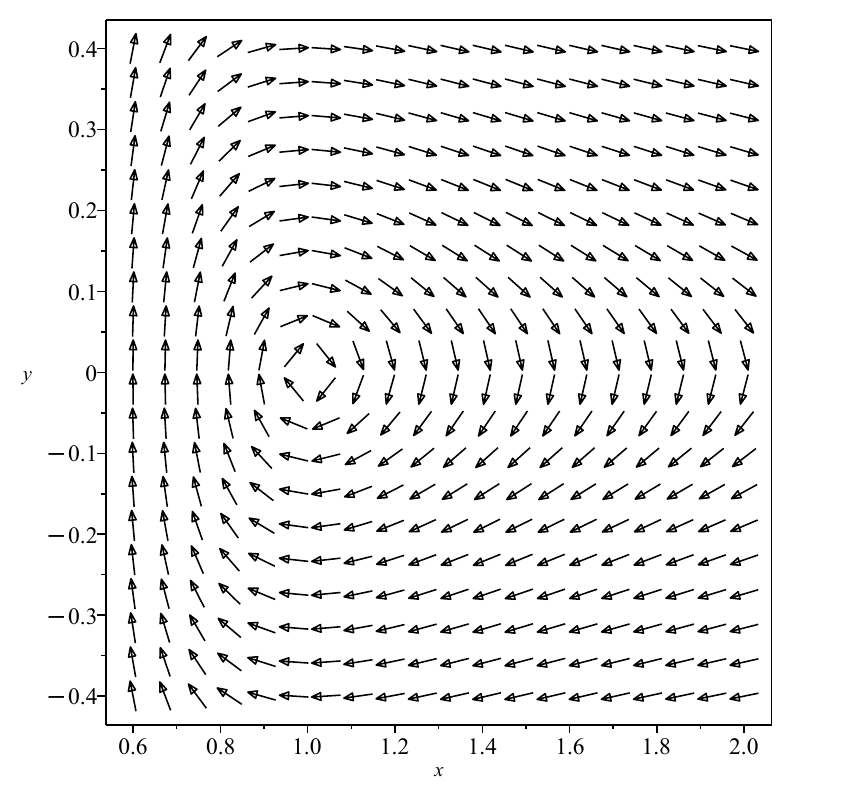}
  \caption{\small  A phase portrait for the stable Einstein state. We assumed $\frac{4\pi G_0}{3}\rho_0^\text{(GDE)}=1$, $\omega=1$, and $\omega_\text{GDE}=-0.3$.}\label{FigL1}
\end{figure}

Regarding that $\omega_\text{GDE}<0$ and $\omega$ has a positive value, the sign $\lambda^2$ depends on the sign of $(1+3\omega_\text{GDE})$ which determines the stability. Trajectories starting in the vicinity of a saddle point $(1+3\omega_\text{GDE})>0$ experience exponential divergence from it, indicating that the Einstein static state is unstable. In contrast, when $(1+3\omega_\text{GDE})<0$, the Einstein static solution turns into a center equilibrium point $(x=1,\dot y=0)$ that is circularly stable. This means that little deviations from the fixed point cause oscillations around it as opposed to an exponential departure from it; see Fig. \ref{FigL1}. 
In this instance, the universe permanently remains (oscillates) in the vicinity of the Einstein static solution. Hence, the requirement for stability is provided by
\begin{equation}
    -\frac{1}{3}<\omega_\text{GDE}<0,~~~~\text{and}~~~~\rho_0<0.
\end{equation}
Note that the stability of the Einstein state requires a negative energy density for the matter component. The second law of thermodynamics and maybe other general relativity energy conditions may be broken in a world mostly composed of negative energy density components \cite{hawking2023large}. However, some extensions of quantum field theory and recent analyses of Casimir force experiments suggest that the same energy conditions in relativity need only be satisfied globally, or on an average measure, leaving open the possibility of local or brief violations of the same energy conditions \cite{Epstein:1965zza, Roman:1986tp, Ford:1995wg, Graham:2002yr}. Therefore, one can imagine the realization of this kind of matter field in the early Einstein state of the universe. In addition, the cosmological applications of negative phantom energy, negative domain walls, negative cosmic strings, negative mass,
negative radiation, and negative ultralight are presented in \cite{Nemiroff:2014gea}. In cosmology, the conceptualization of a matter field with negative energy density encompasses various theoretical constructs and occurrences.

In contrast to the positive energy density condition, it is intriguing to observe numerous authors presenting the notion of negative energy density with compelling reasons in its favor. Ijjas and Steinhardt \cite{Ijjas:2019pyf} examine negative energy density, wherein models undergo a bounce evolution. The authors suggested that subsequent rebounds may also occur. The examination of negative vacuum energy density within the framework of rainbow gravity is presented in \cite{Wong:2019xlj}. Ref. \cite{Nemiroff:2014gea} demonstrates that, under some conditions, negative energy density can provide repulsive gravitational pressure. Fay in Ref. \cite{Fay:2014fta} asserts that the universe undergoes evolution through inflation when the linked fluid possesses negative energy density throughout the beginning epoch. Sawicki and Vikman \cite{Sawicki:2012pz} examine an accelerating world with negative energy density. Macorra and German \cite{delaMacorra:2004et} provided an explanation of energy density with a negative value, characterized by an equation of state parameter $\omega<-1$. In \cite{1990JMP}, an examination of models developed with negative energy density in the infinite past. 

Note that in the model under investigation, the effective energy density in the Friedmann equation (\ref{eq37}) is $\rho+\rho^\text{(GDE)}$. Therefore, using (\ref{L3}), the effective energy density in Einstein's static epoch is given by
\begin{eqnarray}
    \rho_0+\rho^\text{(GDE)}_0=\frac{3(\omega-\omega_\text{GED})}{1+3\omega}\rho^\text{(GDE)}_0.
\end{eqnarray}
Regarding the facts $\omega>0,~\omega_\text{GED}<0$, and $\rho^\text{(GDE)}_0>0$, one observes that the effective matter energy density is positive, and consequently, the positive energy condition is preserved.  However, it is worth mentioning that although this keeps negative energy density under control, its violation in any of the matter sectors may lead to potential pathologies, such as instability or violations of unitarity at the quantum level.
\begin{figure}[ht]
  \centering
   \includegraphics[width=6cm]{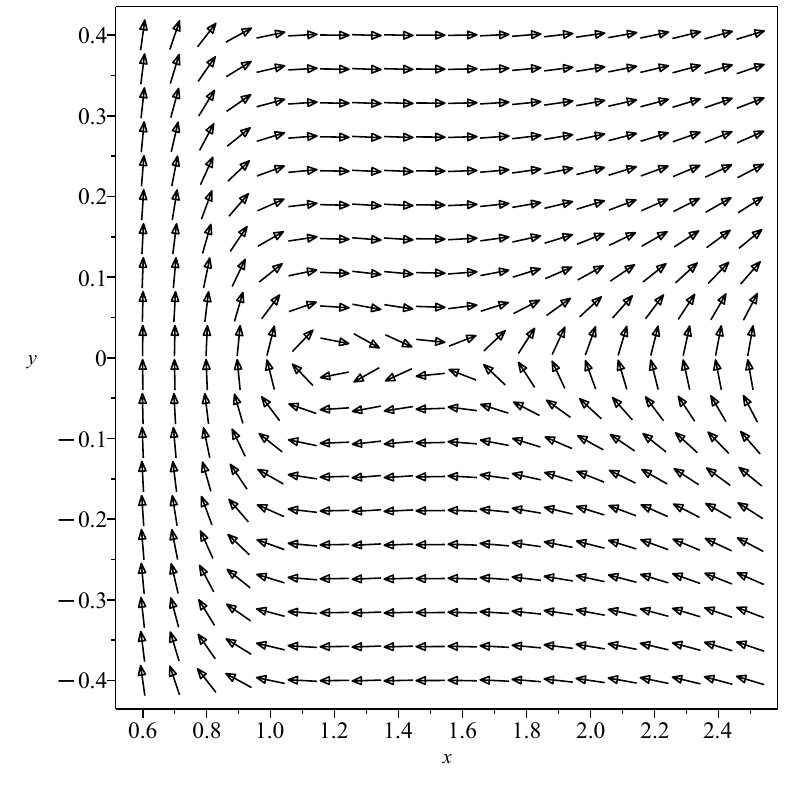}
  \caption{\small  A phase portrait for the transition from stable Einstein state to the inflation area. We assumed $\frac{4\pi G_0}{3}\rho_0^\text{(GDE)}=1$, $\omega=1$, $\omega_\text{GDE}=-0.3$, and $\rho^{(\Phi)}_0/\rho_0^\text{(GDE)}=1/3$.}\label{FigL2}
\end{figure}

To realize the transition from the Einstein state to the inflationary area, we assume the usual inflation scalar field, $\Phi$, with a potential term $V(\Phi)$. Also, regarding the symmetries of FLRW spacetime, we assume the scalar field is a function of cosmic time, $t$. Then, we have the equation of motion of the inflaton field
\begin{equation}
    \label{L6}
    \ddot\Phi+3H\dot\Phi+\frac{dV}{d\Phi}=0.
\end{equation}
In Einstein's static universe, $x=1$, and $H=0$. Let us assume the scalar field is switched on at transition time, $t=t_T$. If we assume the scalar potential is flat enough, the energy density of the scalar field, $\rho^{(\Phi)}=\dot\Phi^2/2+V\simeq V$, is constant. Therefore, one can assume the scalar field in the slow-rolling approximation plays the role of a cosmological constant, and
dynamical system (\ref{L4}) changes to the following form
\begin{equation}
    \label{L8}
    \begin{split}
    \dot x&=y,\\
    \dot y&=\frac{4\pi G_0}{3}(1+3\omega_\text{GDE})\rho_0^\text{(GDE)}\left(x^{-3(1+\omega)}- x^{-3(1+\omega_\text{GDE})}\right)x+\frac{8\pi G_0}{3}\rho_0^{(\Phi)}x.
 \end{split}
\end{equation}
For small values of the energy density of the scalar field, specifically when $\rho^{(\Phi)}_0/\rho_0^\text{(GDE)}<1$, upon activating the scalar field, the initial focus is on the dominance of the first two terms in the second equation of the dynamic system, causing the scale factor to experience a notably slow growth rate from the initial value of the scale factor $a_{ES}$. 
The condition for prolonged inflation is that the Hubble friction is almost canceled, with the potential gradient term being larger. One can ignore the two first terms in (\ref{L8}) by dominating the scalar field. Based on the illustration in Fig. \ref{L2}, the dominance of the scalar field will lead to the departure of the universe from its static Einstein state. 
Therefore, the dynamical system (\ref{L8}) will reduce to
\begin{equation}
    \label{L8a}
    \begin{split}
    \dot x&=y,\\
    \dot y&=\frac{8\pi G_0}{3}\rho_0^{(\Phi)}x,
 \end{split}
\end{equation}
with an inflationary solution $x=\exp(Ht)$.
The resulting Hubble parameter, given by $H=\sqrt{8\pi G_0V/3}$, will lead to nearly exponential inflation. Hence, the universe is anticipated to undergo a transition into the traditional inflationary phase.
\section{The possible future singularities}\label{FS}

In general relativity and its associated modified metric theories of gravity, it has been discussed that the universe may face a catastrophic end at a finite future, and intriguingly, this devastating outcome can be triggered by the existence of certain matter fields that contravene the dominant energy condition. Barrow demonstrated in his study \cite{Barrow:2004xh} that such singularities can emerge even in scenarios where the energy density, $\rho$, and the pressure of the matter content of the universe satisfy $\rho+3p>0$ and $\rho>0$ conditions. In this section, we will use Barrow's approach to investigate potential future singularities within the CEG theory.

Here, regarding the results in the previous sections of the article, we follow the following scenario: we assume the following three possible phases of the model universe: The initial state is an Einstein static universe in the presence of extra dimensions and a perfect fluid with a barotropic EoS (\ref{Baro}). Then, in the second phase, the universe enters the inflationary phase, and finally, after the inflationary stage, we have a classical universe with extra dimensions and a perfect fluid with generally independent pressure and matter density.

Let us rewrite the explicit form of the Friedmann equations (\ref{eq37}) with a general matter field with a pressure $p$, and an energy density $\rho$ as
\begin{subequations}
    \label{eq37new}
    \begin{align}
& H^2+\frac{k}{a^2}=\frac{8\pi G_0}{3}\left(\frac{a}{a_0}\right)^{-\frac{3{\mathfrak N}(1+\omega_\text{(GDE)})}{2}}\rho +\frac{{\mathfrak N}}{\phi_0^2}\left(\frac{a}{a_0}\right)^{-3\left(1+\omega_\text{GDE}\right)},\label{eq37anew}\\
 &\frac{\ddot a}{a}=-\frac{4\pi G_0}{3}\left(\frac{a}{a_0}\right)^{-\frac{3{\mathfrak N}(1+\omega_\text{(GDE)})}{2}}\left(\rho+3p\right)-\frac{\mathfrak N}{2\phi_0^2}(1+3\omega_\text{GDE})\left(\frac{a}{a_0}\right)^{-3\left(1+\omega_\text{GDE}\right)},\label{eq37bnew}
    \end{align}
\end{subequations}
where $a_0$ refers to an initial value of the scale factor, and in obtaining these equations, we substituted $G_N$, $\rho_\text{(GDE)}$ and $p_\text{(GDE)}$ from equations (\ref{eq30}) and (\ref{eq36}) into the Friedmann equations (\ref{eq37}). The matter fluid here differs from the pre-inflation perfect fluid associated with the stability of the Einstein static state, and we do not assume a barotropic equation of state for it. One notes that solving the above system of highly non-linear differential equations for the unknowns $a(t),~ \rho(t)$ and $p(t)$ without implementing an equation of state or an ansatz for the scale factor is not possible. Hence, as in Barrow \cite{Barrow:2004xh, Barrow:2004hk}, here we will study all possible kinds of finite-time future singularities by considering the scale factor as a power function of time. 

A classification of potential future singularities based on the behavior of the scale factor, matter density, and pressure over finite cosmic time is given in Table \ref{table2}.
  \begin{table}[ht]
    \centering
    \begin{tabular}{|c|c|c|c|c|}
    \hline
      {Sing. type} & t &a(t)& $\rho (t)$ & p(t)\\
      \hline
      I& $t\longrightarrow t_s $& $a\longrightarrow \infty$ & $\rho \longrightarrow \infty$ & $p \longrightarrow \infty$\\
      \hline
      II & $t\longrightarrow t_s $& $a\longrightarrow a_s $ & $\rho \longrightarrow \rho _s$ & $p \longrightarrow \infty$\\
      \hline
      III & $t\longrightarrow t_s $& $a\longrightarrow a_s $ & $\rho \longrightarrow \infty$ & $p \longrightarrow \infty$\\
      \hline
     IV & $t\longrightarrow t_s $& $a\longrightarrow a_s$ & $\rho \longrightarrow 0 $ & $p \longrightarrow 0 $\\ [1ex]
           \hline
          \end{tabular}
    \caption{The possible finite-time future singularities for an FLRW cosmology. }
    \label{table2}
\end{table}
\\
Here are some additional detail regarding these singularities:
\begin{itemize}
\item  Type I Singularity (Big Rip/Cosmic Doomsday): Within this particular type of singularity, the scale factor, effective energy density, and pressure diverge. The presented scenario depicts an event of universal demise in which all entities within the universe will be subject to progressive destruction.
\item Type II (Sudden/Quiescent singularity): Only the effective pressure density diverges. Therefore, scale factor derivatives diverge from the second derivative. Notably, Big Brake and Big Boost are two types of this singularity.
\item  Type III (Big Freeze): This kind is identified by the scale factor derivative diverging starting with the first derivative. It has been noted that generalized Chaplygin gas models exhibit this kind of divergence.
\item  Type IV (Big Brake/Big Separation): The scale factor, the effective pressure, and energy densities, in this case, are all finite, however, the higher derivatives of $H$ diverge.
\end{itemize}
As a remark, we would like to mention that new cosmological singularities, called `quiescent' singularities characterized by finite matter density and Hubble parameter but divergent higher derivatives of the scale factor, are reported in \cite{Shtanov:2002ek}. It is discussed that those singularities are the consequence of embedding a $(3+ 1)$ dimensional brane in a bulk and can exist even for an empty, isotropic, and homogeneous FLRW spacetime. 

According to Barrow \cite{Barrow:2004xh, Barrow:2004hk}, to investigate the future singularities in a model with Big Bang singularity, one can assume 
\begin{equation}\label{eq51}
a(t)=1+Bt^q+C(t_s -t)^n.
\end{equation}
Then, $a(0) =0$ results $C = \frac{1}{t_s^n}$ and the scale factor (\ref{eq51}) takes the following form
\begin{equation}\label{eq52}
    a(t)=1+(a_s -1)\left(\frac{t}{t_s}\right) ^q -\left(1-\frac{t}{t_s}\right) ^n.
\end{equation}
However, following the previous sections of our study, the initial state of our model is a non-singular Einstein static state $a(0)=a_0$. Thus, a modification to Barrows' suggestion is required (\ref{eq51}). 
Following \cite{Heydarzade:2019dpf}, we consider the following form for the scale factor
\begin{equation}\label{eq53}
    a(t)=1+(a_s-1)\left(\frac{t}{t_s}\right) ^q +(a_{0}-1)\left(1-\frac{t}{t_s}\right) ^n,
\end{equation}
 where $q>0,\, n>0,\, 0<a_{0}<<1<a_s$, in which $a(0)=a_{0}$ and $a(t_s)=a_s$ represent the values of the scale factor at $t=0$ and $t=t_s$, respectively\footnote{Here we note that one can always do a Taylor expansion of the scale factor in terms of the observable physical quantities such as Hubble $H$, deceleration $q$, jerk $j$ and snap
$s$ parameters, respectively \cite{Visser:2003vq}. The power law expansion used in Eq. (\ref{eq53}) represents a  continuous scale factor function subject to both an initial $a_{0}$ and a final state $a_s$, which can be
either singular or non-singular.}.
The first and second derivatives of the scale factor are
\begin{equation}\label{eq54}
    \begin{split}
\dot a(t)=&\frac{q(a_s -1)}{t_s}\left(\frac{t}{t_s}\right) ^{q-1} -\frac{n(a_{0} -1)}{t_s}\left(1-\frac{t}{t_s}\right) ^{n-1},\\
\ddot a(t)=&\frac{q(q-1)(a_s -1)}{t_s^2}\left(\frac{t}{t_s}\right) ^{q-2} +\frac{n(n-1)(a_{0} -1)}{t_s^2}\left(1-\frac{t}{t_s}\right) ^{n-2}.
    \end{split}
\end{equation}
By substituting(\ref{eq53}) and (\ref{eq54}) in (\ref{eq37bnew}), we obtain the following relations for energy density and isotropic pressure
\begin{equation}\label{eq55}
\frac{8\pi G_0}{3}\rho =\left(\frac{a_{0}}{a}\right)^\gamma\Bigg\{\left(\frac{\dot a}{a}\right)^2+ \frac{k}{(a)^2}-\frac{3{\mathfrak N}}{\phi_0^2}\left(\frac{a_{0}}{a} \right)^{3(1+\omega_\text{GDE})}\Bigg\},
\end{equation}

\begin{equation}\label{eq56}
8\pi G_0p=-\left(\frac{a_{0}}{a}\right)^\gamma\Bigg\{ \frac{2\ddot a}{a}+\left(\frac{\dot a}{a}\right)^2+\frac{k}{a^2}+\frac{3{\mathfrak N\omega_{GDE} }}{\phi_0^2}\left(\frac{a_{0}}{a}\right)^{3(1+\omega_{GDE})}\Bigg\}.
\end{equation}


In the following, we shall  categorize the possible finite-time singularities according to $q$, $n$, and $\gamma$ values, as shown in the following classifications. Our work will be based on $8\pi G_0=1$ for the remainder of this paper.

\bigskip
\textbf{Class I:}

\bigskip 
\textbf{A:} $t\longrightarrow t_s$ with $\gamma=0$, $\omega_\text{GDE}=-1$,~~$0 < n < 1$, and $0 < q \le1$. Which leads to

\begin{equation}
\begin{split}
a({t_s}) \longrightarrow {a_s},~~~\dot a({t_s}) \longrightarrow +\infty,~~~H({t_s}) \longrightarrow +\infty,&\\ \ddot a({t_s})\longrightarrow +\infty,~~~|\rho({t_s})| \longrightarrow \infty, ~~~|p({t_s)}|\longrightarrow \infty.
\end{split}
\end{equation}

\begin{figure*}[hbt!]
  \centering
 \subfigure[]{\includegraphics[width=0.30\textwidth]{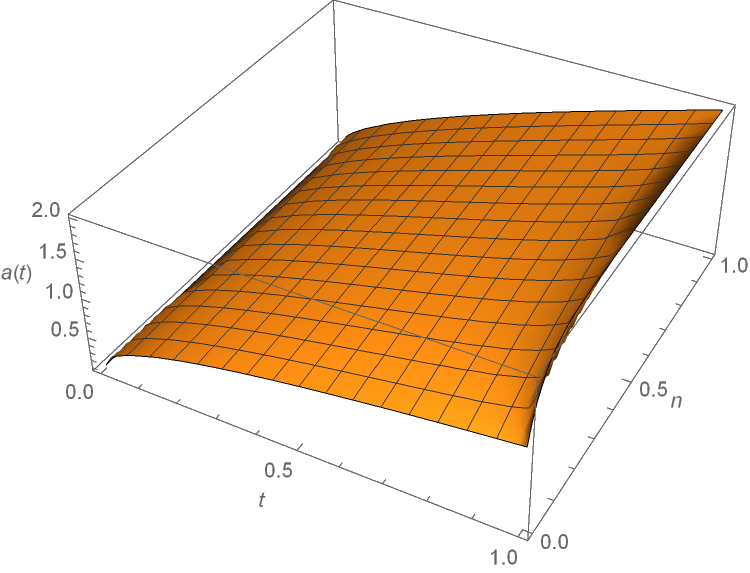}}\subfigure[]{\includegraphics[width=0.34\textwidth]{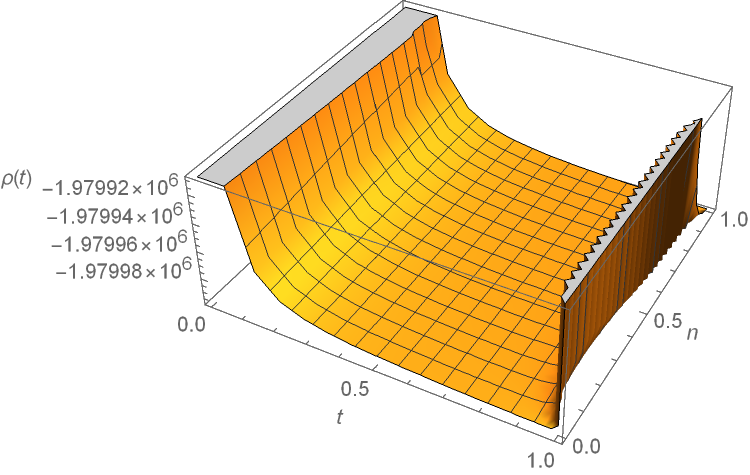}}\subfigure[]{\includegraphics[width=0.32\textwidth]{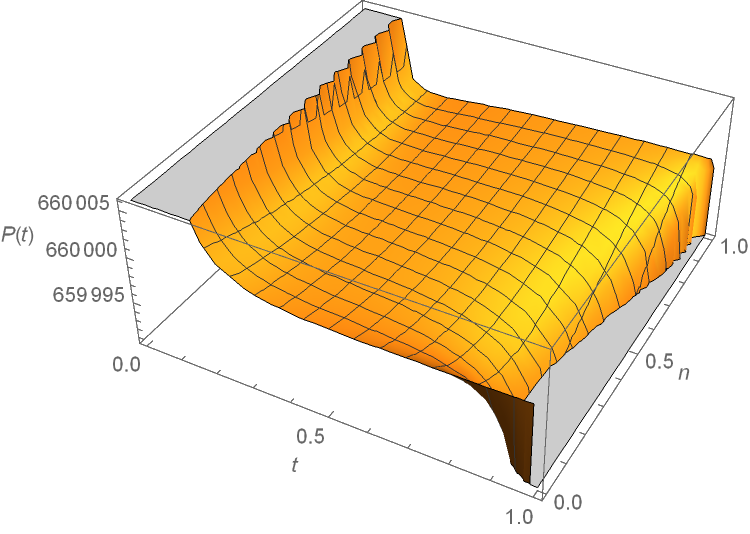}}
  \caption{\small The figures show the evolution of the scale factor (a), the energy density $\rho(t)$ (b), and the pressure $p(t)$ (c) for $0 < q \le1$, $0<n<1$, and $\gamma=0$. We have fixed the values of $q = 0.5$, $a_s=2$, $a_{0}=0.1$, $t_s=1$, $\phi_0=0.01$, $\mathfrak N=22$, and $k=1$.}\label{fig3}
\end{figure*}

Fig. \ref{fig3} shows the evolution of the scale factor $a(t)$, pressure $p(t)$, and density $\rho(t)$ for this case.

\bigskip
\textbf{B:} $t\longrightarrow  ts$ with $0<\gamma<1$,~$\omega_\text{GDE}<-1$,~~$0 < n < 1$, and $0 < q \le1$. Which leads to

\begin{equation}
\begin{split}
a({t_s}) \longrightarrow {a_s},~~~\dot a({t_s}) \longrightarrow +\infty,~~~H({t_s}) \longrightarrow +\infty,&\\ \ddot a({t_s})\longrightarrow +\infty,~~~|\rho({t_s})| \longrightarrow \infty, ~~~|p({t_s)}|\longrightarrow \infty.
\end{split}
\end{equation}

\begin{figure*}[hbt!]
  \centering
  \subfigure[]{\includegraphics[width=0.37\textwidth]
  {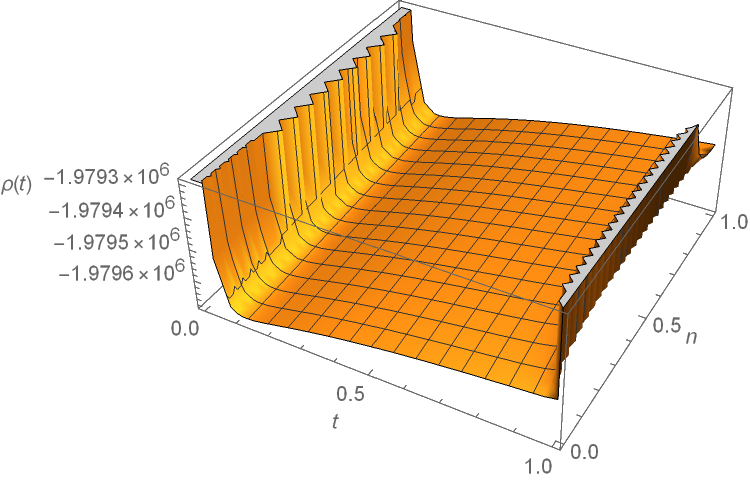}}\subfigure[]{\includegraphics[width=0.32\textwidth]{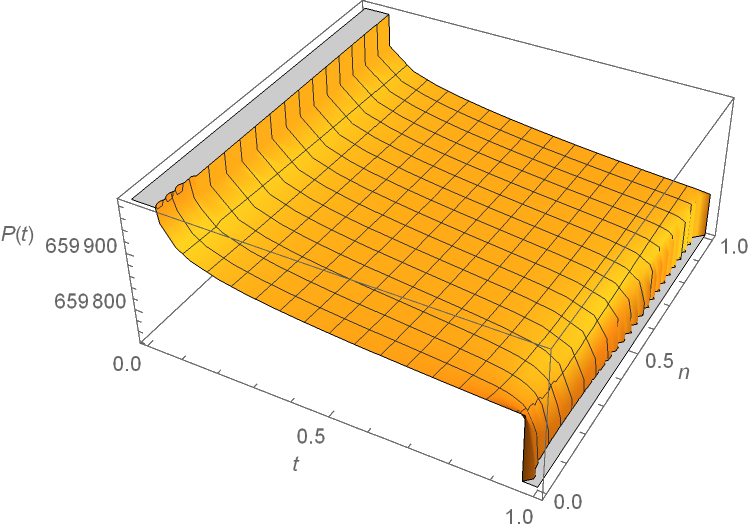}}
  \caption{\small The evolution of the 
 energy density $\rho(t)$ (a), and the pressure $p(t)$ (b), for $0 < q \le1$, $0<n<1$, and $0<\gamma<1$. We have fixed the values of $q = 0.5$, $\gamma=0.0001$, $a_s=2$, $a_{0}=0.1$, $t_s=1$, $\phi_0=0.01$, $\mathfrak N=22$, and $k=1$. The scale factor's evolution is as Fig. \ref{fig3} (a).}\label{fig4}
\end{figure*}

For this particular case, Fig. \ref{fig4} illustrates the evolution of the density $\rho(t)$, and pressure $p(t)$.

\bigskip
\textbf{C:} $t\longrightarrow  ts$ with $-1<\gamma<0$,~$\omega_\text{GDE}>-1$,~~$0 < n < 1$, and $0 < q \le1$. Which leads to 

\begin{equation}
\begin{split}
a({t_s}) \longrightarrow {a_s},~~~\dot a({t_s}) \longrightarrow +\infty,~~~H({t_s}) \longrightarrow +\infty,&\\ \ddot a({t_s})\longrightarrow +\infty,~~~|\rho({t_s})| \longrightarrow \infty, ~~~|p({t_s)}|\longrightarrow \infty.
\end{split}
\end{equation}

\begin{figure*}[hbt!]
  \centering
  \subfigure[]{\includegraphics[width=0.37\textwidth]
  {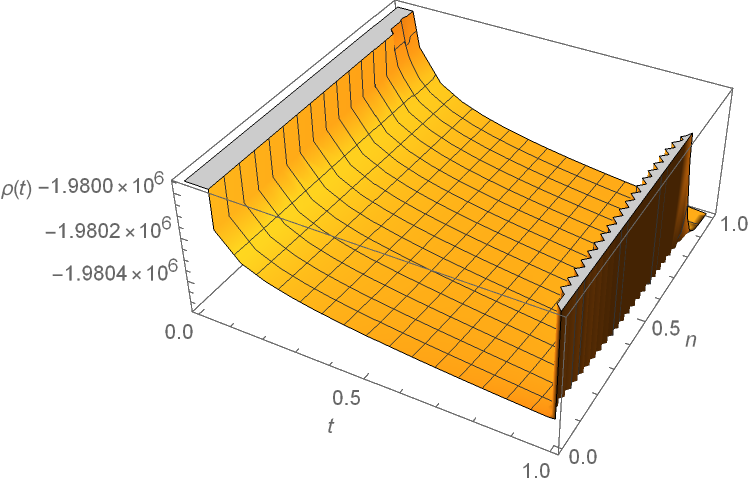}}\subfigure[]{\includegraphics[width=0.32\textwidth]{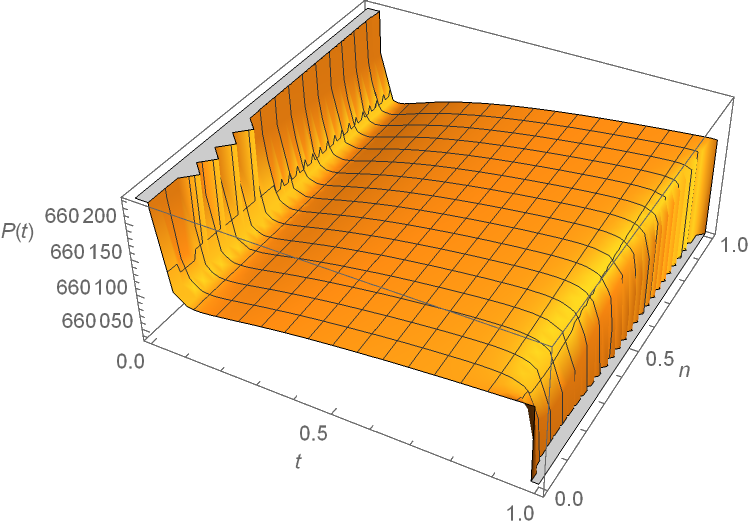}}
  \caption{\small The evolution of the 
  energy density $\rho(t)$ (a), and the pressure $p(t)$ (b), for $0 < q \le1$, $0<n<1$, and -$1<\gamma<0$. We have fixed the values of $q = 0.5$, $\gamma=-0.0001$, $a_s=2$, $a_{0}=0.1$, $t_s=1$, $\phi_0=0.01$, $\mathfrak N=22$, and $k=1$. The scale factor's evolution is as Fig. \ref{fig3} (a).}\label{fig5}
\end{figure*}

Fig. \ref{fig5} shows the evolution of the pressure $p(t)$, and the density $\rho(t)$ for the current case.

In all these cases (Cases A, B, and C from class I), the energy density, the pressure, and
the derivative of the Hubble parameter diverge. Therefore, there will be a singularity of the third type or the same Big Freeze.

\bigskip
\textbf{Class II:}

\bigskip
\textbf{A:} $t\longrightarrow t_s$ with $\gamma=0$, $\omega_\text{GDE}=-1$, $n=1$, and $0 < q \le1$. Which leads to

\begin{equation}
\begin{split}
a({t_s}) \longrightarrow {a_s},~~~\dot a({t_s}) \longrightarrow \dot a_s >0,~~~H({t_s})>0,&\\ \ddot a({t_s})\longrightarrow \ddot a_s \le 0,~~~|\rho({t_s})| \longrightarrow \rho_s, ~~~|p({t_s)}|\longrightarrow p_s.
\end{split}
\end{equation}

\begin{figure*}[hbt!]
  \centering
  \subfigure[]{\includegraphics[width=0.30\textwidth]{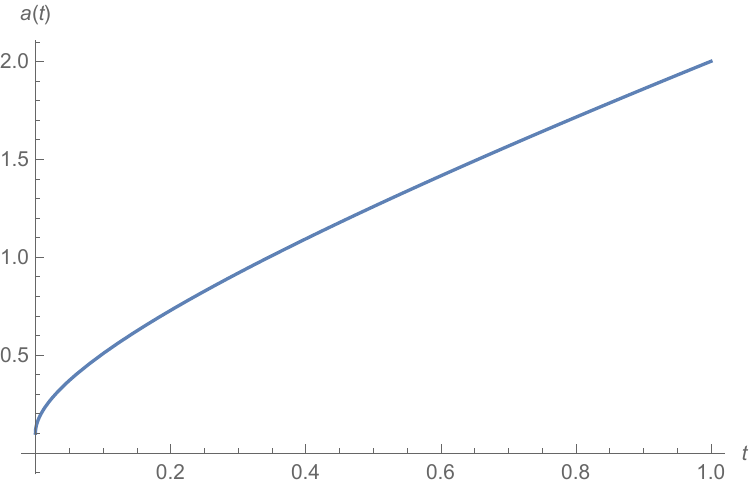}}\subfigure[]{\includegraphics[width=0.34\textwidth]{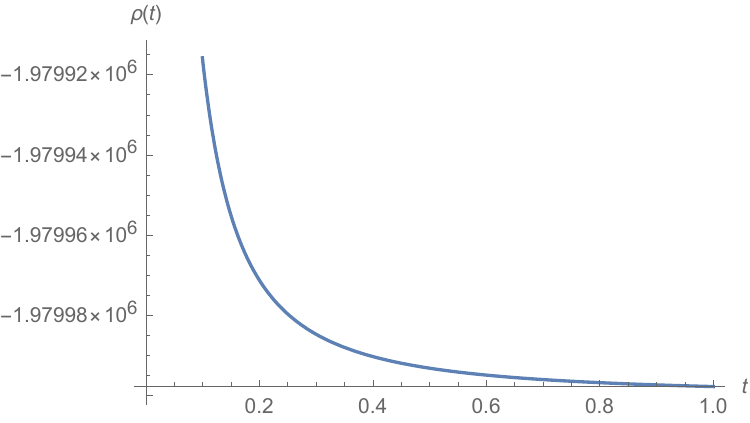}}\subfigure[]{\includegraphics[width=0.32\textwidth]{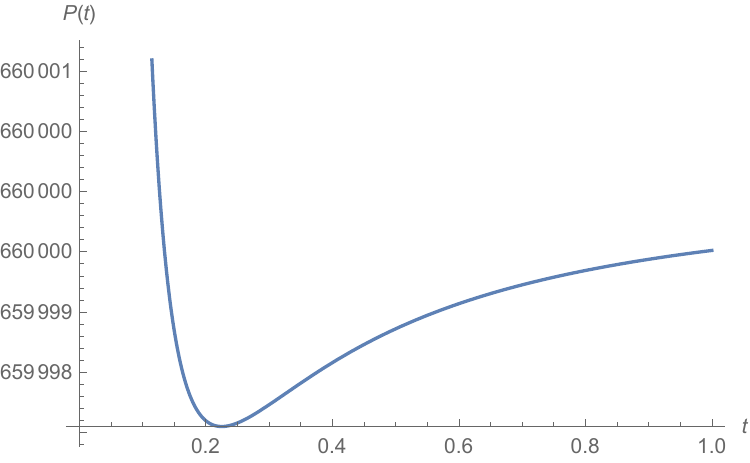}}
  \caption{\small  The scale factor's evolution (a), the energy density $\rho(t)$ (b), and the pressure $p(t)$ (c), for $0 < q \le1$, $n=1$, and, $\gamma=0$. We have fixed the value of $q = 0.5$, $a_s=2$, $a_{0}=0.1$, $t_s=1$, $\phi_0=0.01$, $\mathfrak N=22$, and $k=1$.}\label{fig6}
\end{figure*}

Fig. \ref{fig6} shows the evolution of pressure $p(t)$, the density $\rho(t)$, and the scale factor $a(t)$ for the present case.

\bigskip
\textbf{B:} $t\longrightarrow  ts$ with $0<\gamma<1$, $\omega_\text{GDE}<-1$,~$n=1$ and, $0 < q \le1$. Which leads to

\begin{equation}
\begin{split}
a({t_s}) \longrightarrow {a_s},~~~\dot a({t_s}) \longrightarrow \dot a_s >0,~~~H({t_s})>0,&\\ \ddot a({t_s})\longrightarrow \ddot a_s \le 0,~~~|\rho({t_s})| \longrightarrow \rho_s, ~~~|p({t_s)}|\longrightarrow p_s.
\end{split}
\end{equation}

\begin{figure*}[hbt!]
  \centering
  \subfigure[]{\includegraphics[width=0.38\textwidth]
  {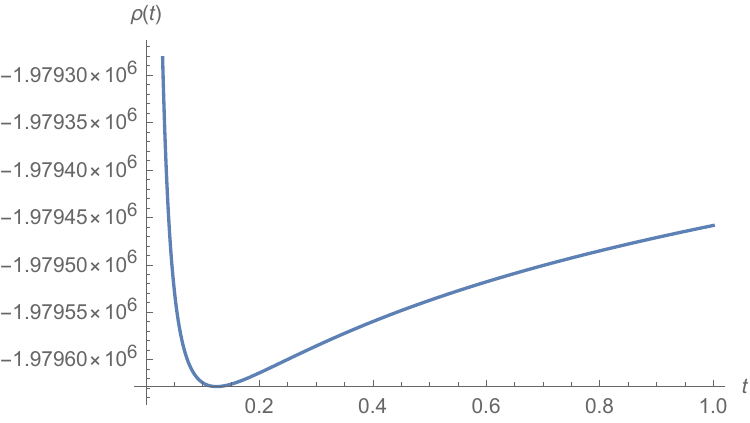}}\subfigure[]{\includegraphics[width=0.32\textwidth]{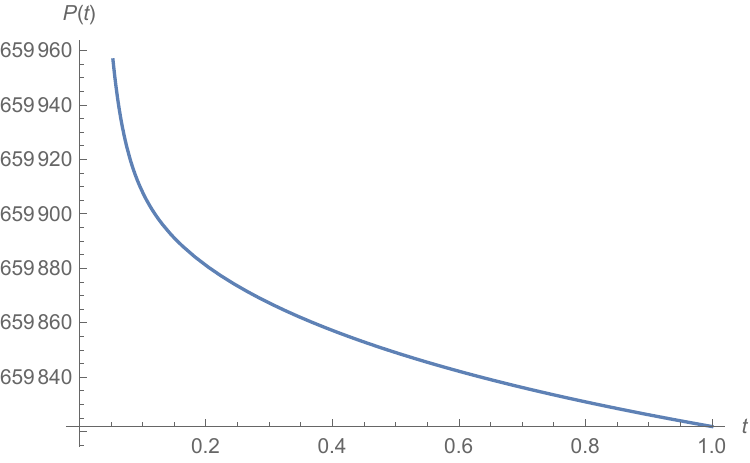}}
  \caption{\small The evolution of the energy density $\rho(t)$ (a), and the pressure $p(t)$ (b) for $0 < q \le1$, $n=1$, and $0<\gamma<1$. We have fixed the value of $q = 0.5$, $\gamma=0.0001$, $a_s=2$, $a_{0}=0.1$, $t_s=1$, $\phi_0=0.01$, $\mathfrak N=22$, and $k=1$. The scale factor's evolution is as Fig. \ref{fig6} (a).}\label{fig7}
\end{figure*}

Fig. \ref{fig7} shows the evolution of pressure $p(t)$ and density $\rho(t)$ for the present case.

\bigskip
\textbf{C:} $t\longrightarrow  ts$ with $-1<\gamma<0$,~$\omega_\text{GDE}>-1$,~$n=1$, and $0 < q \le1$. Which leads to

\begin{equation}
\begin{split}
a({t_s}) \longrightarrow {a_s},~~~\dot a({t_s}) \longrightarrow \dot a_s >0,~~~H({t_s})>0,&\\ \ddot a({t_s})\longrightarrow \ddot a_s \le 0,~~~|\rho({t_s})| \longrightarrow \rho_s, ~~~|p({t_s)}|\longrightarrow p_s.
\end{split}
\end{equation}

\begin{figure*}[hbt!]
  \centering
  \subfigure[]{\includegraphics[width=0.36\textwidth]
  {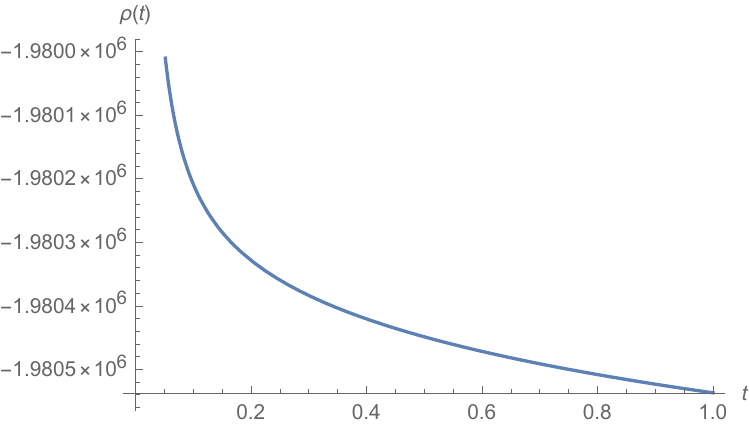}}\subfigure[]{\includegraphics[width=0.32\textwidth]{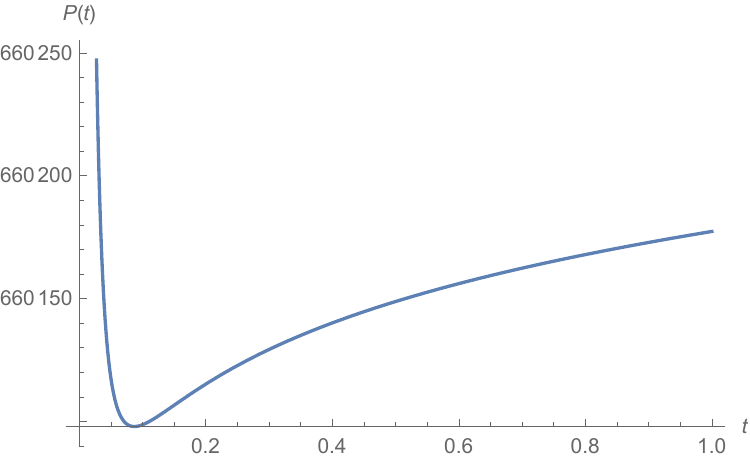}}
  \caption{\small  The evolution of the energy density $\rho(t)$ (a), and the pressure $p(t)$ (b), for $0 < q \le1$, $n=1$, and $-1<\gamma<0$. We have fixed the value of $q = 0.5$, $\gamma=-0.0001$, $a_s=2$, $a_{0}=0.1$, $t_s=1$, $\phi_0=0.01$, $\mathfrak N=22$, and $k=1$. The scale factor's evolution is as Fig. \ref{fig6} (a).}\label{fig8}
\end{figure*}

Fig. \ref{fig8} shows the evolution of pressure $p(t)$ and density $\rho(t)$ for the present case.

The energy density, pressure, Hubble parameter, and derivative of the Hubble parameter are all finite in these situations (situations A, B, and C from class II). Thus, under these particular conditions, there is no finite-time future singularity.

\bigskip
\textbf{Class III:}

\bigskip
\textbf{A:} $t\longrightarrow t_s$ with $\gamma=0$, $\omega_\text{GDE}=-1$, $1 < n < 2$, and $0 < q \le1$. Which leads to

\begin{equation}
\begin{split}
a({t_s}) \longrightarrow {a_s},~~~\dot a({t_s}) \longrightarrow \dot a_s >0,~~~H({t_s})>0,&\\ \ddot a({t_s})\longrightarrow -\infty,~~~|\rho({t_s})| \longrightarrow \rho_s, ~~~|p({t_s)}|\longrightarrow \infty.
\end{split}
\end{equation}

\begin{figure*}[hbt!]
  \centering
  \subfigure[]{\includegraphics[width=0.30\textwidth]{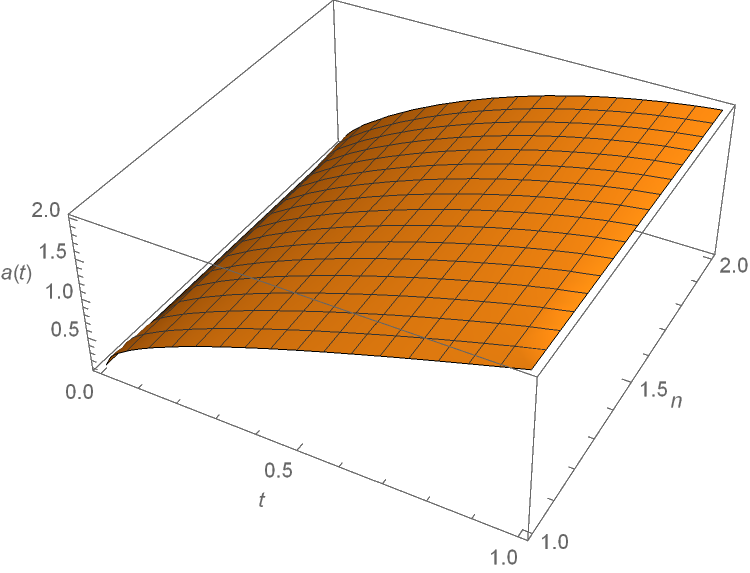}}\subfigure[]{\includegraphics[width=0.34\textwidth]{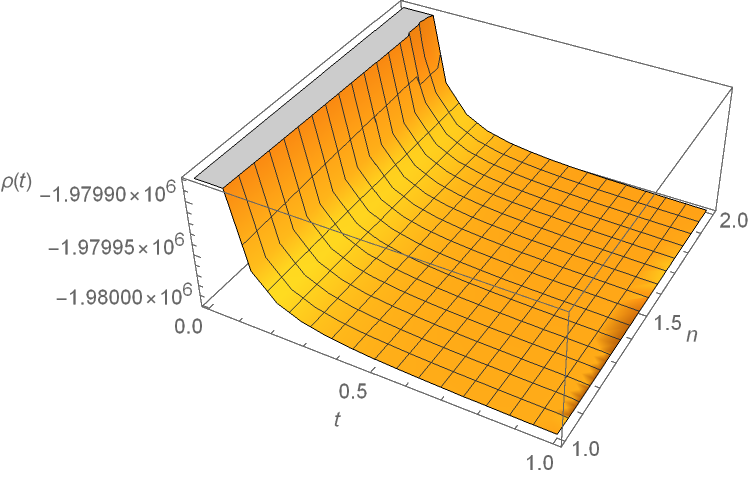}}\subfigure[]{\includegraphics[width=0.32\textwidth]{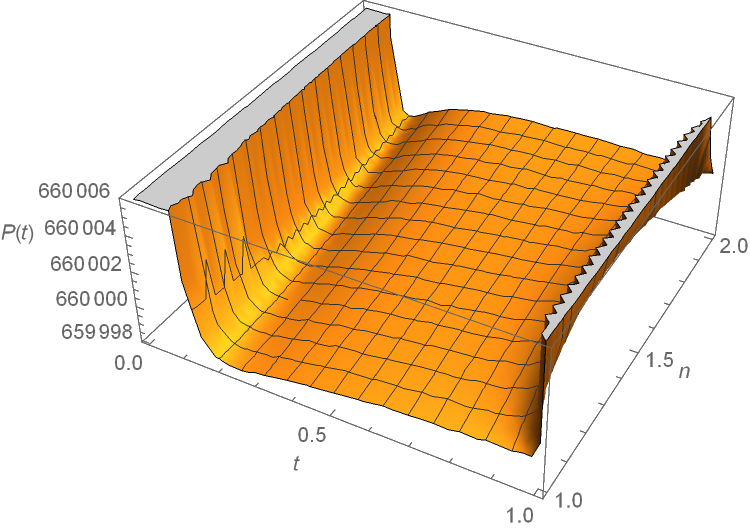}}
  \caption{\small The scale factor's evolution (a), the energy density $\rho(t)$ (b), and the pressure $p(t)$ (c), for $0 < q \le1$, $1<n<2$, and $\gamma=0$. We have fixed the value of $q = 0.5$, $a_s=2$, $a_{0}=0.1$, $t_s=1$, $\phi_0=0.01$, $\mathfrak N=22$, and $k=1$.}\label{fig9}
\end{figure*}

Fig. \ref{fig9} shows the evolution of pressure $p(t)$, density $\rho(t)$, and the scale factor $a(t)$  for the present case.

\bigskip
\textbf{B:} $t\longrightarrow  ts$ with $0<\gamma<1$, $\omega_\text{GDE}<-1$,~$1 < n < 2$, and $0 < q \le1$. Which leads to

\begin{equation}
\begin{split}
a({t_s}) \longrightarrow {a_s},~~~\dot a({t_s}) \longrightarrow \dot a_s >0,~~~H({t_s})>0,&\\ \ddot a({t_s})\longrightarrow -\infty,~~~|\rho({t_s})| \longrightarrow \rho_s, ~~~|p({t_s)}|\longrightarrow \infty.
\end{split}
\end{equation}

\begin{figure*}[hbt!]
  \centering
  \subfigure[]{\includegraphics[width=0.38\textwidth]
  {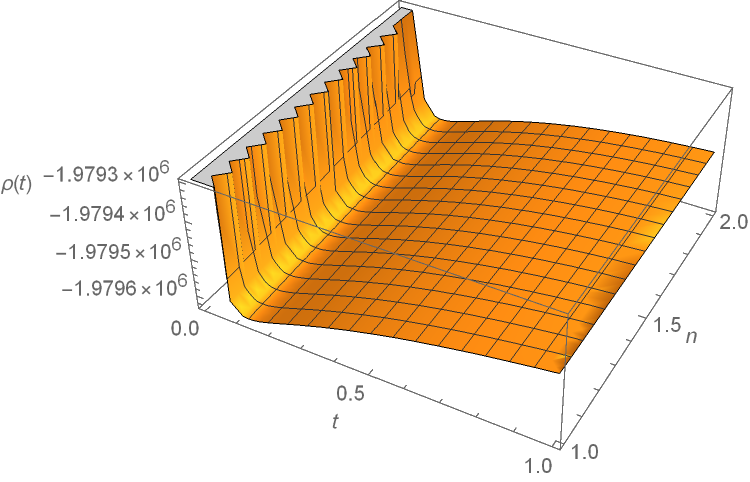}}\subfigure[]{\includegraphics[width=0.32\textwidth]{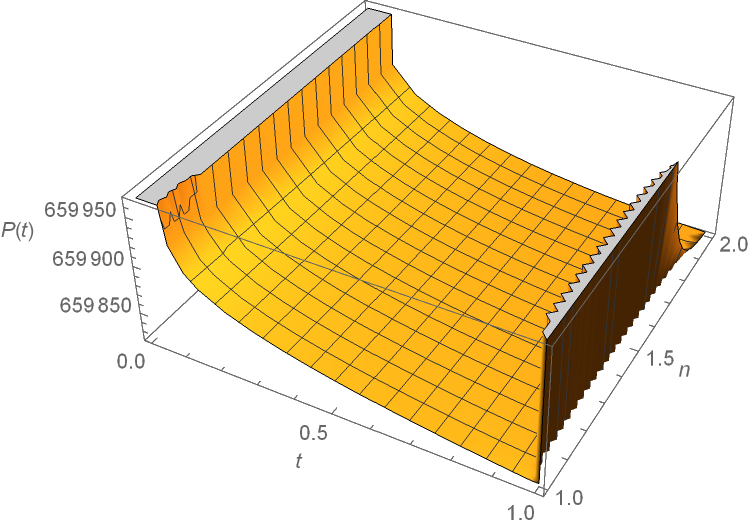}}
  \caption{\small The evolution of the energy density $\rho(t)$ (a), and the pressure $p(t)$ (b) for $0 < q \le1$, $1<n<2$, and $0<\gamma<1$. We have fixed the value of $q = 0.5$, $\gamma=0.0001$, $a_s=2$, $a_{0}=0.1$, $t_s=1$, $\phi_0=0.01$, $\mathfrak N=22$, and $k=1$. The scale factor's evolution is as Fig. \ref{fig9} (a).}\label{fig10}
\end{figure*}

Fig. \ref{fig10} shows the evolution of pressure $p(t)$ and density $\rho(t)$ for the present case.

\bigskip
\textbf{C:} $t\longrightarrow  ts$ with $-1<\gamma<0$,~$\omega_\text{GDE}>-1$,~$1 < n < 2$, and $0 < q \le1$. Which leads to

\begin{equation}
\begin{split}
a({t_s}) \longrightarrow {a_s},~~~\dot a({t_s}) \longrightarrow \dot a_s >0,~~~H({t_s})>0,&\\ \ddot a({t_s})\longrightarrow -\infty,~~~|\rho({t_s})| \longrightarrow \rho_s, ~~~|p({t_s)}|\longrightarrow \infty.
\end{split}
\end{equation}

\begin{figure*}[hbt!]
  \centering
    \subfigure[]{\includegraphics[width=0.38\textwidth]
  {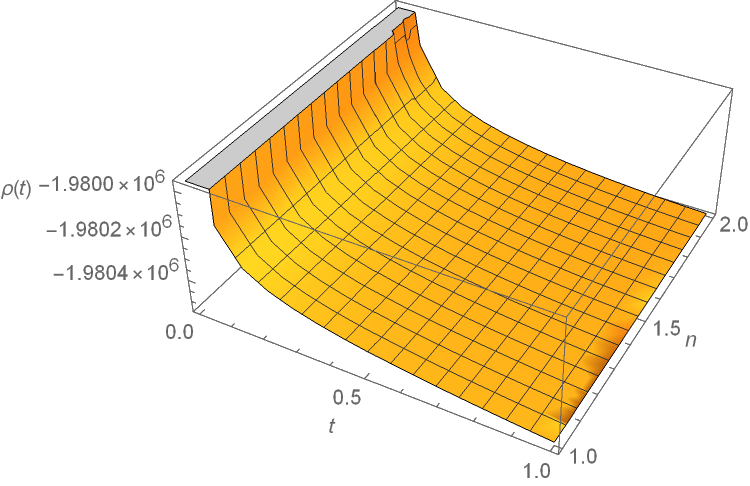}}\subfigure[]{\includegraphics[width=0.32\textwidth]{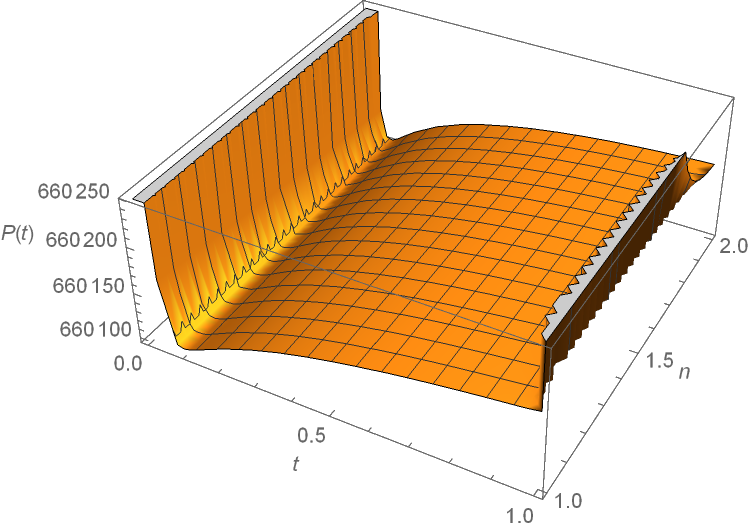}}
  \caption{\small The evolution of the energy density $\rho(t)$ (a), and the pressure $p(t)$ (b), for $0 < q \le1$, $1<n<2$ and $-1<\gamma<0$. We have fixed the value of $q = 0.5$, $\gamma=-0.0001$, $a_s=2$, $a_{0}=0.1$, $t_s=1$, $\phi_0=0.01$, $\mathfrak N=22$, and $k=1$. The scale factor's evolution is as Fig. \ref{fig9} (a).}\label{fig11}
\end{figure*}

Fig. \ref{fig11} shows the evolution of pressure $p(t)$ and density $\rho(t)$ for the present case.

The scale factor and energy density are finite in all these cases (Cases A, B, and C from class III). But the pressure diverges. It represents a cosmological sudden singularity.

\bigskip
\textbf{Class IV:}

\bigskip
\textbf{A:} $t\longrightarrow t_s$ with $\gamma=0$, $\omega_\text{GDE}=-1$, $n\ge 2$, and $0 < q \le1$. Which leads to

\begin{equation}
\begin{split}
a({t_s}) \longrightarrow {a_s},~~~\dot a({t_s}) \longrightarrow \dot a_s >0,~~~H({t_s})>0,&\\ \ddot a({t_s})\longrightarrow \ddot a_s \le 0,~~~|\rho({t_s})| \longrightarrow \rho_s, ~~~|p({t_s)}|\longrightarrow p_s.
\end{split}
\end{equation}

\begin{figure*}[hbt!]
  \centering
  \subfigure[]{\includegraphics[width=0.30\textwidth]{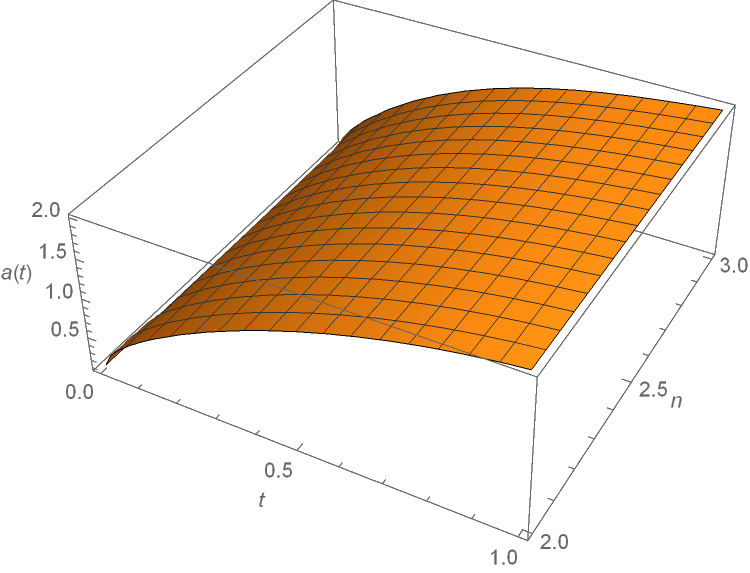}}\subfigure[]{\includegraphics[width=0.34\textwidth]{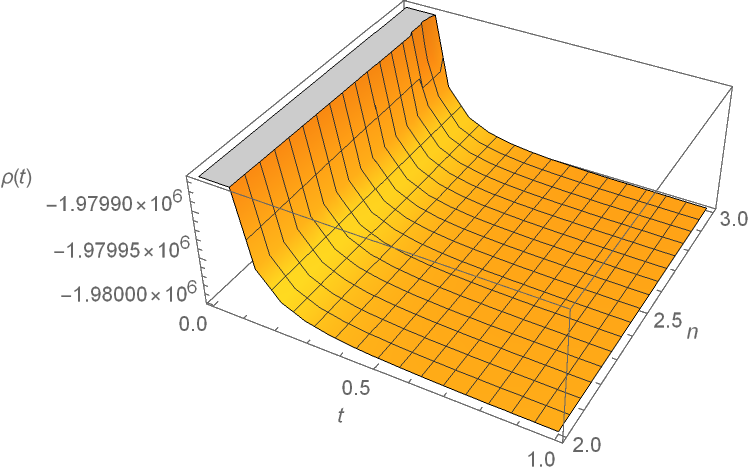}}\subfigure[]{\includegraphics[width=0.32\textwidth]{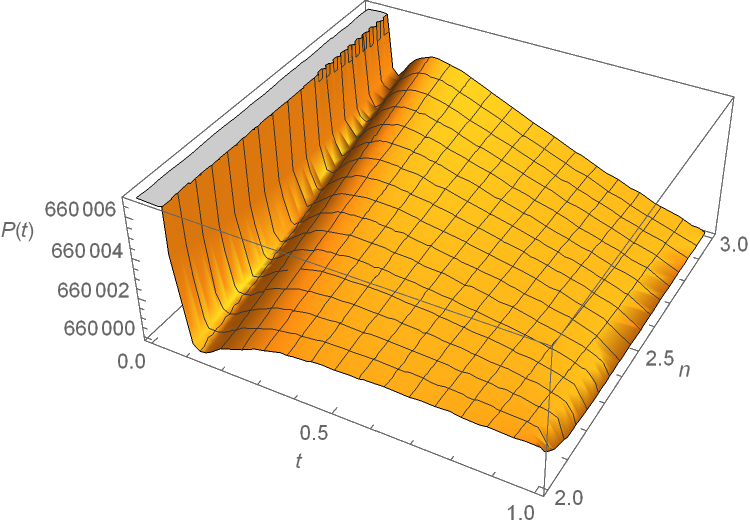}}
  \caption{\small The scale factor's evolution (a), the energy density $\rho(t)$ (b), and the pressure $p(t)$ (c), for $0 < q \le1$, $n>2$, and $\gamma=0$. We have fixed the value of $q = 0.5$, $a_s=2$, $a_{0}=0.1$, $t_s=1$, $\phi_0=0.01$, $\mathfrak N=22$, and $k=1$.}\label{fig12}
\end{figure*}

Fig. \ref{fig12} shows the evolution of pressure $p(t)$, density $\rho(t)$, and the scale factor for the present case.

\bigskip
\textbf{B:} $t\longrightarrow  t_s$ with $0<\gamma<1$, $\omega_\text{GDE}<-1$, ~$n\ge 2$, and $0 < q \le1$. Which leads to

\begin{equation}
\begin{split}
a({t_s}) \longrightarrow {a_s},~~~\dot a({t_s}) \longrightarrow \dot a_s >0,~~~H({t_s})>0,&\\ \ddot a({t_s})\longrightarrow \ddot a_s \le 0,~~~|\rho({t_s})| \longrightarrow \rho_s, ~~~|p({t_s)}|\longrightarrow p_s.
\end{split}
\end{equation}

\begin{figure*}[hbt!]
  \centering
  \subfigure[]{\includegraphics[width=0.38\textwidth]
 {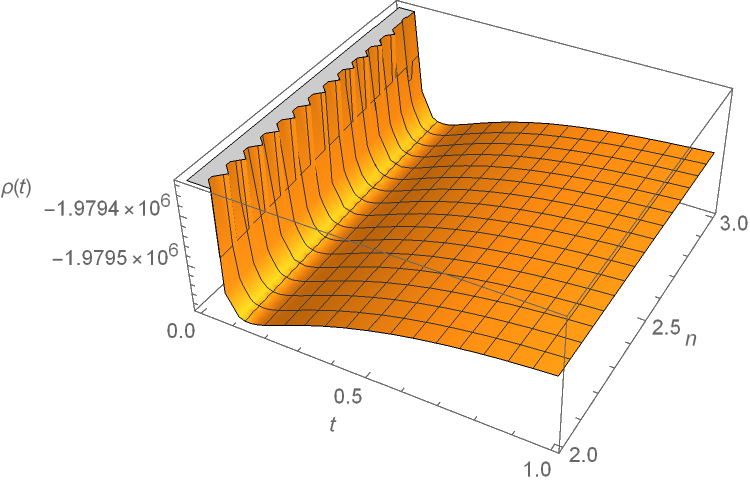}}\subfigure[]{\includegraphics[width=0.32\textwidth]{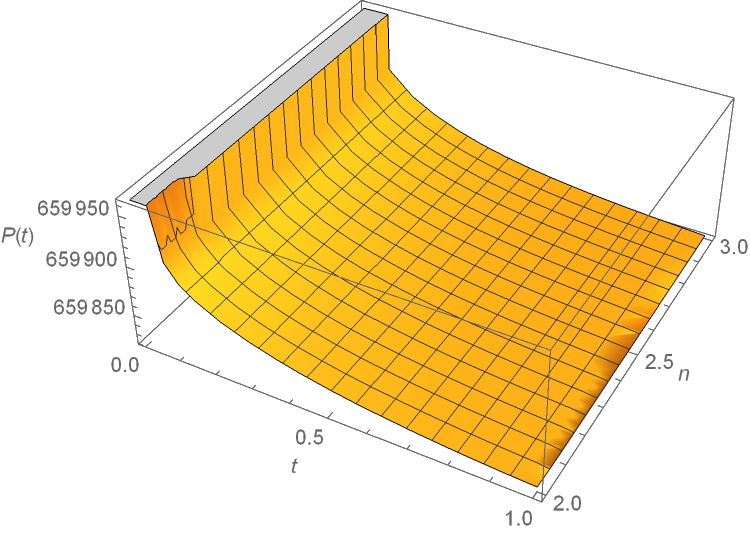}}
  \caption{\small The evolution of the energy density $\rho(t)$ (a), and the pressure $p(t)$ (b), for $0 < q \le1$, $n>2$, and $0<\gamma<1$. We have fixed the value of $q = 0.5$, $\gamma=0.0001$, $a_s=2$, $a_{0}=0.1$, $t_s=1$, $\phi_0=0.01$, $\mathfrak N=22$, and $k=1$. The scale factor's evolution is as Fig. \ref{fig12} (a).}\label{fig13}
\end{figure*}

Fig. \ref{fig13} shows the evolution of pressure $p(t)$, and density $\rho(t)$ for the present case.

\bigskip
\textbf{C:} $t\longrightarrow  t_s$ with $-1<\gamma<0$,~$\omega_\text{GDE}>-1$,~$n\ge 2$, and $0 < q \le1$. Which leads to

\begin{equation}
\begin{split}
a({t_s}) \longrightarrow {a_s},~~~\dot a({t_s}) \longrightarrow \dot a_s >0,~~~H({t_s})>0,&\\ \ddot a({t_s})\longrightarrow \ddot a_s \le 0,~~~|\rho({t_s})| \longrightarrow \rho_s, ~~~|p({t_s)}|\longrightarrow p_s.
\end{split}
\end{equation}

\begin{figure*}[hbt!]
  \centering
  \subfigure[]{\includegraphics[width=0.38\textwidth]
  {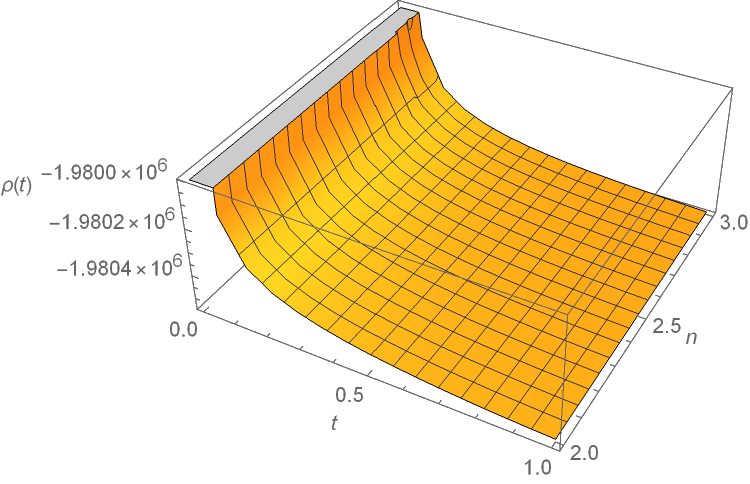}}\subfigure[]{\includegraphics[width=0.32\textwidth]{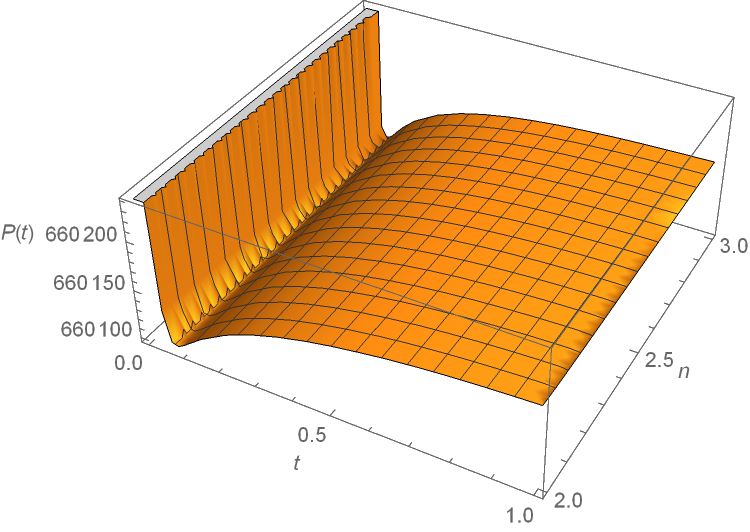}}
  \caption{\small The evolution of the energy density $\rho(t)$ (a), and the pressure $p(t)$ (b), for $0 < q \le1$, $n>2$, and $-1<\gamma<0$. We have fixed the value of $q = 0.5$, $\gamma=-0.0001$, $a_s=2$, $a_{0}=0.1$, $t_s=1$, $\phi_0=0.01$, $\mathfrak N=22$, and $k=1$. The scale factor's evolution is as Fig. \ref{fig12} (a).}\label{fig14}
\end{figure*}

Fig. \ref{fig14} shows the evolution of pressure $p(t)$, density $\rho(t)$, and the scale factor $a(t)$ for the present case.

In each of these scenarios (Cases A, B, and C from Class IV), the scale factor, energy density, and pressure are finite. Thus, under these particular conditions, there is no finite-time future singularity.

\section{Conclusion}\label{Con}We conduct a thorough investigation of both past and finite-time future cosmological singularities within the framework of Covariant Extrinsic Gravity (CEG). To address the initial singularity problem, we employ the emergent universe scenario proposed by Ellis et al., which posits a nonsingular origin for the cosmos.
First, we derive the necessary conditions for the existence of an Einstein static state in this framework. We then perform a detailed stability analysis of this static solution by considering: $(i)$ Homogeneous linear perturbations in the scale factor and matter density, and $(ii)$ Inhomogeneous perturbations, including vector and tensor modes.
Our results demonstrate that, for specific parameter ranges in CEG theory, a stable, nonsingular emergent universe can be realized within a FLRW spacetime with positive spatial curvature. This indicates that the initial singularity problem inherent in General Relativity can be naturally resolved in the CEG framework. Furthermore, we establish that a simple inflationary mechanism can facilitate a smooth transition from this nonsingular initial state to subsequent cosmic epochs, thereby generating a complete and viable cosmological history.
Additionally, we explore the possible finite-time future singularities following the classification by Barrow et al. Our analysis reveals that, depending on the model parameters, an FLRW universe in CEG theory may either encounter or avoid future singularities.
These findings highlight the rich dynamical possibilities within CEG and its potential to address key cosmological challenges beyond the limitations of standard GR.

\bmhead{Acknowledgments}
S.J. acknowledges financial support from the National Council for Scientific and Technological Development, Brazil--CNPq, Grant no. 308131/2022-3.

\section*{Data Availability Statement}
No Data is associated with the manuscript.

\bibliographystyle{sn-mathphys}
\bibliography{sn-bibliography}


\begin{thebibliography}{108}
\ifx \bisbn   \undefined \def \bisbn  #1{ISBN #1}\fi
\ifx \binits  \undefined \def \binits#1{#1}\fi
\ifx \bauthor  \undefined \def \bauthor#1{#1}\fi
\ifx \batitle  \undefined \def \batitle#1{#1}\fi
\ifx \bjtitle  \undefined \def \bjtitle#1{#1}\fi
\ifx \bvolume  \undefined \def \bvolume#1{\textbf{#1}}\fi
\ifx \byear  \undefined \def \byear#1{#1}\fi
\ifx \bissue  \undefined \def \bissue#1{#1}\fi
\ifx \bfpage  \undefined \def \bfpage#1{#1}\fi
\ifx \blpage  \undefined \def \blpage #1{#1}\fi
\ifx \burl  \undefined \def \burl#1{\textsf{#1}}\fi
\ifx \doiurl  \undefined \def \doiurl#1{\url{https://doi.org/#1}}\fi
\ifx \betal  \undefined \def \betal{\textit{et al.}}\fi
\ifx \binstitute  \undefined \def \binstitute#1{#1}\fi
\ifx \binstitutionaled  \undefined \def \binstitutionaled#1{#1}\fi
\ifx \bctitle  \undefined \def \bctitle#1{#1}\fi
\ifx \beditor  \undefined \def \beditor#1{#1}\fi
\ifx \bpublisher  \undefined \def \bpublisher#1{#1}\fi
\ifx \bbtitle  \undefined \def \bbtitle#1{#1}\fi
\ifx \bedition  \undefined \def \bedition#1{#1}\fi
\ifx \bseriesno  \undefined \def \bseriesno#1{#1}\fi
\ifx \blocation  \undefined \def \blocation#1{#1}\fi
\ifx \bsertitle  \undefined \def \bsertitle#1{#1}\fi
\ifx \bsnm \undefined \def \bsnm#1{#1}\fi
\ifx \bsuffix \undefined \def \bsuffix#1{#1}\fi
\ifx \bparticle \undefined \def \bparticle#1{#1}\fi
\ifx \barticle \undefined \def \barticle#1{#1}\fi
\bibcommenthead
\ifx \bconfdate \undefined \def \bconfdate #1{#1}\fi
\ifx \botherref \undefined \def \botherref #1{#1}\fi
\ifx \url \undefined \def \url#1{\textsf{#1}}\fi
\ifx \bchapter \undefined \def \bchapter#1{#1}\fi
\ifx \bbook \undefined \def \bbook#1{#1}\fi
\ifx \bcomment \undefined \def \bcomment#1{#1}\fi
\ifx \oauthor \undefined \def \oauthor#1{#1}\fi
\ifx \citeauthoryear \undefined \def \citeauthoryear#1{#1}\fi
\ifx \endbibitem  \undefined \def \endbibitem {}\fi
\ifx \bconflocation  \undefined \def \bconflocation#1{#1}\fi
\ifx \arxivurl  \undefined \def \arxivurl#1{\textsf{#1}}\fi
\csname PreBibitemsHook\endcsname

\bibitem[\protect\citeauthoryear{Penrose}{1965}]{Penrose:1964wq}
\begin{barticle}
\bauthor{\bsnm{Penrose}, \binits{R.}}:
\batitle{{Gravitational collapse and space-time singularities}}.
\bjtitle{Phys. Rev. Lett.}
\bvolume{14},
\bfpage{57}--\blpage{59}
(\byear{1965})
\doiurl{10.1103/PhysRevLett.14.57}
\end{barticle}
\endbibitem

\bibitem[\protect\citeauthoryear{Hawking}{1967}]{Hawking:1967ju}
\begin{barticle}
\bauthor{\bsnm{Hawking}, \binits{S.}}:
\batitle{{The occurrence of singularities in cosmology. III. Causality and singularities}}.
\bjtitle{Proc. Roy. Soc. Lond. A}
\bvolume{300},
\bfpage{187}--\blpage{201}
(\byear{1967})
\doiurl{10.1098/rspa.1967.0164}
\end{barticle}
\endbibitem

\bibitem[\protect\citeauthoryear{Borde and Vilenkin}{1996}]{Borde:1996pt}
\begin{barticle}
\bauthor{\bsnm{Borde}, \binits{A.}},
\bauthor{\bsnm{Vilenkin}, \binits{A.}}:
\batitle{{Singularities in inflationary cosmology: A Review}}.
\bjtitle{Int. J. Mod. Phys. D}
\bvolume{5},
\bfpage{813}--\blpage{824}
(\byear{1996})
\doiurl{10.1142/S0218271896000497}
{\href{https://arxiv.org/abs/gr-qc/9612036}{{arXiv:gr-qc/9612036}}}
\end{barticle}
\endbibitem

\bibitem[\protect\citeauthoryear{Borde et~al.}{2003}]{Borde:2001nh}
\begin{barticle}
\bauthor{\bsnm{Borde}, \binits{A.}},
\bauthor{\bsnm{Guth}, \binits{A.H.}},
\bauthor{\bsnm{Vilenkin}, \binits{A.}}:
\batitle{{Inflationary space-times are incompletein past directions}}.
\bjtitle{Phys. Rev. Lett.}
\bvolume{90},
\bfpage{151301}
(\byear{2003})
\doiurl{10.1103/PhysRevLett.90.151301}
{\href{https://arxiv.org/abs/gr-qc/0110012}{{arXiv:gr-qc/0110012}}}
\end{barticle}
\endbibitem

\bibitem[\protect\citeauthoryear{Ellis and Maartens}{2004}]{Ellis:2002we}
\begin{barticle}
\bauthor{\bsnm{Ellis}, \binits{G.F.R.}},
\bauthor{\bsnm{Maartens}, \binits{R.}}:
\batitle{{The emergent universe: Inflationary cosmology with no singularity}}.
\bjtitle{Class. Quant. Grav.}
\bvolume{21},
\bfpage{223}--\blpage{232}
(\byear{2004})
\doiurl{10.1088/0264-9381/21/1/015}
{\href{https://arxiv.org/abs/gr-qc/0211082}{{arXiv:gr-qc/0211082}}}
\end{barticle}
\endbibitem

\bibitem[\protect\citeauthoryear{Ellis et~al.}{2004}]{Ellis:2003qz}
\begin{barticle}
\bauthor{\bsnm{Ellis}, \binits{G.F.R.}},
\bauthor{\bsnm{Murugan}, \binits{J.}},
\bauthor{\bsnm{Tsagas}, \binits{C.G.}}:
\batitle{{The Emergent universe: An Explicit construction}}.
\bjtitle{Class. Quant. Grav.}
\bvolume{21}(\bissue{1}),
\bfpage{233}--\blpage{250}
(\byear{2004})
\doiurl{10.1088/0264-9381/21/1/016}
{\href{https://arxiv.org/abs/gr-qc/0307112}{{arXiv:gr-qc/0307112}}}
\end{barticle}
\endbibitem

\bibitem[\protect\citeauthoryear{Khoury et~al.}{2001}]{Khoury:2001wf}
\begin{barticle}
\bauthor{\bsnm{Khoury}, \binits{J.}},
\bauthor{\bsnm{Ovrut}, \binits{B.A.}},
\bauthor{\bsnm{Steinhardt}, \binits{P.J.}},
\bauthor{\bsnm{Turok}, \binits{N.}}:
\batitle{{The Ekpyrotic universe: Colliding branes and the origin of the hot big bang}}.
\bjtitle{Phys. Rev. D}
\bvolume{64},
\bfpage{123522}
(\byear{2001})
\doiurl{10.1103/PhysRevD.64.123522}
{\href{https://arxiv.org/abs/hep-th/0103239}{{arXiv:hep-th/0103239}}}
\end{barticle}
\endbibitem

\bibitem[\protect\citeauthoryear{Steinhardt and Turok}{2002}]{Steinhardt:2001st}
\begin{barticle}
\bauthor{\bsnm{Steinhardt}, \binits{P.J.}},
\bauthor{\bsnm{Turok}, \binits{N.}}:
\batitle{{Cosmic evolution in a cyclic universe}}.
\bjtitle{Phys. Rev. D}
\bvolume{65},
\bfpage{126003}
(\byear{2002})
\doiurl{10.1103/PhysRevD.65.126003}
{\href{https://arxiv.org/abs/hep-th/0111098}{{arXiv:hep-th/0111098}}}
\end{barticle}
\endbibitem

\bibitem[\protect\citeauthoryear{Khoury et~al.}{2004}]{Khoury:2003rt}
\begin{barticle}
\bauthor{\bsnm{Khoury}, \binits{J.}},
\bauthor{\bsnm{Steinhardt}, \binits{P.J.}},
\bauthor{\bsnm{Turok}, \binits{N.}}:
\batitle{{Designing cyclic universe models}}.
\bjtitle{Phys. Rev. Lett.}
\bvolume{92},
\bfpage{031302}
(\byear{2004})
\doiurl{10.1103/PhysRevLett.92.031302}
{\href{https://arxiv.org/abs/hep-th/0307132}{{arXiv:hep-th/0307132}}}
\end{barticle}
\endbibitem

\bibitem[\protect\citeauthoryear{Barrow et~al.}{2004}]{Barrow:2004ad}
\begin{barticle}
\bauthor{\bsnm{Barrow}, \binits{J.D.}},
\bauthor{\bsnm{Kimberly}, \binits{D.}},
\bauthor{\bsnm{Magueijo}, \binits{J.}}:
\batitle{{Bouncing universes with varying constants}}.
\bjtitle{Class. Quant. Grav.}
\bvolume{21},
\bfpage{4289}--\blpage{4296}
(\byear{2004})
\doiurl{10.1088/0264-9381/21/18/001}
{\href{https://arxiv.org/abs/astro-ph/0406369}{{arXiv:astro-ph/0406369}}}
\end{barticle}
\endbibitem

\bibitem[\protect\citeauthoryear{Mukherjee et~al.}{2006}]{Mukherjee:2006ds}
\begin{barticle}
\bauthor{\bsnm{Mukherjee}, \binits{S.}},
\bauthor{\bsnm{Paul}, \binits{B.C.}},
\bauthor{\bsnm{Dadhich}, \binits{N.K.}},
\bauthor{\bsnm{Maharaj}, \binits{S.D.}},
\bauthor{\bsnm{Beesham}, \binits{A.}}:
\batitle{{Emergent Universe with Exotic Matter}}.
\bjtitle{Class. Quant. Grav.}
\bvolume{23},
\bfpage{6927}--\blpage{6934}
(\byear{2006})
\doiurl{10.1088/0264-9381/23/23/020}
{\href{https://arxiv.org/abs/gr-qc/0605134}{{arXiv:gr-qc/0605134}}}
\end{barticle}
\endbibitem

\bibitem[\protect\citeauthoryear{Mulryne et~al.}{2005}]{Mulryne:2005ef}
\begin{barticle}
\bauthor{\bsnm{Mulryne}, \binits{D.J.}},
\bauthor{\bsnm{Tavakol}, \binits{R.}},
\bauthor{\bsnm{Lidsey}, \binits{J.E.}},
\bauthor{\bsnm{Ellis}, \binits{G.F.R.}}:
\batitle{{An Emergent Universe from a loop}}.
\bjtitle{Phys. Rev. D}
\bvolume{71},
\bfpage{123512}
(\byear{2005})
\doiurl{10.1103/PhysRevD.71.123512}
{\href{https://arxiv.org/abs/astro-ph/0502589}{{arXiv:astro-ph/0502589}}}
\end{barticle}
\endbibitem

\bibitem[\protect\citeauthoryear{Parisi et~al.}{2007}]{Parisi:2007kv}
\begin{barticle}
\bauthor{\bsnm{Parisi}, \binits{L.}},
\bauthor{\bsnm{Bruni}, \binits{M.}},
\bauthor{\bsnm{Maartens}, \binits{R.}},
\bauthor{\bsnm{Vandersloot}, \binits{K.}}:
\batitle{{The Einstein static universe in Loop Quantum Cosmology}}.
\bjtitle{Class. Quant. Grav.}
\bvolume{24},
\bfpage{6243}--\blpage{6254}
(\byear{2007})
\doiurl{10.1088/0264-9381/24/24/007}
{\href{https://arxiv.org/abs/0706.4431}{{arXiv:0706.4431}}}
{[gr-qc]}
\end{barticle}
\endbibitem

\bibitem[\protect\citeauthoryear{Goheer et~al.}{2009}]{Goheer:2008tn}
\begin{barticle}
\bauthor{\bsnm{Goheer}, \binits{N.}},
\bauthor{\bsnm{Goswami}, \binits{R.}},
\bauthor{\bsnm{Dunsby}, \binits{P.K.S.}}:
\batitle{{Dynamics of f(R)-cosmologies containing Einstein static models}}.
\bjtitle{Class. Quant. Grav.}
\bvolume{26},
\bfpage{105003}
(\byear{2009})
\doiurl{10.1088/0264-9381/26/10/105003}
{\href{https://arxiv.org/abs/0809.5247}{{arXiv:0809.5247}}}
{[gr-qc]}
\end{barticle}
\endbibitem

\bibitem[\protect\citeauthoryear{Khodadi et~al.}{2016}]{Khodadi:2015fav}
\begin{barticle}
\bauthor{\bsnm{Khodadi}, \binits{M.}},
\bauthor{\bsnm{Heydarzade}, \binits{Y.}},
\bauthor{\bsnm{Darabi}, \binits{F.}},
\bauthor{\bsnm{Saridakis}, \binits{E.N.}}:
\batitle{{Emergent universe in Ho\v{r}ava-Lifshitz-like F(R) gravity}}.
\bjtitle{Phys. Rev. D}
\bvolume{93}(\bissue{12}),
\bfpage{124019}
(\byear{2016})
\doiurl{10.1103/PhysRevD.93.124019}
{\href{https://arxiv.org/abs/1512.08674}{{arXiv:1512.08674}}}
{[gr-qc]}
\end{barticle}
\endbibitem

\bibitem[\protect\citeauthoryear{Wu and Yu}{2011}]{Wu:2011xa}
\begin{barticle}
\bauthor{\bsnm{Wu}, \binits{P.}},
\bauthor{\bsnm{Yu}, \binits{H.}}:
\batitle{{The Stability of the Einstein static state in $f(T)$ gravity}}.
\bjtitle{Phys. Lett. B}
\bvolume{703},
\bfpage{223}--\blpage{227}
(\byear{2011})
\doiurl{10.1016/j.physletb.2011.07.087}
{\href{https://arxiv.org/abs/1108.5908}{{arXiv:1108.5908}}}
{[gr-qc]}
\end{barticle}
\endbibitem

\bibitem[\protect\citeauthoryear{Boehmer}{2004}]{Boehmer:2003iv}
\begin{barticle}
\bauthor{\bsnm{Boehmer}, \binits{C.G.}}:
\batitle{{The Einstein static universe with torsion and the sign problem of the cosmological constant}}.
\bjtitle{Class. Quant. Grav.}
\bvolume{21},
\bfpage{1119}--\blpage{1124}
(\byear{2004})
\doiurl{10.1088/0264-9381/21/4/025}
{\href{https://arxiv.org/abs/gr-qc/0310058}{{arXiv:gr-qc/0310058}}}
\end{barticle}
\endbibitem

\bibitem[\protect\citeauthoryear{Parisi et~al.}{2012}]{Parisi:2012cg}
\begin{barticle}
\bauthor{\bsnm{Parisi}, \binits{L.}},
\bauthor{\bsnm{Radicella}, \binits{N.}},
\bauthor{\bsnm{Vilasi}, \binits{G.}}:
\batitle{{On the stability of the Einstein Static Universe in Massive Gravity}}.
\bjtitle{Phys. Rev. D}
\bvolume{86},
\bfpage{024035}
(\byear{2012})
\doiurl{10.1103/PhysRevD.86.024035}
{\href{https://arxiv.org/abs/1207.3922}{{arXiv:1207.3922}}}
{[gr-qc]}
\end{barticle}
\endbibitem

\bibitem[\protect\citeauthoryear{Lidsey and Mulryne}{2006}]{Lidsey:2006md}
\begin{barticle}
\bauthor{\bsnm{Lidsey}, \binits{J.E.}},
\bauthor{\bsnm{Mulryne}, \binits{D.J.}}:
\batitle{{A Graceful entrance to braneworld inflation}}.
\bjtitle{Phys. Rev. D}
\bvolume{73},
\bfpage{083508}
(\byear{2006})
\doiurl{10.1103/PhysRevD.73.083508}
{\href{https://arxiv.org/abs/hep-th/0601203}{{arXiv:hep-th/0601203}}}
\end{barticle}
\endbibitem

\bibitem[\protect\citeauthoryear{Atazadeh et~al.}{2014}]{Atazadeh:2014xsa}
\begin{barticle}
\bauthor{\bsnm{Atazadeh}, \binits{K.}},
\bauthor{\bsnm{Heydarzade}, \binits{Y.}},
\bauthor{\bsnm{Darabi}, \binits{F.}}:
\batitle{{Einstein Static Universe in Braneworld Scenario}}.
\bjtitle{Phys. Lett. B}
\bvolume{732},
\bfpage{223}--\blpage{227}
(\byear{2014})
\doiurl{10.1016/j.physletb.2014.03.009}
{\href{https://arxiv.org/abs/1401.7638}{{arXiv:1401.7638}}}
{[gr-qc]}
\end{barticle}
\endbibitem

\bibitem[\protect\citeauthoryear{Heydarzade and Darabi}{2015}]{Heydarzade:2015tua}
\begin{barticle}
\bauthor{\bsnm{Heydarzade}, \binits{Y.}},
\bauthor{\bsnm{Darabi}, \binits{F.}}:
\batitle{{Induced Matter Brane Gravity and Einstein Static Universe}}.
\bjtitle{JCAP}
\bvolume{04}(\bissue{04}),
\bfpage{028}
(\byear{2015})
\doiurl{10.1088/1475-7516/2015/04/028}
{\href{https://arxiv.org/abs/1501.02624}{{arXiv:1501.02624}}}
{[gr-qc]}
\end{barticle}
\endbibitem

\bibitem[\protect\citeauthoryear{Ghorani and Heydarzade}{2021}]{Ghorani:2021xrs}
\begin{barticle}
\bauthor{\bsnm{Ghorani}, \binits{E.}},
\bauthor{\bsnm{Heydarzade}, \binits{Y.}}:
\batitle{{On the initial singularity in Kantowski\textendash{}Sachs spacetime}}.
\bjtitle{Eur. Phys. J. C}
\bvolume{81}(\bissue{6}),
\bfpage{557}
(\byear{2021})
\doiurl{10.1140/epjc/s10052-021-09355-7}
{\href{https://arxiv.org/abs/2107.02872}{{arXiv:2107.02872}}}
{[gr-qc]}
\end{barticle}
\endbibitem

\bibitem[\protect\citeauthoryear{Nojiri et~al.}{2005}]{Nojiri:2005sx}
\begin{barticle}
\bauthor{\bsnm{Nojiri}, \binits{S.}},
\bauthor{\bsnm{Odintsov}, \binits{S.D.}},
\bauthor{\bsnm{Tsujikawa}, \binits{S.}}:
\batitle{{Properties of singularities in (phantom) dark energy universe}}.
\bjtitle{Phys. Rev. D}
\bvolume{71},
\bfpage{063004}
(\byear{2005})
\doiurl{10.1103/PhysRevD.71.063004}
{\href{https://arxiv.org/abs/hep-th/0501025}{{arXiv:hep-th/0501025}}}
\end{barticle}
\endbibitem

\bibitem[\protect\citeauthoryear{Lake}{2004}]{Lake:2004fu}
\begin{barticle}
\bauthor{\bsnm{Lake}, \binits{K.}}:
\batitle{{Sudden future singularities in FLRW cosmologies}}.
\bjtitle{Class. Quant. Grav.}
\bvolume{21},
\bfpage{129}
(\byear{2004})
\doiurl{10.1088/0264-9381/21/21/L01}
{\href{https://arxiv.org/abs/gr-qc/0407107}{{arXiv:gr-qc/0407107}}}
\end{barticle}
\endbibitem

\bibitem[\protect\citeauthoryear{Bouhmadi-Lopez et~al.}{2008}]{Bouhmadi-Lopez:2006fwq}
\begin{barticle}
\bauthor{\bsnm{Bouhmadi-Lopez}, \binits{M.}},
\bauthor{\bsnm{Gonzalez-Diaz}, \binits{P.F.}},
\bauthor{\bsnm{Martin-Moruno}, \binits{P.}}:
\batitle{{Worse than a big rip?}}
\bjtitle{Phys. Lett. B}
\bvolume{659},
\bfpage{1}--\blpage{5}
(\byear{2008})
\doiurl{10.1016/j.physletb.2007.10.079}
{\href{https://arxiv.org/abs/gr-qc/0612135}{{arXiv:gr-qc/0612135}}}
\end{barticle}
\endbibitem

\bibitem[\protect\citeauthoryear{Chimento and Richarte}{2016}]{Chimento:2015gum}
\begin{barticle}
\bauthor{\bsnm{Chimento}, \binits{L.P.}},
\bauthor{\bsnm{Richarte}, \binits{M.G.}}:
\batitle{{Big brake singularity is accommodated as an exotic quintessence field}}.
\bjtitle{Phys. Rev. D}
\bvolume{93}(\bissue{4}),
\bfpage{043524}
(\byear{2016})
\doiurl{10.1103/PhysRevD.93.043524}
{\href{https://arxiv.org/abs/1512.02664}{{arXiv:1512.02664}}}
{[gr-qc]}.
\bcomment{[Erratum: Phys.Rev.D 95, 069902 (2017)]}
\end{barticle}
\endbibitem

\bibitem[\protect\citeauthoryear{Boko and Salako}{2021}]{Boko:2021opx}
\begin{barticle}
\bauthor{\bsnm{Boko}, \binits{R.D.}},
\bauthor{\bsnm{Salako}, \binits{I.G.}}:
\batitle{{Infinite-time singularities models and possible avoidance}}.
\bjtitle{Annals Phys.}
\bvolume{432},
\bfpage{168569}
(\byear{2021})
\doiurl{10.1016/j.aop.2021.168569}
\end{barticle}
\endbibitem

\bibitem[\protect\citeauthoryear{Maia et~al.}{2005}]{Maia:2004fq}
\begin{barticle}
\bauthor{\bsnm{Maia}, \binits{M.D.}},
\bauthor{\bsnm{Monte}, \binits{E.M.}},
\bauthor{\bsnm{Maia}, \binits{J.M.F.}},
\bauthor{\bsnm{Alcaniz}, \binits{J.S.}}:
\batitle{{On the geometry of dark energy}}.
\bjtitle{Class. Quant. Grav.}
\bvolume{22},
\bfpage{1623}--\blpage{1636}
(\byear{2005})
\doiurl{10.1088/0264-9381/22/9/010}
{\href{https://arxiv.org/abs/astro-ph/0403072}{{arXiv:astro-ph/0403072}}}
\end{barticle}
\endbibitem

\bibitem[\protect\citeauthoryear{Jalalzadeh and Rostami}{2015}]{Jalalzadeh:2013wza}
\begin{barticle}
\bauthor{\bsnm{Jalalzadeh}, \binits{S.}},
\bauthor{\bsnm{Rostami}, \binits{T.}}:
\batitle{{Covariant extrinsic gravity and the geometric origin of dark energy}}.
\bjtitle{Int. J. Mod. Phys. D}
\bvolume{24}(\bissue{03}),
\bfpage{1550027}
(\byear{2015})
\doiurl{10.1142/S0218271815500273}
{\href{https://arxiv.org/abs/1307.1913}{{arXiv:1307.1913}}}
{[gr-qc]}
\end{barticle}
\endbibitem

\bibitem[\protect\citeauthoryear{Rostami and Jalalzadeh}{2015}]{Rostami:2015ixa}
\begin{barticle}
\bauthor{\bsnm{Rostami}, \binits{T.}},
\bauthor{\bsnm{Jalalzadeh}, \binits{S.}}:
\batitle{{Why the measured cosmological constant is small}}.
\bjtitle{Phys. Dark Univ.}
\bvolume{9-10},
\bfpage{31}--\blpage{36}
(\byear{2015})
\doiurl{10.1016/j.dark.2015.10.001}
{\href{https://arxiv.org/abs/1510.02068}{{arXiv:1510.02068}}}
{[gr-qc]}
\end{barticle}
\endbibitem

\bibitem[\protect\citeauthoryear{Jalalzadeh et~al.}{2023}]{Jalalzadeh:2023upb}
\begin{barticle}
\bauthor{\bsnm{Jalalzadeh}, \binits{R.}},
\bauthor{\bsnm{Jalalzadeh}, \binits{S.}},
\bauthor{\bsnm{Malekolkalami}, \binits{B.}}:
\batitle{{Probing extra dimensions through cosmological observations of dark energy}}.
\bjtitle{Phys. Dark Univ.}
\bvolume{41},
\bfpage{101235}
(\byear{2023})
\doiurl{10.1016/j.dark.2023.101235}
{\href{https://arxiv.org/abs/2304.07229}{{arXiv:2304.07229}}}
{[gr-qc]}
\end{barticle}
\endbibitem

\bibitem[\protect\citeauthoryear{Jalalzadeh et~al.}{2024}]{Jalalzadeh:2024mip}
\begin{barticle}
\bauthor{\bsnm{Jalalzadeh}, \binits{R.}},
\bauthor{\bsnm{Jalalzadeh}, \binits{S.}},
\bauthor{\bsnm{Malekolkalami}, \binits{B.}},
\bauthor{\bsnm{Davari}, \binits{Z.}}:
\batitle{{Observational constraints on FLRW, Bianchi type I and V brane models}}.
\bjtitle{Phys. Dark Univ.}
\bvolume{46},
\bfpage{101591}
(\byear{2024})
\doiurl{10.1016/j.dark.2024.101591}
{\href{https://arxiv.org/abs/2407.15565}{{arXiv:2407.15565}}}
{[gr-qc]}
\end{barticle}
\endbibitem

\bibitem[\protect\citeauthoryear{Kaluza}{1921}]{Kaluza:1921tu}
\begin{barticle}
\bauthor{\bsnm{Kaluza}, \binits{T.}}:
\batitle{Zum unit\"atsproblem der physik}.
\bjtitle{Sitzungsber. Preuss. Akad. Wiss. Berlin (Math. Phys. )}
\bvolume{1921},
\bfpage{966}--\blpage{972}
(\byear{1921})
\doiurl{10.1142/S0218271818700017}
{\href{https://arxiv.org/abs/1803.08616}{{arXiv:1803.08616}}}
{[physics.hist-ph]}
\end{barticle}
\endbibitem

\bibitem[\protect\citeauthoryear{{Klein}}{1926}]{1926ZPhy95K}
\begin{barticle}
\bauthor{\bsnm{{Klein}}, \binits{O.}}:
\batitle{{Quantentheorie und f{\"u}nfdimensionale Relativit{\"a}tstheorie}}.
\bjtitle{Zeitschrift fur Physik}
\bvolume{37}(\bissue{12}),
\bfpage{895}--\blpage{906}
(\byear{1926})
\doiurl{10.1007/BF01397481}
\end{barticle}
\endbibitem

\bibitem[\protect\citeauthoryear{Wesson}{1998}]{wesson-1998}
\begin{bbook}
\bauthor{\bsnm{Wesson}, \binits{P.S.}}:
\bbtitle{{Space-Time-Matter: Modern Kaluza-Klein Theory}}.
\bpublisher{World Scientific Pub Co Inc}, \blocation{???}
(\byear{1998}).
\burl{http://ci.nii.ac.jp/ncid/BA40557473}
\end{bbook}
\endbibitem

\bibitem[\protect\citeauthoryear{Doroud et~al.}{2009}]{Doroud:2009zza}
\begin{barticle}
\bauthor{\bsnm{Doroud}, \binits{N.}},
\bauthor{\bsnm{Rasouli}, \binits{S.M.M.}},
\bauthor{\bsnm{Jalalzadeh}, \binits{S.}}:
\batitle{{A class of cosmological solutions in induced matter theory with conformally flat bulk space}}.
\bjtitle{Gen. Rel. Grav.}
\bvolume{41},
\bfpage{2637}--\blpage{2656}
(\byear{2009})
\doiurl{10.1007/s10714-009-0793-y}
\end{barticle}
\endbibitem

\bibitem[\protect\citeauthoryear{Jalalzadeh and Yazdani}{2008}]{Jalalzadeh:2008xu}
\begin{barticle}
\bauthor{\bsnm{Jalalzadeh}, \binits{S.}},
\bauthor{\bsnm{Yazdani}, \binits{A.M.}}:
\batitle{{Variation of mass in primordial nucleosynthesis as a test of Induced Matter Brane Gravity}}.
\bjtitle{Phys. Lett. B}
\bvolume{664},
\bfpage{229}--\blpage{234}
(\byear{2008})
\doiurl{10.1016/j.physletb.2008.05.041}
{\href{https://arxiv.org/abs/0805.3017}{{arXiv:0805.3017}}}
{[gr-qc]}
\end{barticle}
\endbibitem

\bibitem[\protect\citeauthoryear{Arkani-Hamed et~al.}{1998}]{Arkani-Hamed:1998jmv}
\begin{barticle}
\bauthor{\bsnm{Arkani-Hamed}, \binits{N.}},
\bauthor{\bsnm{Dimopoulos}, \binits{S.}},
\bauthor{\bsnm{Dvali}, \binits{G.R.}}:
\batitle{{The Hierarchy problem and new dimensions at a millimeter}}.
\bjtitle{Phys. Lett. B}
\bvolume{429},
\bfpage{263}--\blpage{272}
(\byear{1998})
\doiurl{10.1016/S0370-2693(98)00466-3}
{\href{https://arxiv.org/abs/hep-ph/9803315}{{arXiv:hep-ph/9803315}}}
\end{barticle}
\endbibitem

\bibitem[\protect\citeauthoryear{Arkani-Hamed et~al.}{1999}]{Arkani-Hamed:1998sfv}
\begin{barticle}
\bauthor{\bsnm{Arkani-Hamed}, \binits{N.}},
\bauthor{\bsnm{Dimopoulos}, \binits{S.}},
\bauthor{\bsnm{Dvali}, \binits{G.R.}}:
\batitle{{Phenomenology, astrophysics and cosmology of theories with submillimeter dimensions and TeV scale quantum gravity}}.
\bjtitle{Phys. Rev. D}
\bvolume{59},
\bfpage{086004}
(\byear{1999})
\doiurl{10.1103/PhysRevD.59.086004}
{\href{https://arxiv.org/abs/hep-ph/9807344}{{arXiv:hep-ph/9807344}}}
\end{barticle}
\endbibitem

\bibitem[\protect\citeauthoryear{Randall and Sundrum}{1999}]{Randall:1999vf}
\begin{barticle}
\bauthor{\bsnm{Randall}, \binits{L.}},
\bauthor{\bsnm{Sundrum}, \binits{R.}}:
\batitle{{An Alternative to compactification}}.
\bjtitle{Phys. Rev. Lett.}
\bvolume{83},
\bfpage{4690}--\blpage{4693}
(\byear{1999})
\doiurl{10.1103/PhysRevLett.83.4690}
{\href{https://arxiv.org/abs/hep-th/9906064}{{arXiv:hep-th/9906064}}}
\end{barticle}
\endbibitem

\bibitem[\protect\citeauthoryear{Dvali et~al.}{2000}]{Dvali:2000hr}
\begin{barticle}
\bauthor{\bsnm{Dvali}, \binits{G.R.}},
\bauthor{\bsnm{Gabadadze}, \binits{G.}},
\bauthor{\bsnm{Porrati}, \binits{M.}}:
\batitle{{4-D gravity on a brane in 5-D Minkowski space}}.
\bjtitle{Phys. Lett. B}
\bvolume{485},
\bfpage{208}--\blpage{214}
(\byear{2000})
\doiurl{10.1016/S0370-2693(00)00669-9}
{\href{https://arxiv.org/abs/hep-th/0005016}{{arXiv:hep-th/0005016}}}
\end{barticle}
\endbibitem

\bibitem[\protect\citeauthoryear{Battye and Carter}{2001}]{Battye:2001pb}
\begin{barticle}
\bauthor{\bsnm{Battye}, \binits{R.A.}},
\bauthor{\bsnm{Carter}, \binits{B.}}:
\batitle{{Generic junction conditions in brane world scenarios}}.
\bjtitle{Phys. Lett. B}
\bvolume{509},
\bfpage{331}--\blpage{336}
(\byear{2001})
\doiurl{10.1016/S0370-2693(01)00495-6}
{\href{https://arxiv.org/abs/hep-th/0101061}{{arXiv:hep-th/0101061}}}
\end{barticle}
\endbibitem

\bibitem[\protect\citeauthoryear{Arkani-Hamed and Schmaltz}{2000}]{Arkani-Hamed:1999ylh}
\begin{barticle}
\bauthor{\bsnm{Arkani-Hamed}, \binits{N.}},
\bauthor{\bsnm{Schmaltz}, \binits{M.}}:
\batitle{{Hierarchies without symmetries from extra dimensions}}.
\bjtitle{Phys. Rev. D}
\bvolume{61},
\bfpage{033005}
(\byear{2000})
\doiurl{10.1103/PhysRevD.61.033005}
{\href{https://arxiv.org/abs/hep-ph/9903417}{{arXiv:hep-ph/9903417}}}
\end{barticle}
\endbibitem

\bibitem[\protect\citeauthoryear{Dzhunushaliev et~al.}{2010}]{Dzhunushaliev:2009va}
\begin{barticle}
\bauthor{\bsnm{Dzhunushaliev}, \binits{V.}},
\bauthor{\bsnm{Folomeev}, \binits{V.}},
\bauthor{\bsnm{Minamitsuji}, \binits{M.}}:
\batitle{{Thick brane solutions}}.
\bjtitle{Rept. Prog. Phys.}
\bvolume{73},
\bfpage{066901}
(\byear{2010})
\doiurl{10.1088/0034-4885/73/6/066901}
{\href{https://arxiv.org/abs/0904.1775}{{arXiv:0904.1775}}}
{[gr-qc]}
\end{barticle}
\endbibitem

\bibitem[\protect\citeauthoryear{Jalalzadeh and Sepangi}{2005a}]{Jalalzadeh:2005ax}
\begin{barticle}
\bauthor{\bsnm{Jalalzadeh}, \binits{S.}},
\bauthor{\bsnm{Sepangi}, \binits{H.R.}}:
\batitle{{Brane gravity and confinement of test particles}}.
\bjtitle{Int. J. Mod. Phys. A}
\bvolume{20},
\bfpage{2275}--\blpage{2281}
(\byear{2005})
\doiurl{10.1142/S0217751X05024493}
\end{barticle}
\endbibitem

\bibitem[\protect\citeauthoryear{Jalalzadeh and Sepangi}{2005b}]{Jalalzadeh:2004uv}
\begin{barticle}
\bauthor{\bsnm{Jalalzadeh}, \binits{S.}},
\bauthor{\bsnm{Sepangi}, \binits{H.R.}}:
\batitle{{Classical and quantum dynamics of confined test particles in brane gravity}}.
\bjtitle{Class. Quant. Grav.}
\bvolume{22},
\bfpage{2035}--\blpage{2048}
(\byear{2005})
\doiurl{10.1088/0264-9381/22/11/008}
{\href{https://arxiv.org/abs/gr-qc/0408004}{{arXiv:gr-qc/0408004}}}
\end{barticle}
\endbibitem

\bibitem[\protect\citeauthoryear{Zeldovich}{1967}]{Zeldovich:1967gd}
\begin{barticle}
\bauthor{\bsnm{Zeldovich}, \binits{Y.B.}}:
\batitle{{Cosmological Constant and Elementary Particles}}.
\bjtitle{JETP Lett.}
\bvolume{6},
\bfpage{316}
(\byear{1967})
\end{barticle}
\endbibitem

\bibitem[\protect\citeauthoryear{Weinberg}{1972}]{Weinberg:1972kfs}
\begin{bbook}
\bauthor{\bsnm{Weinberg}, \binits{S.}}:
\bbtitle{{Gravitation and Cosmology}: {Principles and Applications of the General Theory of Relativity}}.
\bpublisher{John Wiley and Sons},
\blocation{New York}
(\byear{1972})
\end{bbook}
\endbibitem

\bibitem[\protect\citeauthoryear{Labra\~na}{2015}]{Labrana:2013oca}
\begin{barticle}
\bauthor{\bsnm{Labra\~na}, \binits{P.}}:
\batitle{{Emergent universe scenario and the low CMB multipoles}}.
\bjtitle{Phys. Rev. D}
\bvolume{91}(\bissue{8}),
\bfpage{083534}
(\byear{2015})
\doiurl{10.1103/PhysRevD.91.083534}
{\href{https://arxiv.org/abs/1312.6877}{{arXiv:1312.6877}}}
{[astro-ph.CO]}
\end{barticle}
\endbibitem

\bibitem[\protect\citeauthoryear{Bastero-Gil et~al.}{2003}]{Bastero-Gil:2003hfz}
\begin{barticle}
\bauthor{\bsnm{Bastero-Gil}, \binits{M.}},
\bauthor{\bsnm{Freese}, \binits{K.}},
\bauthor{\bsnm{Mersini-Houghton}, \binits{L.}}:
\batitle{{What can WMAP tell us about the very early universe? New physics as an explanation of suppressed large scale power and running spectral index}}.
\bjtitle{Phys. Rev. D}
\bvolume{68},
\bfpage{123514}
(\byear{2003})
\doiurl{10.1103/PhysRevD.68.123514}
{\href{https://arxiv.org/abs/hep-ph/0306289}{{arXiv:hep-ph/0306289}}}
\end{barticle}
\endbibitem

\bibitem[\protect\citeauthoryear{Biswas and Mazumdar}{2014}]{Biswas:2013dry}
\begin{barticle}
\bauthor{\bsnm{Biswas}, \binits{T.}},
\bauthor{\bsnm{Mazumdar}, \binits{A.}}:
\batitle{{Super-Inflation, Non-Singular Bounce, and Low Multipoles}}.
\bjtitle{Class. Quant. Grav.}
\bvolume{31},
\bfpage{025019}
(\byear{2014})
\doiurl{10.1088/0264-9381/31/2/025019}
{\href{https://arxiv.org/abs/1304.3648}{{arXiv:1304.3648}}}
{[hep-th]}
\end{barticle}
\endbibitem

\bibitem[\protect\citeauthoryear{Huang et~al.}{2023}]{Huang:2022hye}
\begin{barticle}
\bauthor{\bsnm{Huang}, \binits{Q.}},
\bauthor{\bsnm{Zhang}, \binits{K.}},
\bauthor{\bsnm{Huang}, \binits{H.}},
\bauthor{\bsnm{Xu}, \binits{B.}},
\bauthor{\bsnm{Tu}, \binits{F.}}:
\batitle{{CMB Power Spectrum in the Emergent Universe with K-Essence}}.
\bjtitle{Universe}
\bvolume{9}(\bissue{5}),
\bfpage{221}
(\byear{2023})
\doiurl{10.3390/universe9050221}
{\href{https://arxiv.org/abs/2210.16166}{{arXiv:2210.16166}}}
{[gr-qc]}
\end{barticle}
\endbibitem

\bibitem[\protect\citeauthoryear{R\'\i{}os et~al.}{2016}]{Rios:2016trs}
\begin{barticle}
\bauthor{\bsnm{R\'\i{}os}, \binits{C.}},
\bauthor{\bsnm{Labra\~na}, \binits{P.}},
\bauthor{\bsnm{Cid}, \binits{A.}}:
\batitle{{The Emergent Universe and the Anomalies in the Cosmic Microwave Background}}.
\bjtitle{J. Phys. Conf. Ser.}
\bvolume{720},
\bfpage{012008}
(\byear{2016})
\doiurl{10.1088/1742-6596/720/1/012008}
\end{barticle}
\endbibitem

\bibitem[\protect\citeauthoryear{Maia and Mecklenburg}{1984}]{Maia:1983zh}
\begin{barticle}
\bauthor{\bsnm{Maia}, \binits{M.D.}},
\bauthor{\bsnm{Mecklenburg}, \binits{W.}}:
\batitle{{Aspects of high dimensional theories in embedding spaces}}.
\bjtitle{J. Math. Phys.}
\bvolume{25},
\bfpage{3047}
(\byear{1984})
\doiurl{10.1063/1.526020}
\end{barticle}
\endbibitem

\bibitem[\protect\citeauthoryear{Monte and Maia}{1996}]{twist}
\begin{barticle}
\bauthor{\bsnm{Monte}, \binits{E.M.}},
\bauthor{\bsnm{Maia}, \binits{M.D.}}:
\batitle{{The twisting connection of space-time}}.
\bjtitle{J. Math. Phys.}
\bvolume{37},
\bfpage{1972}--\blpage{1981}
(\byear{1996})
\doiurl{10.1063/1.531488}
\end{barticle}
\endbibitem

\bibitem[\protect\citeauthoryear{Maia}{1989}]{Yang}
\begin{barticle}
\bauthor{\bsnm{Maia}, \binits{M.D.}}:
\batitle{{On the Integrability Conditions for Extended Objects}}.
\bjtitle{Class. Quant. Grav.}
\bvolume{6},
\bfpage{173}--\blpage{183}
(\byear{1989})
\doiurl{10.1088/0264-9381/6/2/011}
\end{barticle}
\endbibitem

\bibitem[\protect\citeauthoryear{Shiromizu et~al.}{2000}]{Shiromizu:1999wj}
\begin{barticle}
\bauthor{\bsnm{Shiromizu}, \binits{T.}},
\bauthor{\bsnm{Maeda}, \binits{K.-i.}},
\bauthor{\bsnm{Sasaki}, \binits{M.}}:
\batitle{{The Einstein equation on the 3-brane world}}.
\bjtitle{Phys. Rev. D}
\bvolume{62},
\bfpage{024012}
(\byear{2000})
\doiurl{10.1103/PhysRevD.62.024012}
{\href{https://arxiv.org/abs/gr-qc/9910076}{{arXiv:gr-qc/9910076}}}
\end{barticle}
\endbibitem

\bibitem[\protect\citeauthoryear{Williams et~al.}{2004}]{Williams:2003wu}
\begin{barticle}
\bauthor{\bsnm{Williams}, \binits{J.G.}},
\bauthor{\bsnm{Turyshev}, \binits{S.G.}},
\bauthor{\bsnm{Murphy}, \binits{T.W.} \bsuffix{Jr.}}:
\batitle{{Improving LLR tests of gravitational theory}}.
\bjtitle{Int. J. Mod. Phys. D}
\bvolume{13},
\bfpage{567}--\blpage{582}
(\byear{2004})
\doiurl{10.1142/S0218271804004682}
{\href{https://arxiv.org/abs/gr-qc/0311021}{{arXiv:gr-qc/0311021}}}
\end{barticle}
\endbibitem

\bibitem[\protect\citeauthoryear{Merkowitz}{2010}]{Merkowitz:2010kka}
\begin{barticle}
\bauthor{\bsnm{Merkowitz}, \binits{S.M.}}:
\batitle{{Tests of Gravity Using Lunar Laser Ranging}}.
\bjtitle{Living Rev. Rel.}
\bvolume{13},
\bfpage{7}
(\byear{2010})
\doiurl{10.12942/lrr-2010-7}
\end{barticle}
\endbibitem

\bibitem[\protect\citeauthoryear{Gaztanaga et~al.}{2002}]{Gaztanaga:2001fh}
\begin{barticle}
\bauthor{\bsnm{Gaztanaga}, \binits{E.}},
\bauthor{\bsnm{Garcia-Berro}, \binits{E.}},
\bauthor{\bsnm{Isern}, \binits{J.}},
\bauthor{\bsnm{Bravo}, \binits{E.}},
\bauthor{\bsnm{Dominguez}, \binits{I.}}:
\batitle{{Bounds on the possible evolution of the gravitational constant from cosmological type Ia supernovae}}.
\bjtitle{Phys. Rev. D}
\bvolume{65},
\bfpage{023506}
(\byear{2002})
\doiurl{10.1103/PhysRevD.65.023506}
{\href{https://arxiv.org/abs/astro-ph/0109299}{{arXiv:astro-ph/0109299}}}
\end{barticle}
\endbibitem

\bibitem[\protect\citeauthoryear{{{Guenther}, D.~B. and {Krauss}, L.~M. and {Demarque}, P.}}{1998}]{1998ApJ871G}
\begin{barticle}
\bauthor{\bsnm{{{Guenther}, D.~B. and {Krauss}, L.~M. and {Demarque}, P.}}}:
\batitle{{Testing the Constancy of the Gravitational Constant Using Helioseismology}}.
\bjtitle{Astrophys. J.}
\bvolume{498}(\bissue{2}),
\bfpage{871}--\blpage{876}
(\byear{1998})
\doiurl{10.1086/305567}
\end{barticle}
\endbibitem

\bibitem[\protect\citeauthoryear{Damour et~al.}{1988}]{Damour:1988zz}
\begin{barticle}
\bauthor{\bsnm{Damour}, \binits{T.}},
\bauthor{\bsnm{Gibbons}, \binits{G.W.}},
\bauthor{\bsnm{Taylor}, \binits{J.H.}}:
\batitle{{Limits on the Variability of G Using Binary-Pulsar Data}}.
\bjtitle{Phys. Rev. Lett.}
\bvolume{61},
\bfpage{1151}--\blpage{1154}
(\byear{1988})
\doiurl{10.1103/PhysRevLett.61.1151}
\end{barticle}
\endbibitem

\bibitem[\protect\citeauthoryear{Barrow et~al.}{2003}]{Barrow:2003ni}
\begin{barticle}
\bauthor{\bsnm{Barrow}, \binits{J.D.}},
\bauthor{\bsnm{Ellis}, \binits{G.F.R.}},
\bauthor{\bsnm{Maartens}, \binits{R.}},
\bauthor{\bsnm{Tsagas}, \binits{C.G.}}:
\batitle{{On the stability of the Einstein static universe}}.
\bjtitle{Class. Quant. Grav.}
\bvolume{20},
\bfpage{155}--\blpage{164}
(\byear{2003})
\doiurl{10.1088/0264-9381/20/11/102}
{\href{https://arxiv.org/abs/gr-qc/0302094}{{arXiv:gr-qc/0302094}}}
\end{barticle}
\endbibitem

\bibitem[\protect\citeauthoryear{Bruni et~al.}{1992}]{Bruni:1992dg}
\begin{barticle}
\bauthor{\bsnm{Bruni}, \binits{M.}},
\bauthor{\bsnm{Dunsby}, \binits{P.K.S.}},
\bauthor{\bsnm{Ellis}, \binits{G.F.R.}}:
\batitle{{Cosmological perturbations and the physical meaning of gauge invariant variables}}.
\bjtitle{Astrophys. J.}
\bvolume{395},
\bfpage{34}--\blpage{53}
(\byear{1992})
\doiurl{10.1086/171629}
\end{barticle}
\endbibitem

\bibitem[\protect\citeauthoryear{Gibbons}{1987}]{Gibbons:1987jt}
\begin{barticle}
\bauthor{\bsnm{Gibbons}, \binits{G.W.}}:
\batitle{{The Entropy and Stability of the Universe}}.
\bjtitle{Nucl. Phys. B}
\bvolume{292},
\bfpage{784}--\blpage{792}
(\byear{1987})
\doiurl{10.1016/0550-3213(87)90670-5}
\end{barticle}
\endbibitem

\bibitem[\protect\citeauthoryear{Gibbons}{1988}]{Gibbons:1988bm}
\begin{barticle}
\bauthor{\bsnm{Gibbons}, \binits{G.W.}}:
\batitle{{Sobolev's Inequality, Jensen's Theorem and the Mass and Entropy of the Universe}}.
\bjtitle{Nucl. Phys. B}
\bvolume{310},
\bfpage{636}--\blpage{642}
(\byear{1988})
\doiurl{10.1016/0550-3213(88)90096-X}
\end{barticle}
\endbibitem

\bibitem[\protect\citeauthoryear{Dunsby et~al.}{1997}]{Dunsby:1997fyr}
\begin{barticle}
\bauthor{\bsnm{Dunsby}, \binits{P.K.S.}},
\bauthor{\bsnm{Bassett}, \binits{B.A.C.C.}},
\bauthor{\bsnm{Ellis}, \binits{G.F.R.}}:
\batitle{{Covariant analysis of gravitational waves in a cosmological context}}.
\bjtitle{Class. Quant. Grav.}
\bvolume{14},
\bfpage{1215}--\blpage{1222}
(\byear{1997})
\doiurl{10.1088/0264-9381/14/5/023}
{\href{https://arxiv.org/abs/gr-qc/9811092}{{arXiv:gr-qc/9811092}}}
\end{barticle}
\endbibitem

\bibitem[\protect\citeauthoryear{Challinor}{2000}]{Challinor:1999xz}
\begin{barticle}
\bauthor{\bsnm{Challinor}, \binits{A.}}:
\batitle{{Microwave background anisotropies from gravitational waves: The (1+3) covariant approach}}.
\bjtitle{Class. Quant. Grav.}
\bvolume{17},
\bfpage{871}--\blpage{889}
(\byear{2000})
\doiurl{10.1088/0264-9381/17/4/309}
{\href{https://arxiv.org/abs/astro-ph/9906474}{{arXiv:astro-ph/9906474}}}
\end{barticle}
\endbibitem

\bibitem[\protect\citeauthoryear{Maartens et~al.}{2001}]{Maartens:2001zu}
\begin{barticle}
\bauthor{\bsnm{Maartens}, \binits{R.}},
\bauthor{\bsnm{Tsagas}, \binits{C.G.}},
\bauthor{\bsnm{Ungarelli}, \binits{C.}}:
\batitle{{Magnetized gravitational waves}}.
\bjtitle{Phys. Rev. D}
\bvolume{63},
\bfpage{123507}
(\byear{2001})
\doiurl{10.1103/PhysRevD.63.123507}
{\href{https://arxiv.org/abs/astro-ph/0101151}{{arXiv:astro-ph/0101151}}}
\end{barticle}
\endbibitem

\bibitem[\protect\citeauthoryear{Langlois and Vernizzi}{2005}]{Langlois:2005ii}
\begin{barticle}
\bauthor{\bsnm{Langlois}, \binits{D.}},
\bauthor{\bsnm{Vernizzi}, \binits{F.}}:
\batitle{{Evolution of non-linear cosmological perturbations}}.
\bjtitle{Phys. Rev. Lett.}
\bvolume{95},
\bfpage{091303}
(\byear{2005})
\doiurl{10.1103/PhysRevLett.95.091303}
{\href{https://arxiv.org/abs/astro-ph/0503416}{{arXiv:astro-ph/0503416}}}
\end{barticle}
\endbibitem

\bibitem[\protect\citeauthoryear{Unruh and Losic}{2008}]{Unruh:2008zza}
\begin{barticle}
\bauthor{\bsnm{Unruh}, \binits{W.G.}},
\bauthor{\bsnm{Losic}, \binits{B.}}:
\batitle{{Aspects of nonlinear perturbations in cosmological models}}.
\bjtitle{Class. Quant. Grav.}
\bvolume{25},
\bfpage{154012}
(\byear{2008})
\doiurl{10.1088/0264-9381/25/15/154012}
\end{barticle}
\endbibitem

\bibitem[\protect\citeauthoryear{Mondal}{2021}]{Mondal:2021pyz}
\begin{barticle}
\bauthor{\bsnm{Mondal}, \binits{P.}}:
\batitle{{The linear stability of the $n$ + 1 dimensional FLRW spacetimes}}.
\bjtitle{Class. Quant. Grav.}
\bvolume{38}(\bissue{22}),
\bfpage{225009}
(\byear{2021})
\doiurl{10.1088/1361-6382/ac2be2}
\end{barticle}
\endbibitem

\bibitem[\protect\citeauthoryear{Mondal}{2024}]{Mondal:2022oyj}
\begin{barticle}
\bauthor{\bsnm{Mondal}, \binits{P.}}:
\batitle{{The nonlinear stability of (n + 1)-dimensional FLRW spacetimes}}.
\bjtitle{J. Hyperbol. Diff. Equat.}
\bvolume{21}(\bissue{02}),
\bfpage{329}--\blpage{422}
(\byear{2024})
\doiurl{10.1142/S0219891624500115}
{\href{https://arxiv.org/abs/2203.04785}{{arXiv:2203.04785}}}
{[gr-qc]}
\end{barticle}
\endbibitem

\bibitem[\protect\citeauthoryear{Ijjas et~al.}{2019}]{Ijjas:2018cdm}
\begin{barticle}
\bauthor{\bsnm{Ijjas}, \binits{A.}},
\bauthor{\bsnm{Pretorius}, \binits{F.}},
\bauthor{\bsnm{Steinhardt}, \binits{P.J.}}:
\batitle{{Stability and the Gauge Problem in Non-Perturbative Cosmology}}.
\bjtitle{JCAP}
\bvolume{01},
\bfpage{015}
(\byear{2019})
\doiurl{10.1088/1475-7516/2019/01/015}
{\href{https://arxiv.org/abs/1809.07010}{{arXiv:1809.07010}}}
{[gr-qc]}
\end{barticle}
\endbibitem

\bibitem[\protect\citeauthoryear{Camara et~al.}{2010}]{Camara:2010zm}
\begin{barticle}
\bauthor{\bsnm{Camara}, \binits{P.G.}},
\bauthor{\bsnm{Condeescu}, \binits{C.}},
\bauthor{\bsnm{Dudas}, \binits{E.}},
\bauthor{\bsnm{Lennek}, \binits{M.}}:
\batitle{{Non-perturbative Vacuum Destabilization and D-brane Dynamics}}.
\bjtitle{JHEP}
\bvolume{06},
\bfpage{062}
(\byear{2010})
\doiurl{10.1007/JHEP06(2010)062}
{\href{https://arxiv.org/abs/1003.5805}{{arXiv:1003.5805}}}
{[hep-th]}
\end{barticle}
\endbibitem

\bibitem[\protect\citeauthoryear{Shuhmaher and Brandenberger}{2006}]{Shuhmaher:2005mf}
\begin{barticle}
\bauthor{\bsnm{Shuhmaher}, \binits{N.}},
\bauthor{\bsnm{Brandenberger}, \binits{R.}}:
\batitle{{Non-perturbative instabilities as a solution of the cosmological moduli problem}}.
\bjtitle{Phys. Rev. D}
\bvolume{73},
\bfpage{043519}
(\byear{2006})
\doiurl{10.1103/PhysRevD.73.043519}
{\href{https://arxiv.org/abs/hep-th/0507103}{{arXiv:hep-th/0507103}}}
\end{barticle}
\endbibitem

\bibitem[\protect\citeauthoryear{McInnes}{2004}]{McInnes:2004zm}
\begin{barticle}
\bauthor{\bsnm{McInnes}, \binits{B.}}:
\batitle{{APS instability and the topology of the brane-world}}.
\bjtitle{Phys. Lett. B}
\bvolume{593},
\bfpage{10}--\blpage{16}
(\byear{2004})
\doiurl{10.1016/j.physletb.2004.05.004}
{\href{https://arxiv.org/abs/hep-th/0401035}{{arXiv:hep-th/0401035}}}
\end{barticle}
\endbibitem

\bibitem[\protect\citeauthoryear{Yoshida et~al.}{2017}]{Yoshida:2017swb}
\begin{barticle}
\bauthor{\bsnm{Yoshida}, \binits{D.}},
\bauthor{\bsnm{Quintin}, \binits{J.}},
\bauthor{\bsnm{Yamaguchi}, \binits{M.}},
\bauthor{\bsnm{Brandenberger}, \binits{R.H.}}:
\batitle{{Cosmological perturbations and stability of nonsingular cosmologies with limiting curvature}}.
\bjtitle{Phys. Rev. D}
\bvolume{96}(\bissue{4}),
\bfpage{043502}
(\byear{2017})
\doiurl{10.1103/PhysRevD.96.043502}
{\href{https://arxiv.org/abs/1704.04184}{{arXiv:1704.04184}}}
{[hep-th]}
\end{barticle}
\endbibitem

\bibitem[\protect\citeauthoryear{Notari and Riotto}{2002}]{Notari:2002yc}
\begin{barticle}
\bauthor{\bsnm{Notari}, \binits{A.}},
\bauthor{\bsnm{Riotto}, \binits{A.}}:
\batitle{{Isocurvature perturbations in the ekpyrotic universe}}.
\bjtitle{Nucl. Phys. B}
\bvolume{644},
\bfpage{371}--\blpage{382}
(\byear{2002})
\doiurl{10.1016/S0550-3213(02)00765-4}
{\href{https://arxiv.org/abs/hep-th/0205019}{{arXiv:hep-th/0205019}}}
\end{barticle}
\endbibitem

\bibitem[\protect\citeauthoryear{De~Angelis and van~de Bruck}{2023}]{DeAngelis:2023fdu}
\begin{barticle}
\bauthor{\bsnm{De~Angelis}, \binits{M.}},
\bauthor{\bsnm{Bruck}, \binits{C.}}:
\batitle{{Adiabatic and isocurvature perturbations in extended theories with kinetic couplings}}.
\bjtitle{JCAP}
\bvolume{10},
\bfpage{023}
(\byear{2023})
\doiurl{10.1088/1475-7516/2023/10/023}
{\href{https://arxiv.org/abs/2304.12364}{{arXiv:2304.12364}}}
{[hep-th]}
\end{barticle}
\endbibitem

\bibitem[\protect\citeauthoryear{Matsui et~al.}{2019}]{Matsui:2018xwa}
\begin{barticle}
\bauthor{\bsnm{Matsui}, \binits{H.}},
\bauthor{\bsnm{Takahashi}, \binits{F.}},
\bauthor{\bsnm{Yamada}, \binits{M.}}:
\batitle{{Isocurvature Perturbations of Dark Energy and Dark Matter from the Swampland Conjecture}}.
\bjtitle{Phys. Lett. B}
\bvolume{789},
\bfpage{387}--\blpage{392}
(\byear{2019})
\doiurl{10.1016/j.physletb.2018.12.055}
{\href{https://arxiv.org/abs/1809.07286}{{arXiv:1809.07286}}}
{[astro-ph.CO]}
\end{barticle}
\endbibitem

\bibitem[\protect\citeauthoryear{Di~Marco et~al.}{2003}]{DiMarco:2002eb}
\begin{barticle}
\bauthor{\bsnm{Di~Marco}, \binits{F.}},
\bauthor{\bsnm{Finelli}, \binits{F.}},
\bauthor{\bsnm{Brandenberger}, \binits{R.}}:
\batitle{{Adiabatic and isocurvature perturbations for multifield generalized Einstein models}}.
\bjtitle{Phys. Rev. D}
\bvolume{67},
\bfpage{063512}
(\byear{2003})
\doiurl{10.1103/PhysRevD.67.063512}
{\href{https://arxiv.org/abs/astro-ph/0211276}{{arXiv:astro-ph/0211276}}}
\end{barticle}
\endbibitem

\bibitem[\protect\citeauthoryear{Christopherson}{2014}]{Christopherson:2014eoa}
\begin{barticle}
\bauthor{\bsnm{Christopherson}, \binits{A.J.}}:
\batitle{{Cosmological Perturbations: Vorticity, Isocurvature and Magnetic Fields}}.
\bjtitle{Int. J. Mod. Phys. D}
\bvolume{23}(\bissue{11}),
\bfpage{1430024}
(\byear{2014})
\doiurl{10.1142/S0218271814300249}
{\href{https://arxiv.org/abs/1409.4721}{{arXiv:1409.4721}}}
{[astro-ph.CO]}
\end{barticle}
\endbibitem

\bibitem[\protect\citeauthoryear{Langlois et~al.}{2008}]{Langlois:2008vk}
\begin{barticle}
\bauthor{\bsnm{Langlois}, \binits{D.}},
\bauthor{\bsnm{Vernizzi}, \binits{F.}},
\bauthor{\bsnm{Wands}, \binits{D.}}:
\batitle{{Non-linear isocurvature perturbations and non-Gaussianities}}.
\bjtitle{JCAP}
\bvolume{12},
\bfpage{004}
(\byear{2008})
\doiurl{10.1088/1475-7516/2008/12/004}
{\href{https://arxiv.org/abs/0809.4646}{{arXiv:0809.4646}}}
{[astro-ph]}
\end{barticle}
\endbibitem

\bibitem[\protect\citeauthoryear{Hwang and Noh}{2002}]{Hwang:2001fb}
\begin{barticle}
\bauthor{\bsnm{Hwang}, \binits{J.-c.}},
\bauthor{\bsnm{Noh}, \binits{H.}}:
\batitle{{Cosmological perturbations with multiple fluids and fields}}.
\bjtitle{Class. Quant. Grav.}
\bvolume{19},
\bfpage{527}--\blpage{550}
(\byear{2002})
\doiurl{10.1088/0264-9381/19/3/308}
{\href{https://arxiv.org/abs/astro-ph/0103244}{{arXiv:astro-ph/0103244}}}
\end{barticle}
\endbibitem

\bibitem[\protect\citeauthoryear{Gordon et~al.}{2000}]{Gordon:2000hv}
\begin{barticle}
\bauthor{\bsnm{Gordon}, \binits{C.}},
\bauthor{\bsnm{Wands}, \binits{D.}},
\bauthor{\bsnm{Bassett}, \binits{B.A.}},
\bauthor{\bsnm{Maartens}, \binits{R.}}:
\batitle{{Adiabatic and entropy perturbations from inflation}}.
\bjtitle{Phys. Rev. D}
\bvolume{63},
\bfpage{023506}
(\byear{2000})
\doiurl{10.1103/PhysRevD.63.023506}
{\href{https://arxiv.org/abs/astro-ph/0009131}{{arXiv:astro-ph/0009131}}}
\end{barticle}
\endbibitem

\bibitem[\protect\citeauthoryear{Zhang et~al.}{2016}]{Zhang:2016obw}
\begin{barticle}
\bauthor{\bsnm{Zhang}, \binits{K.}},
\bauthor{\bsnm{Wu}, \binits{P.}},
\bauthor{\bsnm{Yu}, \binits{H.}},
\bauthor{\bsnm{Luo}, \binits{L.-W.}}:
\batitle{{Stability of Einstein static state universe in the spatially flat branemodels}}.
\bjtitle{Phys. Lett. B}
\bvolume{758},
\bfpage{37}--\blpage{41}
(\byear{2016})
\doiurl{10.1016/j.physletb.2016.04.049}
\end{barticle}
\endbibitem

\bibitem[\protect\citeauthoryear{Zhang et~al.}{2010}]{Zhang:2010qwa}
\begin{barticle}
\bauthor{\bsnm{Zhang}, \binits{K.}},
\bauthor{\bsnm{Wu}, \binits{P.}},
\bauthor{\bsnm{Yu}, \binits{H.W.}}:
\batitle{{The Stability of Einstein static universe in the DGP braneworld}}.
\bjtitle{Phys. Lett. B}
\bvolume{690},
\bfpage{229}--\blpage{232}
(\byear{2010})
\doiurl{10.1016/j.physletb.2010.05.040}
{\href{https://arxiv.org/abs/1005.4201}{{arXiv:1005.4201}}}
{[gr-qc]}
\end{barticle}
\endbibitem

\bibitem[\protect\citeauthoryear{Guendelman et~al.}{2015}]{Guendelman:2014bva}
\begin{barticle}
\bauthor{\bsnm{Guendelman}, \binits{E.}},
\bauthor{\bsnm{Herrera}, \binits{R.}},
\bauthor{\bsnm{Labrana}, \binits{P.}},
\bauthor{\bsnm{Nissimov}, \binits{E.}},
\bauthor{\bsnm{Pacheva}, \binits{S.}}:
\batitle{{Emergent Cosmology, Inflation and Dark Energy}}.
\bjtitle{Gen. Rel. Grav.}
\bvolume{47}(\bissue{2}),
\bfpage{10}
(\byear{2015})
\doiurl{10.1007/s10714-015-1852-1}
{\href{https://arxiv.org/abs/1408.5344}{{arXiv:1408.5344}}}
{[gr-qc]}
\end{barticle}
\endbibitem

\bibitem[\protect\citeauthoryear{Huang et~al.}{2020}]{Huang:2020oqt}
\begin{barticle}
\bauthor{\bsnm{Huang}, \binits{Q.}},
\bauthor{\bsnm{Xu}, \binits{B.}},
\bauthor{\bsnm{Huang}, \binits{H.}},
\bauthor{\bsnm{Tu}, \binits{F.}},
\bauthor{\bsnm{Zhang}, \binits{R.}}:
\batitle{{Emergent scenario in mimetic gravity}}.
\bjtitle{Class. Quant. Grav.}
\bvolume{37}(\bissue{19}),
\bfpage{195002}
(\byear{2020})
\doiurl{10.1088/1361-6382/aba223}
{\href{https://arxiv.org/abs/2202.04269}{{arXiv:2202.04269}}}
{[gr-qc]}
\end{barticle}
\endbibitem

\bibitem[\protect\citeauthoryear{Guendelman and Herrera}{2024}]{Guendelman:2023jsk}
\begin{barticle}
\bauthor{\bsnm{Guendelman}, \binits{E.}},
\bauthor{\bsnm{Herrera}, \binits{R.}}:
\batitle{{Unification: Emergent universe followed by inflation and dark epochs from multi-field theory}}.
\bjtitle{Annals Phys.}
\bvolume{460},
\bfpage{169566}
(\byear{2024})
\doiurl{10.1016/j.aop.2023.169566}
{\href{https://arxiv.org/abs/2301.10274}{{arXiv:2301.10274}}}
{[gr-qc]}
\end{barticle}
\endbibitem

\bibitem[\protect\citeauthoryear{Hawking and Ellis}{2023}]{hawking2023large}
\begin{bbook}
\bauthor{\bsnm{Hawking}, \binits{S.W.}},
\bauthor{\bsnm{Ellis}, \binits{G.F.}}:
\bbtitle{The Large Scale Structure of Space-time}.
\bpublisher{Cambridge university press}, \blocation{???}
(\byear{2023})
\end{bbook}
\endbibitem

\bibitem[\protect\citeauthoryear{Epstein et~al.}{1965}]{Epstein:1965zza}
\begin{barticle}
\bauthor{\bsnm{Epstein}, \binits{H.}},
\bauthor{\bsnm{Glaser}, \binits{V.}},
\bauthor{\bsnm{Jaffe}, \binits{A.}}:
\batitle{{Nonpositivity of energy density in Quantized field theories}}.
\bjtitle{Nuovo Cim.}
\bvolume{36},
\bfpage{1016}
(\byear{1965})
\doiurl{10.1007/BF02749799}
\end{barticle}
\endbibitem

\bibitem[\protect\citeauthoryear{Roman}{1986}]{Roman:1986tp}
\begin{barticle}
\bauthor{\bsnm{Roman}, \binits{T.A.}}:
\batitle{{Quantum Stress Energy Tensors and the Weak Energy Condition}}.
\bjtitle{Phys. Rev. D}
\bvolume{33},
\bfpage{3526}--\blpage{3533}
(\byear{1986})
\doiurl{10.1103/PhysRevD.33.3526}
\end{barticle}
\endbibitem

\bibitem[\protect\citeauthoryear{Ford and Roman}{1996}]{Ford:1995wg}
\begin{barticle}
\bauthor{\bsnm{Ford}, \binits{L.H.}},
\bauthor{\bsnm{Roman}, \binits{T.A.}}:
\batitle{{Quantum field theory constrains traversable wormhole geometries}}.
\bjtitle{Phys. Rev. D}
\bvolume{53},
\bfpage{5496}--\blpage{5507}
(\byear{1996})
\doiurl{10.1103/PhysRevD.53.5496}
{\href{https://arxiv.org/abs/gr-qc/9510071}{{arXiv:gr-qc/9510071}}}
\end{barticle}
\endbibitem

\bibitem[\protect\citeauthoryear{Graham and Olum}{2003}]{Graham:2002yr}
\begin{barticle}
\bauthor{\bsnm{Graham}, \binits{N.}},
\bauthor{\bsnm{Olum}, \binits{K.D.}}:
\batitle{{Negative energy densities in quantum field theory with a background potential}}.
\bjtitle{Phys. Rev. D}
\bvolume{67},
\bfpage{085014}
(\byear{2003})
\doiurl{10.1103/PhysRevD.69.109901}
{\href{https://arxiv.org/abs/hep-th/0211244}{{arXiv:hep-th/0211244}}}.
\bcomment{[Erratum: Phys.Rev.D 69, 109901 (2004)]}
\end{barticle}
\endbibitem

\bibitem[\protect\citeauthoryear{Nemiroff et~al.}{2015}]{Nemiroff:2014gea}
\begin{barticle}
\bauthor{\bsnm{Nemiroff}, \binits{R.J.}},
\bauthor{\bsnm{Joshi}, \binits{R.}},
\bauthor{\bsnm{Patla}, \binits{B.R.}}:
\batitle{{An exposition on Friedmann Cosmology with Negative Energy Densities}}.
\bjtitle{JCAP}
\bvolume{06},
\bfpage{006}
(\byear{2015})
\doiurl{10.1088/1475-7516/2015/06/006}
{\href{https://arxiv.org/abs/1402.4522}{{arXiv:1402.4522}}}
{[astro-ph.CO]}
\end{barticle}
\endbibitem

\bibitem[\protect\citeauthoryear{Ijjas and Steinhardt}{2019}]{Ijjas:2019pyf}
\begin{barticle}
\bauthor{\bsnm{Ijjas}, \binits{A.}},
\bauthor{\bsnm{Steinhardt}, \binits{P.J.}}:
\batitle{{A new kind of cyclic universe}}.
\bjtitle{Phys. Lett. B}
\bvolume{795},
\bfpage{666}--\blpage{672}
(\byear{2019})
\doiurl{10.1016/j.physletb.2019.06.056}
{\href{https://arxiv.org/abs/1904.08022}{{arXiv:1904.08022}}}
{[gr-qc]}
\end{barticle}
\endbibitem

\bibitem[\protect\citeauthoryear{Wong et~al.}{2019}]{Wong:2019xlj}
\begin{barticle}
\bauthor{\bsnm{Wong}, \binits{W.}},
\bauthor{\bsnm{Ching}, \binits{C.L.}},
\bauthor{\bsnm{Ng}, \binits{W.K.}}:
\batitle{{Rainbow Gravity: Big Bounce in Bianchi Type I Universe}}.
\bjtitle{EPJ Web Conf.}
\bvolume{206},
\bfpage{09012}
(\byear{2019})
\doiurl{10.1051/epjconf/201920609012}
\end{barticle}
\endbibitem

\bibitem[\protect\citeauthoryear{Fay}{2014}]{Fay:2014fta}
\begin{barticle}
\bauthor{\bsnm{Fay}, \binits{S.}}:
\batitle{{From inflation to late time acceleration with a decaying vacuum coupled to radiation or matter}}.
\bjtitle{Phys. Rev. D}
\bvolume{89},
\bfpage{063514}
(\byear{2014})
\doiurl{10.1103/PhysRevD.89.063514}
{\href{https://arxiv.org/abs/1404.3518}{{arXiv:1404.3518}}}
{[gr-qc]}
\end{barticle}
\endbibitem

\bibitem[\protect\citeauthoryear{Sawicki and Vikman}{2013}]{Sawicki:2012pz}
\begin{barticle}
\bauthor{\bsnm{Sawicki}, \binits{I.}},
\bauthor{\bsnm{Vikman}, \binits{A.}}:
\batitle{{Hidden Negative Energies in Strongly Accelerated Universes}}.
\bjtitle{Phys. Rev. D}
\bvolume{87}(\bissue{6}),
\bfpage{067301}
(\byear{2013})
\doiurl{10.1103/PhysRevD.87.067301}
{\href{https://arxiv.org/abs/1209.2961}{{arXiv:1209.2961}}}
{[astro-ph.CO]}
\end{barticle}
\endbibitem

\bibitem[\protect\citeauthoryear{de~la Macorra and German}{2004}]{delaMacorra:2004et}
\begin{barticle}
\bauthor{\bsnm{Macorra}, \binits{A.}},
\bauthor{\bsnm{German}, \binits{G.}}:
\batitle{{Cosmology with negative potentials with w(phi) \ensuremath{<} -1}}.
\bjtitle{Int. J. Mod. Phys. D}
\bvolume{13},
\bfpage{1939}--\blpage{1953}
(\byear{2004})
\doiurl{10.1142/S0218271804006061}
\end{barticle}
\endbibitem

\bibitem[\protect\citeauthoryear{{Huang}}{1990}]{1990JMP}
\begin{barticle}
\bauthor{\bsnm{{Huang}}, \binits{W.-H.}}:
\batitle{{Anisotropic cosmological models with energy density dependent bulk viscosity.}}
\bjtitle{J. Math. Phys.}
\bvolume{31}(\bissue{6}),
\bfpage{1456}--\blpage{1462}
(\byear{1990})
\doiurl{10.1063/1.528736}
{\href{https://arxiv.org/abs/gr-qc/0308059}{{arXiv:gr-qc/0308059}}}
{[gr-qc]}
\end{barticle}
\endbibitem

\bibitem[\protect\citeauthoryear{Barrow}{2004a}]{Barrow:2004xh}
\begin{barticle}
\bauthor{\bsnm{Barrow}, \binits{J.D.}}:
\batitle{{Sudden future singularities}}.
\bjtitle{Class. Quant. Grav.}
\bvolume{21},
\bfpage{79}--\blpage{82}
(\byear{2004})
\doiurl{10.1088/0264-9381/21/11/L03}
{\href{https://arxiv.org/abs/gr-qc/0403084}{{arXiv:gr-qc/0403084}}}
\end{barticle}
\endbibitem

\bibitem[\protect\citeauthoryear{Barrow}{2004b}]{Barrow:2004hk}
\begin{barticle}
\bauthor{\bsnm{Barrow}, \binits{J.D.}}:
\batitle{{More general sudden singularities}}.
\bjtitle{Class. Quant. Grav.}
\bvolume{21},
\bfpage{5619}--\blpage{5622}
(\byear{2004})
\doiurl{10.1088/0264-9381/21/23/020}
{\href{https://arxiv.org/abs/gr-qc/0409062}{{arXiv:gr-qc/0409062}}}
\end{barticle}
\endbibitem

\bibitem[\protect\citeauthoryear{Shtanov and Sahni}{2002}]{Shtanov:2002ek}
\begin{barticle}
\bauthor{\bsnm{Shtanov}, \binits{Y.}},
\bauthor{\bsnm{Sahni}, \binits{V.}}:
\batitle{{Unusual cosmological singularities in brane world models}}.
\bjtitle{Class. Quant. Grav.}
\bvolume{19},
\bfpage{101}--\blpage{107}
(\byear{2002})
\doiurl{10.1088/0264-9381/19/11/102}
{\href{https://arxiv.org/abs/gr-qc/0204040}{{arXiv:gr-qc/0204040}}}
\end{barticle}
\endbibitem

\bibitem[\protect\citeauthoryear{Heydarzade}{2019}]{Heydarzade:2019dpf}
\begin{barticle}
\bauthor{\bsnm{Heydarzade}, \binits{Y.}}:
\batitle{{Cosmological Singularities in Conformal Weyl Gravity}}.
\bjtitle{Eur. Phys. J. C}
\bvolume{79}(\bissue{11}),
\bfpage{923}
(\byear{2019})
\doiurl{10.1140/epjc/s10052-019-7446-4}
{\href{https://arxiv.org/abs/1910.12149}{{arXiv:1910.12149}}}
{[gr-qc]}
\end{barticle}
\endbibitem

\bibitem[\protect\citeauthoryear{Visser}{2004}]{Visser:2003vq}
\begin{barticle}
\bauthor{\bsnm{Visser}, \binits{M.}}:
\batitle{{Jerk and the cosmological equation of state}}.
\bjtitle{Class. Quant. Grav.}
\bvolume{21},
\bfpage{2603}--\blpage{2616}
(\byear{2004})
\doiurl{10.1088/0264-9381/21/11/006}
{\href{https://arxiv.org/abs/gr-qc/0309109}{{arXiv:gr-qc/0309109}}}
\end{barticle}
\endbibitem

\end{thebibliography}

\end{document}